\documentclass[useAMS,usenatbib]{mn2e}

\usepackage{graphicx}
\usepackage{epsfig}
\usepackage{natbib}
\usepackage[section] {placeins}
\newcommand{\fesc}{\ifmmode{f_{\rm esc}}\else{$f_{\rm esc}$}\fi}
\newcommand{\fescs}{\ifmmode{f_{\rm esc}^\star}\else{$f_{\rm esc}^\star$}\fi}
\newcommand{\kms}{\ifmmode{{\rm km~s^{-1}}}\else{km s$^{-1}$}\fi}
\newcommand{\cubecm}{\ifmmode{{\rm cm^{-3}}}\else{cm$^{-3}$}\fi}
\newcommand{\lsim}{\lower0.3em\hbox{$\,\buildrel <\over\sim\,$}}
\newcommand{\gsim}{\lower0.3em\hbox{$\,\buildrel >\over\sim\,$}}

\newcommand{\enzo}{{\sc enzo}}
\newcommand{\moray}{{\sc enzo+moray}}
\newcommand{\Ms}{\ifmmode{M_\odot}\else{$M_\odot$}\fi}

\newcommand{\hh}{H$_2$}

\newcommand{\tvir}{\ifmmode{T_{\rm{vir}}}\else{$T_{\rm{vir}}$}\fi}
\newcommand{\mvir}{\ifmmode{M_{\rm{vir}}}\else{$M_{\rm{vir}}$}\fi}
\newcommand{\rvir}{\ifmmode{r_{\rm{vir}}}\else{$r_{\rm{vir}}$}\fi}

\newcommand{\jj}{\ifmmode{J_{21}}\else{$J_{21}$}\fi}
\newcommand{\flw}{\ifmmode{F_{LW}}\else{$F_{LW}$}\fi}
\newcommand{\kph}{\ifmmode{k_{\rm ph}}\else{$k_{\rm ph}$}\fi}

\newcommand\tento[1]{$10^{#1}$}
\newcommand\step[1]{\textit{Step #1}.--}
\newcommand{\hi}{H {\sc i}~}
\newcommand{\hii}{H {\sc ii}~}

\newcommand\ion[2]{#1$\;${\small\rmfamily\@Roman{#2}}\relax}
\def\eps@scaling{1.0}%
\newcommand\epsscale[1]{\gdef\eps@scaling{#1}}%
\newcommand\plotone[1]{%
 \centering 
 \leavevmode 
 \includegraphics[width={\eps@scaling\columnwidth}]{#1}%
}%
\newcommand\plottwo[2]{%
 \centering 
 \includegraphics[width={\eps@scaling\columnwidth}]{#1}\\%
 \includegraphics[width={\eps@scaling\columnwidth}]{#2}%
}%

\begin{document}

\title[AMR Simulations with Adaptive Ray Tracing] {Enzo+Moray:
  Radiation Hydrodynamics Adaptive Mesh Refinement Simulations with
  Adaptive Ray Tracing}

\author[J. H. Wise and T. Abel] {John H. Wise$^1$\thanks{Hubble
    Fellow; e-mail: jwise@astro.princeton.edu} and Tom Abel$^2$\\
$^{1}$ Department of Astrophysical Sciences, Princeton University,
Peyton Hall, Ivy Lane, Princeton, NJ 08544, USA\\
$^{2}$ Kavli Institute for Particle Astrophysics and Cosmology,
Stanford University, Menlo Park, CA 94025, USA}

\pagerange{\pageref{firstpage}--\pageref{lastpage}} \pubyear{2010}

\maketitle
\label{firstpage}

\begin{abstract}

  We describe a photon-conserving radiative transfer algorithm, using
  a spatially-adaptive ray tracing scheme, and its parallel
  implementation into the adaptive mesh refinement (AMR) cosmological
  hydrodynamics code, \enzo.  By coupling the solver with the energy
  equation and non-equilibrium chemistry network, our radiation
  hydrodynamics framework can be utilised to study a broad range of
  astrophysical problems, such as stellar and black hole (BH)
  feedback.  Inaccuracies can arise from large timesteps and poor
  sampling, therefore we devised an adaptive time-stepping scheme and
  a fast approximation of the optically-thin radiation field with
  multiple sources.  We test the method with several radiative
  transfer and radiation hydrodynamics tests that are given in
  \citet{RT06, Iliev09}.  We further test our method with more
  dynamical situations, for example, the propagation of an ionisation
  front through a Rayleigh-Taylor instability, time-varying
  luminosities, and collimated radiation.  The test suite also
  includes an expanding \hii region in a magnetised medium, utilising
  the newly implemented magnetohydrodynamics module in \enzo.  This
  method linearly scales with the number of point sources and number
  of grid cells.  Our implementation is scalable to 512 processors on
  distributed memory machines and can include radiation pressure and
  secondary ionisations from X-ray radiation.  It is included in the
  newest public release of \enzo.
  
\end{abstract}

\begin{keywords}
  cosmology --- methods: numerical --- hydrodynamics --- radiative
  transfer.
\end{keywords}

\section{Introduction}

Radiation from stars and black holes strongly affects their
surroundings and plays a crucial role in topics such as stellar
atmospheres, the interstellar medium (ISM), star formation, galaxy
formation, supernovae (SNe) and cosmology.  It is a well-studied
problem \citep[e.g.][]{Mathews65, Rybicki, Mihalas84, Yorke86};
however, its treatment in multi-dimensional calculations is difficult
because of the dependence on seven variables -- three spatial, two
angular, frequency, and time.  The non-local nature of the thermal and
hydrodynamical response to radiation sources further adds to the
difficulty.  Depending on the problem of interest some simplifying
assumptions may be made.

An important case was considered by \citet{Stroemgren39} for an
ultraviolet (UV) radiation source photo-ionising a static uniform
neutral medium.  When recombinations balance photo-ionisations, the
radius of a so-called \hii region,
\begin{equation}
  \label{eqn:Rstr}
  R_s = \left(\frac{3\dot{N}_\gamma}{4\pi \alpha_{\rm B} n_{\rm
        H}^2}\right)^{1/3},
\end{equation}
where $\dot{N}_\gamma$ is the ionising photon luminosity, $\alpha_{\rm
  B}$ is the recombination rate, and $n_H$ is the ambient hydrogen
number density.  Furthermore he found that the delineation between the
neutral and ionised medium to be approximately the mean free path of
the ionising radiation.  His seminal work was expanded upon by
\citet{Spitzer48, Spitzer49, Spitzer54} and \citet{Spitzer50}, who
showed that the ionising radiation heated the medium to $T \sim 10^4$
K.  If the density is equal on both sides of the ionisation front,
then this over-pressurised region would expand and drive a shock
outwards \citep[e.g.][]{Oort54, Schatzman55}.  These early works
provided the basis for the modern topic of radiation hydrodynamics of
the ISM.  A decade later, the first radiation hydrodynamical numerical
models of \hii regions in spherical symmetry and plane-parallel
ionisation fronts were developed \citep[e.g.][]{Mathews65, Lasker66,
  Hjellming66}.  They described the expansion of the ionisation front
and the evolution of its associated shock wave that carries most of
the gas away from the source.  At the same time, theoretical models of
ionisation fronts matured and were classified by \citet{Kahn54} and
\citet{Axford61} as either R-type (rare) or D-type (dense).  In R-type
fronts, the ionised gas density is higher than the neutral gas
density, and in D-type fronts, the opposite is true.  R-type fronts
travel supersonically with respect to the neutral gas, whereas D-type
fronts are subsonic.  Furthermore ``weak'' and ``strong'' R-type
fronts move supersonically and subsonically with respect to the
ionised gas, respectively.  The same terminology conversely applies to
D-type fronts.  ``Critical'' fronts are defined as moving exactly at
the sound speed.  These works established the evolutionary track of an
expanding \hii illuminated by a massive star in a uniform medium:
\begin{enumerate}
\item Weak R-type:  When the star (gradually) starts to shine, the
  ionisation front will move supersonically through the ambient
  medium.  The gas is heated and ionised, but otherwise left
  undisturbed.  This stage continues until $r \sim 0.02R_s$.
\item Critical R-type:  As the ionisation front moves outwards, it
  begins to slow because of the geometric dilution of the radiation.
  It becomes a critical R-type front, which is equivalent to an
  isothermal shock in the neutral gas.
\item Strong and weak D-type:  The front continues to slow, becoming a
  strong D-type front, and then a critical D-type front.  From this
  point forward, it is moving subsonically with respect to the ionised
  gas, i.e. a weak D-type front.  Thus sound waves can travel across
  the ionisation front and form a shock.  The ionisation front
  detaches from the shock, putting the shock ahead of the ionisation
  front.
\item Expansion phase:  After the shock forms, the \hii region
  starts to expand, lowering the interior density and thus the
  recombination rates.  This increases the number of photons available
  for ionising the gas.  The sphere expands until it reaches pressure
  equilibrium with the ambient medium at $r \sim 5R_s$.
\end{enumerate}

In the 1970's and 1980's, algorithmic and computational advances
allowed numerical models to be expanded to two dimensions, mainly
using axi-symmetric to simplify the problem
\citep[e.g.][]{Bodenheimer79, Sandford82, Yorke83}.  One topic that was
studied extensively were champagne flows.  Here the source is embedded
in an overdense region, and the \hii region escapes from this
region in one direction.  The interface between the ambient and dense
medium was usually set up to be a constant pressure boundary.  When
the ionisation front passes this boundary, the dense, ionised gas is
orders of magnitude out of pressure equilibrium as the temperatures on
both sides of initial boundary are within a factor of a few.  In
response, the gas is accelerated outwards in this direction, creating
a fan-shaped outflow.

Only in the past 15 years, computational resources have become large
enough, along with further algorithmic advances, to cope with the
requirements of three-dimensional calculations.  There are two popular
methods to solve the radiative transfer equation in three-dimensions:
\begin{itemize}
\item Moment methods: The angular moments of the radiation field can
  describe its angular structure, which are related to the energy
  energy, flux, and radiation pressure \citep{Auer70, Norman98}.
  These have been implemented in conjunction with short
  characteristics \citep[2D]{Stone92_RHD}, with long characteristics
  \citep{Finlator09}, with a variable Eddington tensor in the
  optically thin limit \citep[OTVET;][]{Gnedin01_OTVET, Petkova09},
  and with an M1 closure relation \citep{Gonzalez07, Aubert08}.
  Moment methods have the advantage of being fast and independent of
  the number of radiation sources.  However, they are diffusive and
  result in incorrect shadows in some situations.
\item Ray tracing:  Radiation can be propagated along rays that extend
  through the computational grid \citep[e.g.][]{Razoumov99, Abel99_RT,
    Ciardi01, Sokasian01, Whalen06, Rijkhorst06, Mellema06, Alvarez06,
    Trac07, Krumholz07_ART, Paardekooper10} or particle set
  \citep[e.g.][]{Susa06, Johnson07, Pawlik08, Pawlik10, Altay08,
    Hasegawa09}.  In general, these methods are very accurate but
  computationally expensive because the radiation field must be well
  sampled by the rays with respect to the spatial resolution of the
  grid or particles.
\end{itemize}
Until the mid-2000's the vast majority of the three-dimensional
calculations were performed with static density distributions.  One
example is calculating cosmological reionisation by post-processing of
density fields from N-body simulations \citep{Ciardi01, Sokasian01,
  McQuinn07, Iliev06, Iliev07}.  Any hydrodynamical response to the
radiation field was thus ignored.  Several radiative transfer codes
were compared in four purely radiative transfer tests in
\citet[hereafter RT06]{RT06}.  Only recently has the radiative
transfer equation been coupled to the hydrodynamics in
three-dimensions \citep[e.g.][]{Krumholz07_FLD}.  In the second
comparison paper \citep[hereafter RT09]{Iliev09}, results from these
radiation hydrodynamics codes were compared.  Even more rare are ones
that couple it with magneto-hydrodynamics
\citep[e.g.][]{Krumholz07_ART}.  The tests in RT06 and RT09 were kept
relatively simple to ease the comparison.

In this paper, we present our implementation, \moray, of adaptive ray
tracing \citep{Abel02_RT} in the cosmological hydrodynamics adaptive
mesh refinement (AMR) code, \enzo~\citep{BryanNorman1997, OShea2004}.
The radiation field is coupled to the hydrodynamics solver at small
time-scales, enabling it to study radiation hydrodynamical problems.
We have used this code to investigate the growth of an \hii region
from a 100\Ms~Population III (Pop III) star \citep{Abel07}, the early
stages of reionisation from Pop III stars \citep{Wise08_Reion},
radiative feedback on the formation of high redshift dwarf galaxies
\citep{Wise08_Gal, Wise10_Gal}, ultraviolet radiation escape fractions
from dwarf galaxies before reionisation \citep{Wise09}, negative
radiative feedback from accreting Pop III seed black holes
\citep{Alvarez09}, and radiative feedback in accreting supermassive
black holes \citep[][in prep.]{Kim11}.  We have included \moray~in the
latest public release of \enzo\footnote{http://enzo.googlecode.com},
and it is also coupled with the newly added MHD solver in
\enzo~\citep{Wang09}.

We have structured this paper as follows.  In Section 2, we describe
the mathematical connections between adaptive ray tracing and the
radiative transfer equation.  Furthermore, we detail how physics other
than photo-ionisation and photo-heating are included.  We then derive a
geometric correction factor to any ray tracing method to improve
accuracy.  We end the section by describing a new computational
technique to approximate an optically-thin radiation field with ray
tracing and multiple sources.  In Section 3, we cover the details of
our radiation hydrodynamics implementation in \enzo, specifically (1)
the ray tracing algorithms, (2) coupling with the hydrodynamics
solver, (3) several methods to calculate the radiative transfer
timestep, and (4) our parallelisation strategy.  We present our
results from the RT06 radiative transfer tests in Section 4.
Afterwards in Section 5, we show the results from the RT09 radiation
hydrodynamics tests.  In Section 6, we expand on these tests to
include more dynamical and complex setups to demonstrate the
flexibility and high fidelity of \moray.  Section 7 gives the results
from spatial, angular, frequency, and temporal resolution tests.  In
Section 8, we illustrate the improvements from the geometric
correction factor and our optically-thin approximation.  We also show
the effects of X-ray radiation and radiation pressure in this section.
Finally in Section 9, we demonstrate the parallel scalability of
\moray.  Last Section 10 summarises our method and results.

\section{Treatment of Radiative Transfer}
\label{sec:rt}

Radiation transport is a well-studied topic, and we begin by
describing our approach in solving the radiative transfer equation,
which in comoving coordinates \citep{Gnedin97} is
\begin{equation}
  \label{eqn:rteqn}
  \frac{1}{c} \; \frac{\partial I_\nu}{\partial t} + 
  \frac{\hat{n} \cdot \nabla I_\nu}{\bar{a}} -
  \frac{H}{c} \; \left( \nu \frac{\partial I_\nu}{\partial \nu} -
  3 I_\nu \right) = -\kappa_\nu I_\nu + j_\nu .
\end{equation}
Here $I_\nu \equiv I(\nu, \mathbf{x}, \Omega, t)$ is the radiation
specific intensity in units of energy per time $t$ per solid angle per
unit area per frequency $\nu$.  $H = \dot{a}/a$ is the Hubble
constant, where $a$ is the scale factor.  $\bar{a} = a/a_{em}$ is the
ratio of scale factors at the current time and time of emission.  The
second term represents the propagation of radiation, where the factor
$1/\bar{a}$ accounts for cosmic expansion.  The third term describes
both the cosmological redshift and dilution of radiation.  On the
right hand side, the first term considers the absorption coefficient
$\kappa_\nu \equiv \kappa_\nu(\mathbf{x},\nu,t)$, and the second term
$j_\nu \equiv j_\nu(\mathbf{x},\nu,t)$ is the emission coefficient
that includes any point sources of radiation or diffuse radiation.  We
neglect any $(v/c)$ terms in equation (\ref{eqn:rteqn}) that become
important in the dynamic diffusion limit ($l \kappa (v/c) \gg 1$),
where $l$ is the characteristic size of the system.  This occurs in
relativistic flows or very optically thick systems, such as stellar
interiors or radiation-dominated shocks \citep[see][for a rigorous
derivation that includes $(v/c)$ terms to
second-order]{Krumholz07_FLD}.

Solving this equation is difficult because of its high dimensionality;
however, we can make some appropriate approximations to reduce its
complexity in order to include radiation transport in numerical
calculations.  Typically timesteps in dynamic calculations are small
enough so that $\Delta a/a \ll 1$, therefore $\bar{a} = 1$ in any
given timestep, reducing the second term to $\hat{n} \partial
I_\nu/\partial \mathbf{x}$.  To determine the importance of the third
term, we evaluate the ratio of the third term to the second term.
This is $HL/c$, where $L$ is the simulation box length.  If this ratio
is $\ll 1$, we can ignore the third term.  For example at $z=5$, this
ratio is 0.1 when $L = c/H(z=5)$ = 53 proper Mpc.  In large boxes
where the light crossing time is comparable to the Hubble time, then
it becomes important to consider cosmological redshifting and dilution
of the radiation.  Thus equation (\ref{eqn:rteqn}) reduces to the
non-cosmological form in this local approximation,
\begin{equation}
  \frac{1}{c} \frac{\partial I_\nu}{\partial t} + 
  \hat{n} \frac{\partial I_\nu}{\partial \mathbf{x}} =
  -\kappa_\nu I_\nu + j_\nu .
\end{equation}
We choose to represent the source term $j_\nu$ as point sources of
radiation (e.g. stars, quasars) that emit radial rays that are
propagated along the direction $\hat{n}$.  Next we describe this
discretisation and its contribution to the radiation field.

\begin{table}
\caption{Variable definitions used in Sections \ref{sec:rt} and
  \ref{sec:numerical}}
\label{tab:notation}
\begin{tabular}{ll}
\hline
Variable & Description \\
\hline
$A_{\rm cell}$ & Cell face area \\
$a$ & Scale factor \\
$C_{\rm H}$ & Collisional ionization rate of hydrogen \\
$C_{\rm RT,cfl}$ & CFL safety factor in timestep calculation \\
$D_{\rm c,i}$ & Distance from ray segment center to cell center \\
$D_{\rm edge}$ & Distance from ray segment center to cell edge \\
$dP$ & Photon loss from absorption \\
$dP_C$ & Photon loss from Compton Scattering \\
$d\mathbf{p}_\gamma$ & Momentum change from radiation pressure \\
$dt_P$ & Photon timestep \\
$E_{\rm ph}$ & Photon energy \\
$E_i$ & Ionization energy of absorber \\
$f_c$ & Geometric correction factor \\
$f_{\rm shield}$ & Shielding function for \hh \\
$H$ & Hubble constant \\
$I_\nu$ & Specific intensity \\
$j_\nu$ & Emission coefficient \\
$k_{\rm ph}$ & Photo-ionization rate \\
$k_{\rm diss}$ & Photo-dissociation rate of \hh \\
$L_{\rm pix}$ & Linear width of a HEALPix pixel \\
$l$ & HEALPix level \\
$N_{\rm H2}$ & Column density of \hh \\
$n_{\rm abs}$ & Number density of absorber \\
$N_{\rm pix}(l)$ & HEALPix pixels on level $l$ \\
$N_{\rm ray}$ & Rays per cell \\
$\dot{N}_\gamma$ & Photon luminosity \\
$\hat{n}$ & Normal direction of radiation \\
$n_x$ & Number density of absorber $x$ \\
$P$ & Photon flux \\
$r$ & Radius \\
$r_{\rm cell}$ & Distance from the radiation source to the \\
& cell center\\
$r_{\rm ray}$ & Distance from the radiation source to the \\
& next cell boundary crossing \\
$V_{\rm cell}$ & Cell volume \\
$v_{\rm IF}$ & Ionization front velocity \\
$x_{\rm 0,i}$ & Cell center coordinates \\
$x_{\rm cell,i}$ & Next cell boundary crossing in the \\
& $i$-th dimension\\
$x_{\rm src,i}$ & Source position in the $i$-th dimension \\
$Y_{\rm x}$ & Secondary ionization factors \\
$\alpha_B$ & Case-B recombination rate \\
$\Gamma_{\rm ph}$ & Photo-heating rate \\
$\Delta x$ & Cell width \\
$\Phi_c$ & Angular resolution in units of rays per cell area \\
$\kappa_\nu$ & Absorption coefficient \\
$\lambda_{\rm mfp}$ & Mean free path \\
$\Omega_{\rm ray}$ & Solid angle associated with a ray \\
$\sigma_{\rm abs}$ & Cross-section of absorber \\
$\theta_{\rm ray}$ & Angle associated with a ray \\
$\tau$ & Optical depth \\
\hline
\end{tabular}
\end{table}

\subsection{Adaptive Ray Tracing}
\label{sec:ART}

Ray tracing is an accurate method to propagate radiation from point
sources on a computational grid, given that there are a sufficient
number of rays passing through each cell.  Along a ray, the radiative
transfer equation reads
\begin{equation}
\label{eqn:rtray}
\frac{1}{c} \frac{\partial P}{\partial t} + \frac{\partial P}{\partial
  r} = -\kappa P,
\end{equation}
where $P$ is the photon number flux along the ray.  To sample the
radiation field at large radii, ray tracing requires at least $N_{ray}
= 4\pi R^2 / (\Delta x)^2$ rays to sample each cell with one ray,
where $R$ is the radius from the source to the cell and $\Delta x$ is
the cell width.  If one were to trace $N_{ray}$ rays out to $R$, the
cells at a smaller radius $r$ would be sampled by, on average,
$(r/R)^2$ rays, which is computationally wasteful because only a few
rays per cell, as we will show later, provides an accurate calculation
of the radiation field.

We avoid this inefficiency by utilising adaptive ray tracing
\citep{Abel02_RT}, which progressively splits rays when the sampling
becomes too coarse and is based on Hierarchical Equal Area isoLatitude
Pixelation \citep[HEALPix;][]{HEALPix}.  In this scheme, the rays are
traced along normal directions of the centres of HEALPix pixels, which
evenly divides a sphere into equal areas.  The rays are initialised at
each point source with the photon luminosity (ph s$^{-1}$) equally
spread across $N_{\rm pix} = 12 \times 4^l$ rays, where $l$ is the
initial HEALPix level.  We usually find $l$ = 0 or 1 is sufficient
because these coarse rays will usually be split before traversing
the first cell.

The rays are traced through the grid in a typical fashion
\citep[e.g.][]{Abel99_RT}, in which we calculate the next cell
boundary crossing.  The ray segment length crossing the cell is
\begin{equation}
  \label{eqn:trace}
  dr = R_0 - \min_{i=1 \rightarrow 3} \left[(x_{\rm cell,i} - x_{\rm src,i}) /
    \hat{n}_{\rm ray,i} \right],
\end{equation}
where $R_0$, $\hat{n}_{\rm ray}$, $x_{\rm cell,i}$, and $x_{\rm
  src,i}$ are the initial distance travelled by the ray, normal
direction of the ray, the next cell boundary crossing in the $i$-th
dimension, and the position of the point source that emitted the ray,
respectively.  However before the ray travels across the cell, we
evaluate the ratio of the face area $A_{\rm cell}$ of the current cell
and the solid angle $\Omega_{\rm ray}$ of the ray,
\begin{equation}
  \label{eqn:split}
  \Phi_c = \frac{A_{\rm cell}} {\Omega_{\rm ray}} = 
  \frac{N_{\rm pix} (\Delta x)^2} {4\pi R_0^2}.
\end{equation}
If $\Phi_c$ is less than a pre-determined value (usually $>3$), the
ray is split into 4 child rays.  We investigate the variations in
solutions with $\Phi_c$ in \S\ref{sec:ang_dep}.  The pixel numbers
of the child rays $p^\prime$ are given by the ``nested'' scheme of
HEALPix at the next level, i.e. $p^\prime = 4 \times p + [0,1,2,3]$,
where $p$ is the original pixel number.  The child rays (1) acquire
the new normal vectors of the pixels, (2) retain the same radius of
the parent ray, and (3) gets a quarter of the photon flux of the
parent ray.  Afterwards the parent ray is discontinued.

A ray propagates and splits until 
\begin{enumerate}
\item the photon has travelled $c \times dt_P$, where $dt_P$ is the
  radiative transfer timestep,
\item its photon flux is almost fully absorbed ($>99.9\%$) in a single
  cell, which significantly reduces the computational time if the
  radiation volume filling fraction is small,
\item the photon leaves the computational domain with isolated
  boundary conditions, or
\item the photon travels $\sqrt{3}$ of the simulation box length with
  periodic boundary conditions.
\end{enumerate}
In the first case, the photon is halted at that position and saved,
where it will be considered in the solution of $I_\nu$ at the next
timestep.  In the next timestep, the photon will encounter a different
hydrodynamical and ionisation state, hence $\kappa$, in its path.
Furthermore any time variations of the luminosities will be retained
in the radiation field.  This is how this method retains the time
derivative of the radiative transfer equation.  The last restriction
prevents our method from considering sources external to the
computational domain, but a uniform radiation background can be used
in conjunction with ray tracing in \moray~that adds the local
radiation field to the background intensity.



\subsection{Radiation Field}

The radiation field is calculated by integrating equation
(\ref{eqn:rtray}) along each ray, which is done by considering the
discretisation of the ray into segments.  In the following section, we
assume the rays are monochromatic.  For convenience, we express the
integration in terms of optical depth $\tau = \int \kappa(r,t) \; dr$,
and for a ray segment,
\begin{equation}
  \label{eqn:dtau}
  d\tau = \sigma_{\rm abs}(\nu) n_{\rm abs} dr.
\end{equation}
Here $\sigma_{\rm abs}$ and $n_{\rm abs}$ are the cross section and
number density of the absorbing medium, respectively.  We use the
cell-centred density in our calculations.  Using trilinearly
interpolated densities \citep[see][]{Mellema06} did not produce
improved results.  In the static case, equation (\ref{eqn:rtray}) has
a simple exponential analytic solution, and the photon flux of a ray
is reduced by
\begin{equation}
  \label{eqn:flux}
  dP = P \times (1 - e^{-\tau})
\end{equation}
as it crosses a cell.  We equate the photo-ionisation rate to the
absorption rate, resulting in photon conservation \citep{Abel99_RT,
  Mellema06}.  Thus the photo-ionisation $k_{\rm ph}$ and
photo-heating $\Gamma_{\rm ph}$ rates associated with a single ray are
\begin{equation}
  \label{eqn:kph}
  k_{\rm ph} = \frac{P (1 - e^{-\tau})}{n_{\rm abs} \; V_{\rm cell} \; dt_P},
\end{equation}
\begin{equation}
  \label{eqn:gamma}
  \Gamma_{\rm ph} = k_{\rm ph} \; (E_{\rm ph} - E_i),
\end{equation}
where $V_{\rm cell}$ is the cell volume, $E_{\rm ph}$ is the photon
energy, and $E_i$ is the ionisation energy of the absorbing material.
In each cell, the photo-ionisation and photo-heating rates from each
ray in the calculation are summed, and after the ray tracing is
complete, these rates can be used to update the ionisation state and
energy of the cells.  Considering a system with only hydrogen
photo-ionisations and radiative recombinations, these changes are very
straightforward and is useful for illustrative purposes.  The change
in neutral hydrogen is
\begin{equation}
\label{eqn:dndt}
\frac{dn_{\rm H}}{dt} = \alpha_B n_e n_p - C_{\rm H} n_e n_{\rm
  H} - k_{\rm ph},
\end{equation}
where $\alpha_B = 2.59 \times 10^{-13} \mathrm{cm}^3 \;
\mathrm{s}^{-1}$ is the recombination coefficient at 10$^4$ K in the
Case B on-the-spot approximation in which all recombinations are
locally reabsorbed, \citep{Spitzer78}, and $C_{\rm H}$ is the
collisional ionisation rate.  However, for more accurate solutions in
calculations that consider several chemical species, the
photo-ionisation rates are terms in the relevant chemical networks
\citep[e.g.][]{Abel97}.

\subsection{Geometric Corrections}
\label{sec:meth_fc}

\begin{figure}
  \plotone{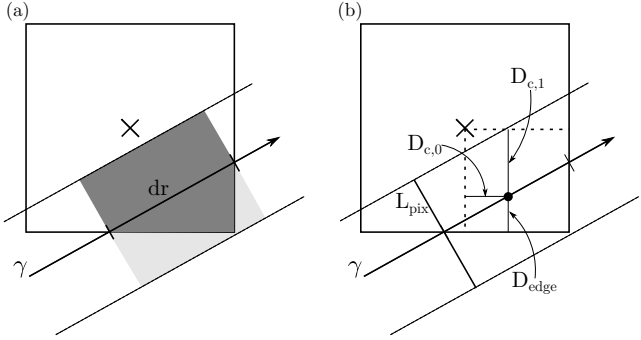}
  \caption{\label{fig:covering} (a) A two-dimensional illustration of
    the overlap between the beam associated with a ray $\gamma$ and a
    computational cell.  The ray has a segment length of $dr$ passing
    through the cell.  The covering area is denoted by dark grey,
    where the full area $(dr \times L_{\rm pix})$ is colored by the
    dark and light grey.  The photo-ionisation and photo-heating rates
    should be corrected by the ratio of these areas, i.e. the overlap
    fraction $f_c$.  (b) Annotation of quantities used in this
    geometric correction.}
\end{figure}

For a ray tracing method to accurately, i.e. without non-spherical
artifacts, compute the radiation field, the computational grid must be
well-sampled by the rays.  The main source of potential artifacts is
the geometrical difference between the cell and the HEALPix pixel
relevant in the angular integration of the intensity over the cell.
In this section, we devise a correction scheme to account for these
differences.  Consider the solid angle $\Omega_{\rm ray}$ and photon
flux $P$ associated with a single ray, and assume the flux is constant
across $\Omega_{\rm ray}$.  There exists a discrepancy between the
geometry cell face and HEALPix pixel when the pixel does not cover the
entire cell face, which is illustrated in Fig. \ref{fig:covering}.
This mismatch causes non-spherical artifacts and is most apparent in
the optically thin case, where the area of the pixel is dominant over
$(1 - e^{-\tau})$ when calculating $k_{\rm ph}$.  One can avoid these
artifacts by increasing the sampling $\Phi_c$ to high values ($>10$)
but we have formulated a simple geometric correction to the
calculation of the radiation field.  This correction is not unique to
the HEALPix formalism but can be applied to any type of pixelisation.

The contribution to $k_{\rm ph}$ and $\Gamma_{\rm ph}$ must be
corrected by the covering factor $f_c$ of the pixel with respect to
the cell.  When the pixel is fully contained within the cell face,
$f_c \equiv 1$.  Because the geometry of the pixel can be complex with
curved edges, we approximate $f_c$ by assuming the pixel is square.
The covering factor is thus related to the width of a pixel, $L_{\rm
  pix} = R_0 \, \theta_{\rm pix}$, and the distance from the ray
segment midpoint to the closest cell boundary $D_{\rm c,i}$, which is
depicted in Fig. \ref{fig:covering}.  To estimate $f_c$, we first
find the distance $d_{\rm centre,i}$ from the midpoint of the ray
segment to the cell centre $x_{\rm 0,i}$ in orthogonal directions,
\begin{equation}
  \label{eqn:midpoint}
  D_{\rm c,i} = \left| R_{\rm 0,i} + \hat{n}_i\,\frac{dr}{2} - x_{\rm
      0,i} \right|,
\end{equation}
where $R_{\rm 0,i}$ is the distance travelled by the ray in each
orthogonal direction.  The distance to the closest cell boundary is
$D_{\rm edge} = dx/2 - \max_{i=1 \rightarrow 3} (D_{\rm c,i})$.  Thus
the covering factor is related to the square of the ratio between the
$L_{\rm pix}$ and $D_{\rm edge}$ by
\begin{equation}
  \label{eqn:fc}
  f_c = \left( \frac{1}{2} + \frac{D_{\rm edge}}{L_{\rm pix}} \right)^2
\end{equation}
One half of the pixel is always contained within the cell, which
results in the factor of 1/2.  Finally we multiply $k_{\rm ph}$ and
$\Gamma_{\rm ph}$ by $f_c$ but leave the absorbed radiation $dP$
untouched because this would underestimate the attenuation of the
incoming radiation.  Using $f_c$ calculated like above, the method is
no longer photon conserving.  In our implementation, we felt that the
spherical symmetry obtained outweighed the loss of photon
conservation.  However we show that there are no perceptible
deviations from photon conservation in \S\ref{sec:test1} and
\S\ref{sec:dx_dep}.

We briefly next describe how to retain photon conservation with a
geometric correction.  Notice that we compute $f_c$ by only
considering the distances in orthogonal directions.  A better estimate
would consider the distance between the cell boundary and ray segment
midpoint in the direction between the midpoint and cell centre of
$\mathbf{x}_{\rm mid} - \mathbf{x}_0$.  We find that the method
outlined here provides a sufficient correction factor to avoid any
non-spherical artifacts and deviations from photon conservation.
Furthermore in principle, the ray should also contribute to any
neighboring cells that overlap with $\Omega_{\rm ray}$, which is the
key to be photon conservative with such a geometric correction.

\subsection{Optically Thin Approximation}

In an optically thin medium, radiation is only attenuated by geometric
dilution in the local approximation to Equation (\ref{eqn:rteqn}),
i.e. the inverse square law.  With such a simple solution, the tracing
of rays is wasteful, however these rays must be propagated because the
the medium farther away can be optically thick.  Here we describe a
method that minimizes the computational work of ray tracing in the
optically thin regime by exploiting this fact.  Each ray tracks the
total column density $N_{\rm abs}$ and the equivalent total optical
depth $\tau$ traversed by the photon.  If $\tau < \tau_{\rm thin} \sim
0.1$ after the ray exits the cell, we calculate the photo-ionisation
and photo-heating rates directly from the incoming ray instead of the
luminosity of the source.
\begin{equation}
  \label{eqn:optthin}
  k_{\rm ph} = \frac{\sigma_{\rm abs} \; P}{dt_P \; \theta_{\rm pix}}
  \; \frac{r_{\rm cell}}{r_{\rm ray}}.
\end{equation}
Note that the photon number $P$ in the ray has already been
geometrically diluted by ray splitting.  Here $r_{\rm cell}$ and
$r_{\rm ray}$ are the radii from the radiation source to the cell
centre and where the ray exits the cell.  Thus the last factor
corrects the flux to a value appropriate for the cell centre.  The
photo-heating ray is calculated in the same manner as the general
case, $\Gamma_{\rm ph} = k_{\rm ph} (E_{\rm ph} - E_i)$.  This should
only be evaluated once per cell per radiation source.  No photons are
removed from the ray.  With this method, we only require one ray
travel through each cell where the gas is optically thin, thus
reducing the computational expense.

We must be careful not the overestimate the radiation when multiple
rays enter a single cell.  In the case of a single radiation source,
the solution is simple -- only assign the cell the photo-ionisation
and photo-heating rates when $k_{\rm ph} = 0$.  However in the case
with multiple sources, this is no longer valid, and we must sum the
flux from all optically thin sources.  Only one ray per source must
contribute to a single cell in this framework.  We create a flagging
field that marks whether a cell has already been touched by an
optically thin photon from a particular radiation source.  Naively, we
would be restricted to tracing rays from a single source at a time if
we use a boolean flagging field.  However we can trace rays for 32
sources at a time by using bitwise operations on a 32-bit integer
field.  For example in \texttt{C}, we would check if an optically thin
ray from the \texttt{n}-th source has propagated through cell
\texttt{i} by evaluating \texttt{(MarkerField[i] $\gg$ n \& 1)}.  If
false, then we can add the optically thin approximation [equation
(\ref{eqn:optthin})] to the cell and set \texttt{MarkerField[i] |= (1
  $\gg$ n);} to mark the cell.

\subsection{Additional Physics}
\label{sec:addphysics}

Other radiative processes can also be important in some situations,
such as attenuation of radiation in the Lyman-Werner bands, secondary
ionisations from X-ray radiation, Compton heating of from scattered
photons, and radiation pressure.  We describe our implementation of
these physics next.

\subsubsection{Absorption of Lyman-Werner Radiation}

Molecular hydrogen can absorb photons in the Lyman-Werner bands
through the two-step Solomon process, which for the lowest
ro-vibrational states already consists of 76 absorption lines ranging
from 11.1 to 13.6 eV \citep{Stecher67, Dalgarno70, Haiman00}.  Each of
these spectral lines can be modelled with a typical exponential
attenuation equation \citep{Ricotti01}, but \citet{Draine96} showed
that this self-shielding is well modeled with the following relation
to total \hh~column density
\begin{equation}
  \label{eqn:lwband}
  f_{\rm shield} (N_{\rm H2}) =
  \left\{ \begin{array}{l@{\quad}l}
      1 & (N_{\rm H2} \le 10^{14})\\
      (N_{\rm H2}/10^{14})^{-0.75} & (N_{\rm H2} >
      10^{14})
    \end{array} \right. ,
\end{equation}
where $N_{\rm H2}$ is in units of cm$^{-2}$.  To incorporate this
shielding function into the ray tracer, we store the total \hh~column
density and calculate the \hh~dissociation rate by summing the
contribution of all rays,
\begin{equation}
  \label{eqn:lwRT}
  k_{\rm diss} = \sum_{\rm rays} \frac{P \; \sigma_{\rm LW} \;
    \Omega_{\rm ray} \; r^2 \; dr}{A_{\rm cell} \; dV \; dt_P},
\end{equation}
where $\sigma_{\rm LW} = 3.71 \times 10^{18} \mathrm{cm}^2$ is the
effective cross-section of \hh~\citep{Abel97}.  To account for
absorption, we attenuate the photon number flux by
\begin{equation}
  \label{eqn:LWdP}
  dP = P [f_{\rm shield}(N_{\rm H2} + dN_{\rm H2}) - f_{\rm shield}(N_{\rm H2})],
\end{equation}
where $dN_{\rm H2}$ is the \hh~column density in the current cell.

\subsubsection{Secondary Ionisations from X-rays}
\label{sec:xrays}

At the other end of the spectrum, a high-energy ($E_{\rm ph} \gsim
100$ eV) photon can ionise multiple neutral hydrogen and helium atoms,
and this should be considered in such radiation fields.
\citet{Shull85} studied this effect with Monte Carlo calculations over
varying electron fractions and photon energies up to 3 keV.  They find
that the excitation of hydrogen and helium and the ionisation of
He {\sc ii} is negligible.  The number of secondary ionisations of H
and He is reduced from the ratio of the photon and ionisation energies
($E_{\rm ph} / E_i$) by a factor of
\begin{equation}
  Y_{\rm k,H} = 0.3908 (1 - x^{0.4092})^{1.7592},
\end{equation}
\begin{equation}
  Y_{\rm k,He} = 0.0554 (1 - x^{0.4614})^{1.6660},
\end{equation}
where $x$ is the electron fraction.  The remainder of the photon
energy is deposited into thermal energy that is approximated by
\begin{equation}
  Y_{\rm \Gamma} = 0.9971 [ 1 - (1 - x^{0.2663})^{1.3163} ]
\end{equation}
and approaches one as $x \rightarrow 1$.  Thus in gas with low
electron fractions, most of the energy results in ionisations of
hydrogen and helium, and in nearly ionised gas, the energy goes into
photo-heating.

\subsubsection{Compton Heating from Photon Scattering}

High energy photons can also cause Compton heating by scattering off
free electrons.  During a scattering, a photon loses $\Delta E(T_e) =
4kT_e \times (E_{ph} / m_e c^2)$ of energy, where $T_e$ is the
electron temperature.  For the case of monochromatic energy groups, we
model this process by considering that the photons are absorbed by a
factor of 
\begin{equation}
  \label{eqn:compton}
  \frac{dP_C}{P} = (1 - e^{-\tau_e}) \frac{\Delta(T_e)}{E_{ph}},
\end{equation}
which is the equivalent of the photon energy decreasing.  Here $\tau_e
= n_e \sigma_{\rm KN} dl$ is the optical depth to Compton scattering,
and $\sigma_{\rm KN}$ is the non-relativistic Klein-Nishina cross
section \citep{Rybicki}.  The Compton heating rate is thus
\begin{equation}
  \Gamma_{\rm ph,C} = \frac{dP_C}{n_e \; V_{\rm cell} \; dt_P},
\end{equation}
which has been used in \citet{Kim11}.

\subsubsection{Radiation Pressure}

Another relevant process is radiation pressure, where the absorption
of radiation transfers momentum from photons to the absorbing medium.
This is easily computed by considering the momentum
\begin{equation}
  \label{eqn:radpres}
  d\mathbf{p}_\gamma = \frac{dP \; E_{\rm ph}}{c} \; \hat{r}
\end{equation}
of the absorbed radiation from the incoming ray, where $\hat{r}$ is
the normal direction of the ray.  We do not include radiation pressure
on dust currently.  The resulting acceleration of the gas because of
radiation pressure is
\begin{equation}
  d\mathbf{a} = \frac{d\mathbf{p}_\gamma} {dt_P\; \rho \; V_{\rm cell}},
\end{equation}
where $\rho$ is the gas density inside the cell.  This acceleration is
then added to the other forces, e.g. gravity, in the calculation in an
operator split fashion.

\section{Numerical Implementation in Enzo}
\label{sec:numerical}

In this section, we describe our parallel implementation of the
adaptive ray tracing method into \enzo.  \enzo~is a parallel
block-structured AMR \citep{BergerAMR} code that is publicly available
\citep{BryanNorman1997, OShea2004}.  First we explain the programming
design of handling the ``photon packages'' that are traced along the
adaptive rays.  We use the terms photon packages and rays
interchangeably.  Next we focus on the details of the radiation
hydrodynamics and then the importance of correct time-stepping.  Last
we give our parallelisation strategy of tracing rays through an AMR
hierarchy.  This implementation is included in the \texttt{v2.0}
public version of \enzo.

\subsection{Programming Design}
\label{sec:design}

Each photon package is stored in the AMR grid with the finest
resolution that contains its current position.  The photon packages
keep track of their (1) photon flux, (2) photon type, (3) photon
energy, (4) the time interval of its emission, (5) emission time, (6)
current time, (7) radius, (8) total column density, (9) HEALPix pixel
number, (10) HEALPix level, and (11) position of the originating
source, totaling 60 (88) bytes for single (double) precision.  When
\enzo~uses double precision for grid and particle positions and time,
items 4-7 and 11 are double precision.

We only treat point sources of radiation in our implementation;
therefore all base level photon packages originate from them.  As they
travel away from the source, they generally pass through many AMR
grids, especially if the simulation has a high dynamic range.  This is
a challenging programming task as rays are constantly entering and
exiting grids.  Before any computation, the number of rays in a
particular grid is highly unpredictable because the intervening medium
is unknown.  Furthermore, the splitting of parent rays into child rays
and a dynamic AMR hierarchy add to the complexity.  Because of this,
we store the photon packages as a doubly linked list
\citep{Abel02_RT}.  Thus we can freely add and remove them from grids
without the concern of allocating enough memory before the tracing
commences.

\begin{figure*}
  \epsscale{2}
  \plotone{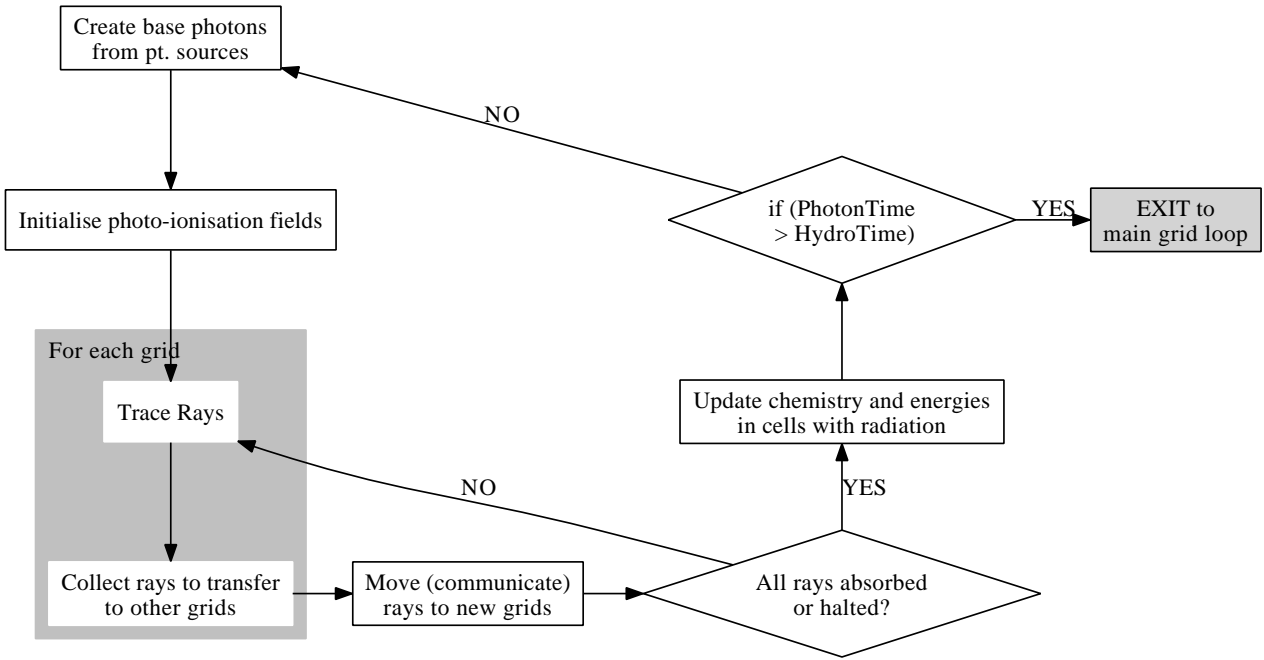}
  \epsscale{1}
  \caption{\label{fig:MainAlgorithm} Flow chart for the overall
    algorithm of the radiative transfer module in \enzo~that
    illustrates (1) the creation of photon packages, (2) ray tracing,
    (3) the transport of photon packages between AMR grids, and (4)
    coupling with the hydrodynamics.  The ray tracing algorithm, which
    is contained in the ``Trace Rays'' is detailed in Fig.
    \ref{fig:RTalgorithm}.}
\end{figure*}

\begin{figure*}
  \epsscale{2}
  \plotone{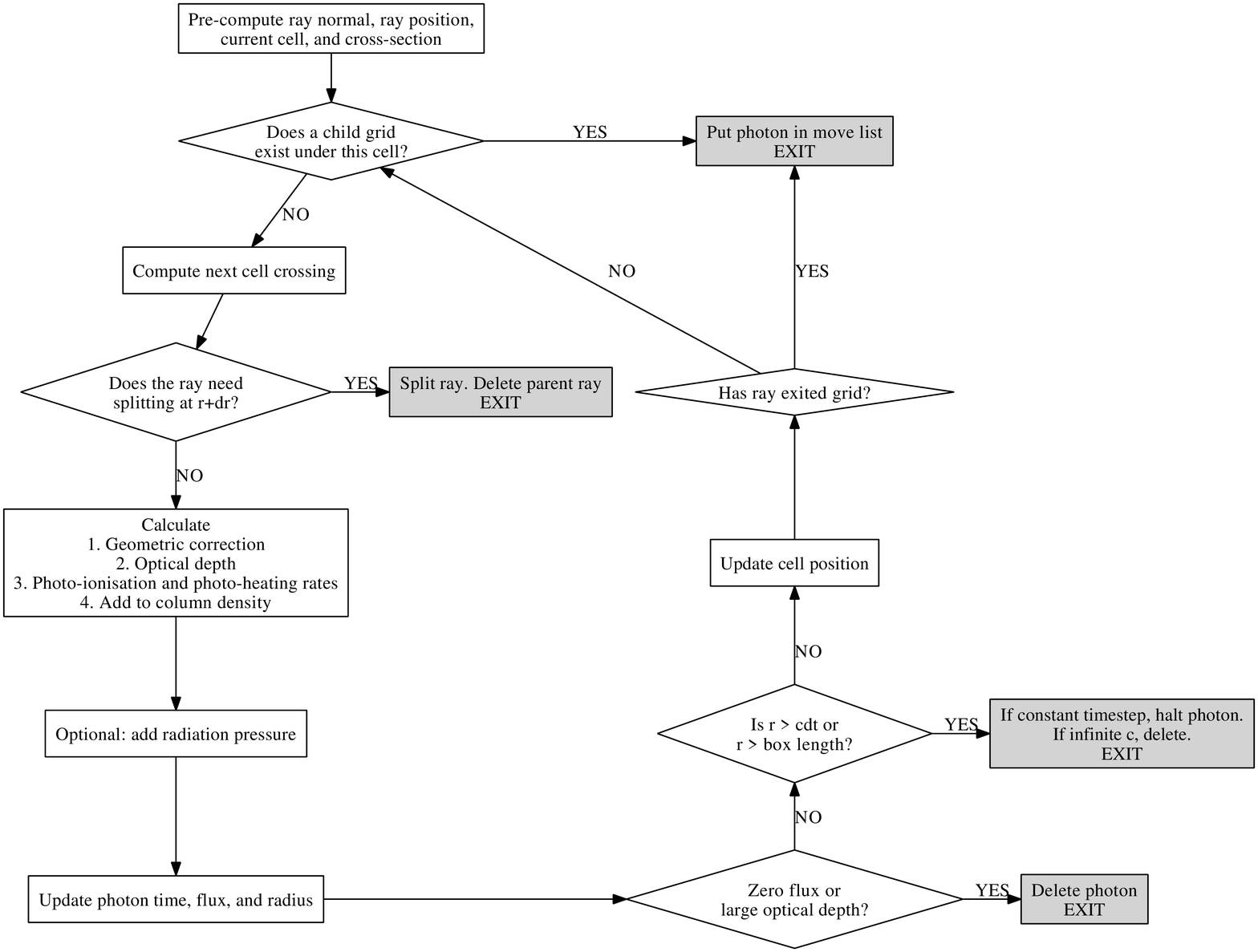}
  \epsscale{1}
  \caption{\label{fig:RTalgorithm} Flow chart for the ray tracing
    algorithm for one photon passing through a grid.  Note that only
    one step is needed in the routine to adaptively split rays.  The
    remainder is a typical ray tracing method.}
\end{figure*}

We illustrate the underlying algorithm of the ray tracing module in
\enzo~in Fig. \ref{fig:MainAlgorithm} and the ray tracing algorithm
is shown in Fig. \ref{fig:RTalgorithm}.  The module is only called
when advancing the finest AMR level.  We describe its steps below.

\step{1} Create $N_{\rm pix}$ new photon packages on the initial
HEALPix level from point sources.  Place the new rays in the highest
resolution AMR grid that contains the source.

\step{2} Initialise all radiation fields to zero.

\step{3} Loop through all AMR grids, tracing any rays that exist in
it.  For each ray, the following substeps are taken.

\step{3a} Compute the ray normal based on the HEALPix level and pixel
number of the photon package with the HEALpix routine
\texttt{pix2vec\_nest}.  One strategy to accelerate the computation is
to store ray segment paths in memory \citep{Abel02_RT,
  Krumholz07_ART}; however this must be recomputed if the grid
structure or point source position changes.  We do not restrict these
two aspects and cannot employ this acceleration method.

\step{3b} Compute the position of the ray ($\mathbf{r}_{\rm src} + r
\mathbf{\hat{n}}$), the current cell coordinates in floating point and
its corresponding integer indices.  Here $\mathbf{r}_{\rm src}$ is the
position of the point source, $r$ is the distance travelled by the
ray, and $\mathbf{\hat{n}}$ is the ray normal.

\step{3c} Check if a subgrid exists under the current cell.  If so,
move the ray to a linked list that contains all rays that should be
moved to other grids.  We call this variable \texttt{PhotonMoveList}.
Store the destination grid number and level.  Continue to the next ray
in the grid (step 3a).  We determine whether a subgrid exists by
creating a temporary 3D field of pointers that either equals the
pointer of the current grid if no subgrid exists under the current
cell or the child grid pointer that exists under the current cell.
This provides a significant speedup when compared to a simple
comparison of a photon package position and all of the child grid
boundaries.  Note that this is the same algorithm used in \enzo~when
moving collisionless particles to child grids.

\step{3d} Compute the next cell crossing of the ray and the ray
segment length across the cell (Equation \ref{eqn:trace}).

\step{3e} Compare the solid angle associated with the ray at radius
$r+dr$ with a user-defined splitting criterion (Equation
\ref{eqn:split}).  If the solid angle is larger than the desired
minimum sampling, split the ray into 4 child rays (\S\ref{sec:ART}).
These rays are inserted into the linked list after the parent ray,
which is subsequently deleted.  Continue to the next ray (step 3a),
which will be the first child ray.

\step{3f} Calculate the geometric correction (Equation \ref{eqn:fc}),
the optical depth of the current cell (Equation \ref{eqn:dtau}),
photo-ionisation and photo-heating rates (Equations \ref{eqn:kph} and
\ref{eqn:gamma}), and add the column density of the cell to the total
column density of the ray.

\step{3g} Add the effects of any optional physics modules
(\S\ref{sec:addphysics}) -- secondary ionisations from X-rays, Compton
heating from scattering, and radiation pressure.

\step{3h} Update the current time ($t = t + cdr$), photon flux ($P = P
- dP$, Equation \ref{eqn:flux}), and radius of the ray ($r = r + dr$).

\step{3i} If the photon flux is zero or the total optical depth is
large ($>20$), delete the ray.

\step{3j} Check if the ray has travelled a total distance of $c dt_P$
in the last timestep.  If we are keeping the time-derivative of the
radiative transfer equation, halt the photon.  If not (i.e. infinite
speed of light), delete the photon.

\step{3k} Check if the ray has exited the current grid.  If false,
continue to the next cell (step 3b).  If true, move the ray to the
linked list \texttt{PhotonMoveList}, similar to step 3c.  If the ray
exits the simulation domain, delete it if the boundary conditions are
isolated; otherwise, we change the source position of the ray by a
distance \texttt{-sign(n[i]) * DomainWidth[i]}, where \texttt{n} is
the ray normal, and \texttt{i} is the dimension of the outer boundary
it has crossed.  The radius is kept unchanged.  In essence, this
creates a ``virtual source'' outside the box because the ray will be
moved to the opposite side of the domain, appearing that it originated
from this virtual source.

\step{4} If any rays exist in the linked list \texttt{PhotonMoveList},
move them to their destination grids and return to step 3.  This
requires MPI communication if the destination grid exists on another
processor.

\step{5} If all rays have not been halted (keeping the time-derivative
of the radiative transfer equation), absorbed, or exited the domain,
return to step 3.

\step{6} With the radiation fields updated, call the chemistry and
energy solver and update only the cells with radiation, which is
discussed further in \S\ref{sec:coupling}.

\step{7} Advance the time associated with the photons $t_P$ by the
global timestep $dt_P$ (for its calculation, see
\S\ref{sec:timestepping}).  If $t_P$ does not exceed the time on the
finest AMR level, return to step 1.

\subsection{Energy groups}

In our implementation, photon packages are mono-chromatic, i.e. energy
groups \citep[][Ch. 6]{Mihalas84}, and are assigned a photon type that
corresponds whether it is a photon that (1) ionises hydrogen, (2)
singly ionises helium, (3) doubly ionises helium, (4) has an X-ray
energy, or (5) dissociates molecular hydrogen (Lyman-Werner
radiation).  One disadvantage of mono-chromatic rays is the number of
rays increase with the number of frequency bins.  However this allows
for early termination of rays that are fully absorbed, which are
likely to have high absorption cross-sections (e.g. \hi ionisations
near 13.6 eV) or a low initial intensity (e.g. He {\sc ii} ionising
photons in typical stellar populations).  The other approach used by
some groups \citep[e.g.][]{Trac07} is to store all energy groups in a
single ray.  This reduces the number of the rays generated and the
computation associated with the ray tracing.  Unless the ray
dynamically adjusts its memory allocation for the energy groups as
they become depleted, this method is also memory intensive in the
situation where most of the energy groups are completely absorbed but
a few groups still have significant flux.

In practice, we have found that one energy group per photon type is
sufficient to match expected analytical tests (\S\ref{sec:nu_dep}).
For example when modeling Population III stellar radiation
\citep[e.g.][for hydrogen ionising radiation only]{Abel07,
  Wise08_Gal}, we have 3 energy groups -- H {\sc i}, He {\sc i}, He
{\sc ii} -- each with an energy that equals the average photon energy
above the ionisation threshold.

\subsection{Coupling with Hydrodynamics}
\label{sec:coupling}

Solving the radiative transfer equation is already an intensive task,
but coupling the effects of radiation to the gas dynamics is even more
difficult because the radiation fields must be updated on a time-scale
such that it can react to the radiative heating, i.e. sound-crossing
time.  The frequency of its evaluation will be discussed in the next
section.

\enzo~solves the physical equations in an operator-split fashion over
a loop of AMR grids.  On the finest AMR level, we call our radiation
transport solver before this main grid loop in the following sequence:
\begin{itemize}
\item \textbf{All grids:}
  \begin{enumerate}
  \item Solve for the radiation field with the adaptive ray tracer
  \item Update species fractions and energies for cells with radiation
    with a non-equilibrium chemistry solver on subcycles (Equation
    \ref{eqn:rate_dt}).
  \end{enumerate}
\item \textbf{For each grid:}
  \begin{enumerate}
  \item Solve for the gravitational potential with the particle mesh
    method
  \item Solve hydrodynamics
  \item Update species fractions and energies for cells without
    radiation with a non-equilibrium chemistry solver on subcycles
    (Equation \ref{eqn:rate_dt}).
  \item Update particle positions
  \item Star particle formation
  \end{enumerate}
\item \textbf{All grids:} Update solution from children grids
\end{itemize}

Since the solver must be called many times, the efficiency of the
radiation solver is paramount.  After every radiation timestep, we
call the non-equilibrium chemistry and energy solver in \enzo.  This
solves both the energy equation and the network of stiff chemical
equations on small timesteps, i.e. subcycles \citep{Anninos97}.  The
timestep is
\begin{equation}
  \label{eqn:rate_dt}
  dt = \min\left(
    \frac{0.1n_e}{|dn_e/dt|}, 
    \frac{0.1n_{\rm HI}}{|dn_{\rm HI}/dt|}, 
    \frac{0.1e}{|de/dt|}, 
    \frac{dt_{\rm hydro}}{2}\right),
\end{equation}
where $n_e$ is the electron number density, $e$ is the specific
energy, and $dt_{\rm hydro}$ is the hydrodynamic timestep.  This
limits the subcycle timestep to a 10\% change in either electron
density, neutral hydrogen density, or specific energy.  In simulations
without radiation, \enzo~calls this solver in a operation-split manner
after the hydrodynamics module for grids only on the AMR level that is
being solved.  In simulations with radiative transfer, the radiation
field can change on much faster time-scales than the normal
hydrodynamical timesteps.

For example, a grid on level $L$ might have no radiation in its
initial evaluation, but the ionisation front exists just outside its
boundary.  Then radiation permeates the grid in the time between
$t_{\rm L} \rightarrow t_{\rm L}+dt_{\rm L}$, and the energy and
chemical state of the cells must be updated with each radiation update
to advance the ionisation front accurately.  If one does not update
these cells, it will appear that the ionisation front does not enter
the grid until the next hydrodynamical timestep!  Visually this
appears as discontinuities in the temperature and electron fraction on
grid boundaries.  One may avert this problem by solving the chemistry
and energy equations for every cell on every radiative transfer
timestep, but this is very time consuming and unnecessary, especially
if the radiation filling factor is small.

We choose to dynamically split the problem by cells with and without
radiation.  In every radiation timestep, the chemo-thermal state of
\textit{only} the cells with radiation are updated.  For the solver
subcycling, we replace $dt_{\rm hydro}$ with $dt_P$ in Equation
\ref{eqn:rate_dt} in this case.  Once the radiative transfer solver is
finished with its timesteps, the hydrodynamic module is called, and
then the chemo-thermal state of the cells without radiation are
updated on a subcycle timestep stated in Equation \ref{eqn:rate_dt}.

For cells that transition from zero to non-zero photo-ionisation
rates, the initial state that enters into the chemistry and energy
solver does not correspond to the current time of the radiation
transport solver $t_{\rm RT}$, but either time $t_{\rm L}$ if the grid
level is the finest level because its chemo-thermal state has not been
updated or time $t_{\rm L}+dt_{\rm L}$ on all other levels.  In
principle, one could first revert the cell back to time $t_{\rm L}$
and then update to $t_{\rm RT}$ with the chemistry and energy solver
if the cell is on the finest level.  However in practice, the
time-scales in gas without radiation are small compared to the
ionisation and heating time-scales when radiation is introduced.
Therefore, we do not perform this correction and find that this does
not introduce any inaccuracies in both test problems (see
\S\ref{sec:rt_tests}) and real world applications.

\subsection{Temporal evolution}
\label{sec:timestepping}

There have been several methods of choosing a maximal timestep to
solve radiation transfer equation while retaining stability and
accuracy.  We describe several methods to calculate the radiative
transfer timestep in this section.  With a small enough timestep, the
solution is accurate (ignoring any systematic ones), but the solver is
slow.  Furthermore for very small timesteps, the photon packages only
advance a short distance, and they will exist in every $dx/dt_P$ cells
with radiation and are stored between timesteps, excessively consuming
memory.  On the shortest time-scale, one can safely set the timestep to
the light-crossing time of a cell \citep{Abel99_RT, Trac07} but
encounters the problems stated above.

If the timestep is too large, the solution will become inaccurate;
specifically, ionisation fronts will advance too slowly, as radiation
intensity exponentially drops with a scale length of the mean free
path
\begin{equation}
  \label{eqn:mfp}
  \lambda_{\rm mfp} = \frac{1}{n_{\rm abs} \; \sigma_{\rm abs}}
\end{equation}
past the ionisation front.  For example in our implementation, the
chemo-thermal state of the system remains constant as the rays are
traced through the cells.  In the case of a single \hii region,
the speed of the ionisation front is limited to approximately
$\lambda_{\rm mfp} / dt_P$.

\subsubsection{Minimizing neutral fraction change}
\label{sec:dt_hi}

\begin{figure}
  \plottwo{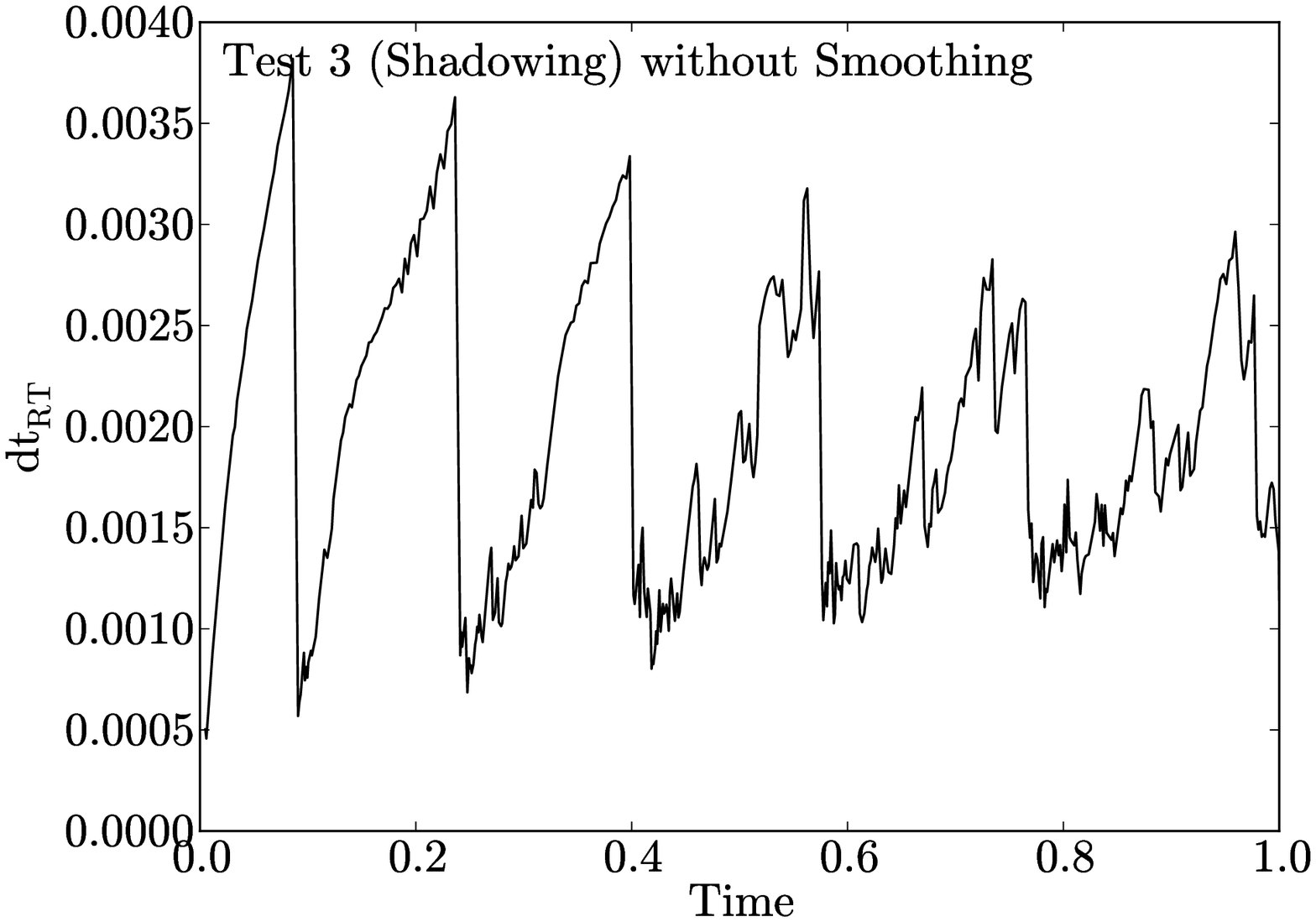}{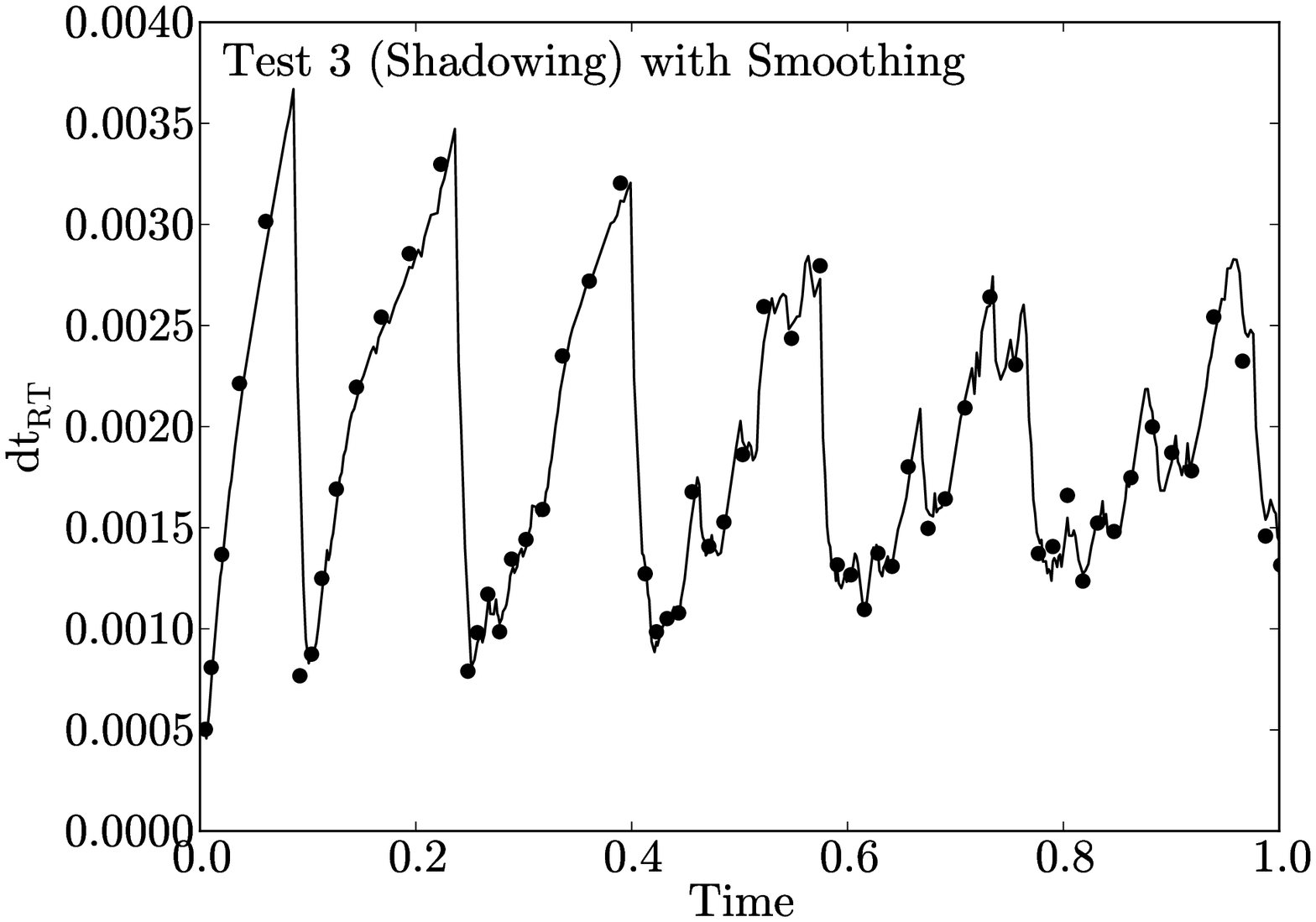}
  \caption{\label{fig:dtsmooth} Radiative transfer adaptive timestep
    in shadowing test (Test 3; \S\ref{sec:test3}) while restricting
    the neutral fraction change to 5\% in the ionisation front.  The
    unmodified timestep (top) is slightly more noisy and the minima
    are more prominent than the timestep computed with a running
    average of the last two timesteps (bottom).  The points show every
    tenth timestep taken into account for the running average.  The
    sawtooth behavior is created by the ionisation front advancing
    into the next neutral cell in the overdensity.}
\end{figure}

Another strategy is restricting the neutral fraction to change a small
amount, i.e. for a single cell,
\begin{equation}
  \label{eqn:dtelec}
  dt_{\rm P,cell} = \epsilon_{\rm ion} \frac{n_{\rm HI}}{|dn_{\rm HI}/dt|} =
  \frac{\epsilon_{\rm ion}}{|k_{\rm ph} + n_e (C_{\rm H} + \alpha_B)|},
\end{equation}
where $\epsilon_{\rm ion}$ is the maximum fraction change in neutral
fraction.  \citet{Shapiro04} found that this limited the speed of the
ionisation front.  We can investigate this further by evaluating the
ionisation front velocity in a growing Str\"{o}mgren sphere without
recombinations, 
\begin{equation}
  \label{eqn:ndot}
  \dot{N}_\gamma = 4 \pi R^2 n_{\rm H} v_{\rm IF}.
\end{equation}
Using $\kph = \dot{N}_\gamma \sigma / 4\pi R^2 A_{\rm cell}$ and $\kph
\propto \dot{n}_{\rm H} / n_{\rm H}$, we can make substitutions
respectively on both sides of Equation (\ref{eqn:ndot}) to arrive at
the ionisation front velocity $v_{\rm IF} \propto n_{\rm H}
/\dot{n}_{\rm H}$.

We have implemented this method but we only consider cells within the
ionisation front (by experiment we choose $\tau > 0.5$) because we are
interested in evolving ionisation fronts at the correct speed.  In the
ionised region, the absolute changes in neutral fraction are small and
will not significantly affect the ionisation front evolution.  In
other words, $n_{\rm HI}/(dn_{\rm HI}/dt)$ may be large but $(dn_{\rm
  HI}/dt) \sim 0$, thus we can safely ignore these cells when
determining the timestep without sacrificing accuracy.

We search for the cell with the smallest $dt_P$ based on Equation
\ref{eqn:dtelec}.  In principle, one could use this value without
modifications as the timestep, but there is considerable noise both
spatially and temporally.  In order to stabilise this technique, we
first spatially smooth the values of $dt_{\rm P,cell}$ with a Gaussian
filter over a $3^3$ cube.  Because we only consider the cells within
the ionisation front, we set $dt_{\rm P,cell}$ to the hydrodynamical
timestep outside the front during the smoothing.  After we have
smoothed $dt_{\rm P,cell}$, we select the minimum value as $dt_P$.
Significant noise in $dt_P$ can exist between time steps as well.
Because the solution can become inaccurate if the timestep is allowed
to be too large, we restrict the next timestep to be less than twice
the previous timestep,
\begin{equation}
  \label{eqn:dtchange}
  dt_{\rm P,1} = \min(2 dt_{\rm P,0},\; dt_{\rm P,1}).
\end{equation}
Fig. \ref{fig:dtsmooth} shows the smooth evolution of $dt_{\rm P}$
in a growing Str\"{o}mgren sphere when compared to the values of
$\min(dt_{\rm P,cell})$.

\subsubsection{Time averaged quantities within a timestep}

\citet{Mellema06} devised an iterative scheme that allows for large
timesteps while retaining accuracy by considering the time-averaged
values ($\tau$, $k_{\rm ph}$, $n_e$, $n_{\rm HI}$) during the
timestep.  Starting with the cells closest to the source, they first
calculate the column density to the cell.  Then they compute the
time-averaged neutral density for the cell and its associated optical
depth, which is added to the total time-averaged optical depth.  With
these quantities, one can compute a photo-ionisation rate and update
the electron density.  This process is repeated until convergence is
found in the neutral number density.  In a test with a Str\"{o}mgren
sphere, they found analytical agreement with $10^{-3}$ less timesteps
than a method without time-averaging.  Another advantage of this
method is the use of pre-calculated tables for the photo-ionisation
rates as a function of optical depth, based on a given spectrum.  This
minimizes the energy groups needed to accurately sample a spectrum.
We are currently implementing this method into \moray.

\subsubsection{Physically motivated}
\label{sec:dt_const}

A constant timestep is necessary when solving the time-dependent
radiative transfer equation in \moray.  It should be small enough
to evolve ionisation fronts accurately, as discussed earlier.  The
timestep can be based on physical arguments, for example, the
sound-crossing time of an ionised region at $T > 10^4$ K.  To be
conservative, one may choose the sound-crossing time of a cell
\citep[e.g.][]{Abel07, Wise08_Gal}.  Alternatively, the diameter of
the smallest relevant system (e.g., an accretion radius, transition
radius to a D-type ionisation front, etc.) in the simulation may be
chosen to calculate the sound-crossing time.

If the timestep is too large, the ionisation front will propagate too
slowly, but it eventually approaches the correct radius at late times
(see \S\ref{sec:dt_dep}).  This does not prevent one from using a
large timestep, particularly if the system is not critically affected
by a slower I-front velocity.  One example is an expanding \hii
region in a power-law density gradient.  After a brief, initial R-type
phase, the I-front becomes D-type phase, where the ionisation and
shock front progress jointly at the sound speed of the ionised region.
A moderately large (0.1 Myr) timestep can accurately follow its
evolution.  However after the I-front passes a critical radius
\citep{Franco90}, the I-front detaches from the shock front,
accelerates, and transitions back to an R-type front.  This can also
occur in champagne flows when the ionisation front passes a density
discontinuity.  The I-front velocities in these two stages differ up
to a factor of $\sim10$.  Although the solution is accurate with a
large timestep in the D-type phase, the I-front may lag behind because
of the constant timestep.  After a few recombination times, the
numerical solution eventually approaches the analytical solution.  If
such a simulation focuses on the density core expansion and any small
scale structures, such as cometary structures and photo-dissociation
regions, one can cautiously sacrifice temporal accuracy at large
scales for computational savings.

\subsubsection{Change of incident radiation}
\label{sec:dt_tau}

Ionisation front velocities can approach significant fractions of the
speed of light in steep density gradients and in the early expansion
of the \hii region.  If the ionisation front position is critical to
the calculation, the radiation transport timestep can be derived from
a non-relativistic estimate of the ionisation front velocity
\begin{equation}
  \label{eqn:dt_incident}
  v_{\rm IF}(\mathbf{r}) \approx \frac{F(\mathbf{r})}
  {n_{\rm abs}(\mathbf{r})},
\end{equation}
based on the incident radiation field at a particular
position\footnote{See \citet{Shapiro06} for the exact calculation of a
  relativistic ionisation front.  Neglecting relativistic terms do not
  affect the solution because the front velocity is only considered in
  the timestep calculation.}.  Alternatively, the propagation of the
ionisation front can be restricted by limiting the change in specific
intensity $I$ to a safety factor $C_{\rm RT,cfl}$, resulting in a
timestep of
\begin{equation}
  \label{eqn:vifront1}
  dt_P = C_{\rm RT,cfl} \frac{I}{|dI/dt|}.
\end{equation}
We consider the specific intensity after the ray travels through the
cell, so $I = I_0 \exp(-\tau)$, where $\tau = n_{\rm H} \sigma dl$ is
the optical depth through the cell.  The time derivative of $I$ is
\begin{equation}
  \label{eqn:vifront2}
  \frac{dI}{dt} = I_0 \exp(-\tau) (-\dot{n}_{\rm H} \sigma dl),
\end{equation}
which can be expressed in terms of local optical depth and neutral
fraction,
\begin{equation}
  \label{eqn:vifront3}
    \frac{dI}{dt} = -I \frac{\tau \dot{n}_{\rm H}}{n_{\rm H}}.
\end{equation}
$\dot{n}_{\rm H}$ is computed with the same formula as equation
(\ref{eqn:dtelec}). Substituting in equation (\ref{eqn:vifront1})
gives
\begin{equation}
  \label{eqn:vifront4}
  dt_P = C_{\rm RT,cfl} \frac{n_{\rm H}}{\tau \dot{n}_{\rm H}}.
\end{equation}
In practice, we have found that a ceiling of 3 can be placed on the
optical depth, so optically thick cells do not create a very small
timestep.  We still find excellent agreement with analytical solutions
with this approximation.  We show the accuracy using this timestep
method in \S\ref{sec:dt_dep}.

\subsection{Parallelisation Strategy}
\label{sec:parallel}

Parallelisation of the ray tracing code is essential when exploring
problems that require high resolution and thus large memory
requirements.  Furthermore, \enzo~is already parallelised and scalable
to $O(10^2)$ processors in AMR simulations, and $O(10^3)$ in unigrid
calculations.  \enzo~stores the AMR grid structure on every processor,
but only one processor contains the actual grid and particle data and
photon packages.  All other processors contain an empty grid
container.  As discussed in \textit{Step 4} in \S\ref{sec:design}, we
store the photon packages that need to be transferred to other grids
in the linked list \texttt{PhotonMoveList}.  In a single processor
(serial) run, moving the rays is trivial by inserting these photons
into the linked list of the destination grid.  For multi-processor
runs, we must send these photons through MPI communication to the
processors that host the data of the destination grids.  We describe
our strategy below.

The easiest case is when the destination grid exists on the same
processor as the source grid, where we move the ray as in the serial
case.  For all other rays, we organise the rays by destination
processors and send them in groups.  We also send the destination grid
level and ID number along with the ray information that is listed at
the beginning of \S\ref{sec:design}.

For maximum overlap of communication and computation, which enables
scaling to large numbers of processors, we must employ
``non-blocking'' MPI communication, where each processor does not wait
for synchronisation with other processors.  We use this technique for
the sending and receiving of rays.  Here we desire to minimize the
idle time of each processor when it is waiting to receive data.  In
the loop shown in Fig. \ref{fig:MainAlgorithm} with the conditional
that checks whether we have traced all of the rays, we aggressively
transport rays that are local on the processor, and process any MPI
receive calls as they arrive, not waiting for their completion in
order to continue to the next iteration.  We describe the steps in
this algorithm next.

\step{1} Before any communication occurs, we count the number of rays
that will be sent to each processor.  The MPI receive calls
(\texttt{MPI\_Irecv}) must have a data buffer that is greater than or
equal to the size of the message.  We choose to send a maximum of
$N_{\rm max}$ (= 10$^5$ in \enzo~\texttt{v2.0}) rays per MPI message.
Therefore, we allocate a buffer of this size for each
\texttt{MPI\_Irecv} call.  We then determine the number of MPI messages
$N_{\rm mesg}$ and send this number in a non-blocking fashion,
i.e. \texttt{MPI\_Isend}.

\step{2} Pack the photon packages into a contiguous array for MPI
communication while the messages from Step 1 completes.

\step{3} Process the number of photon messages that we are expecting
from each processor, sent in Step 1.  Then post this number of
\texttt{MPI\_Irecv} calls for the photon data.  Because we strive to
make the ray tracing routine to be totally non-blocking, the
processors will most likely not be synchronised on the same loop
(Steps 3--5 in \S\ref{sec:design}).  Therefore, there might be
additional $N_{\rm mesg}$ MPI messages waiting to be processed.  We
check for these messages and aggressively drain the message stack to
determine the total number of photon messages that we are expecting
and post their associated \texttt{MPI\_Irecv} calls for the photon
data.

\step{4} Send the grouped photon data with \texttt{MPI\_Isend} with a
maximum size of $N_{\rm max}$ photons.

\step{5} Place any received photon data into the destination grids.
We monitor whether the processor has any rays that were moved to grids
on the same processor.  If so, this processor has rays to transport,
and we do not necessarily have to wait for any MPI receive messages
and thus use \texttt{MPI\_Testsome} to receive any messages that have
already arrived.  If not, we call \texttt{MPI\_Waitsome} to wait for
any MPI receive messages.

\step{6} If all processors have exhausted their workload, then all
rays have been either absorbed, exited the domain, or halted after
travelling a distance $cdt_P$.  We check this in a similar
non-blocking manner as the $N_{\rm mesg}$ calls in Step 1.

Lastly we have experimented with a hybrid OpenMP/MPI version of \enzo,
where workload is partitioned over grids on each MPI process.  We
found that parallelisation over grids for the photon transport does
not scale well, and threading over the rays in each grid is a better
approach.  Because the rays are stored in a linked list in each grid,
we must manually split the list into separate lists and let each
thread work on each list.

\section{Radiative Transfer Tests}
\label{sec:rt_tests}

Tests plays an important role in creating and maintaining
computational tools.  In this section, we present tests drawn from the
Cosmological Radiative Transfer Codes Comparison Project \citep{RT06},
where results from 11 different radiative transfer codes compared
results in four test problems.  The codes use various methods for
radiation transport: ray tracing with short, long, and hybrid
characteristics, Monte Carlo casting; ionisation front tracking
\citep{Alvarez06_IFT}; variable Eddington Tensor formalism
\citep{Gnedin01_OTVET}.  They conducted tests that investigated (1)
the growth of a single Str\"{o}mgren sphere enforcing isothermal
conditions, (2) the same test with an evolving temperature field, (3)
shadowing created by a dense, optically thick clump, and (4) multiple
\hii regions in a cosmological density field.  In all of the
tests presented here, we use the method of restricted neutral fraction
changes (\S\ref{sec:dt_hi}) for choosing a radiative transfer
timestep.  We cast 48 rays (HEALPix level 1) from the point source and
require a sampling of at least $\Phi_c = 5.1$ rays per cell.

\subsection{Test 1. Pure hydrogen isothermal \hii region
  expansion}
\label{sec:test1}

\begin{figure}
  \plottwo{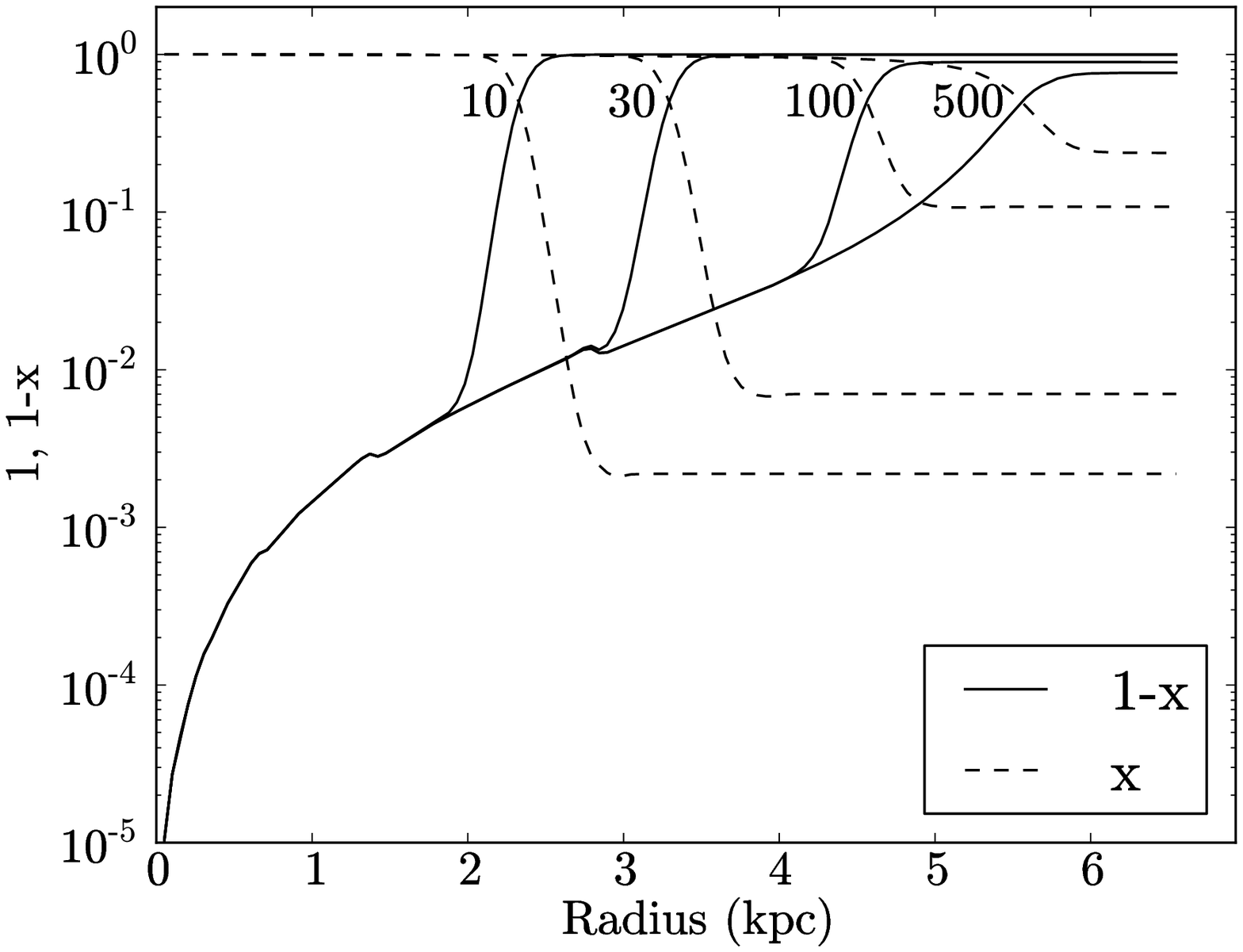}{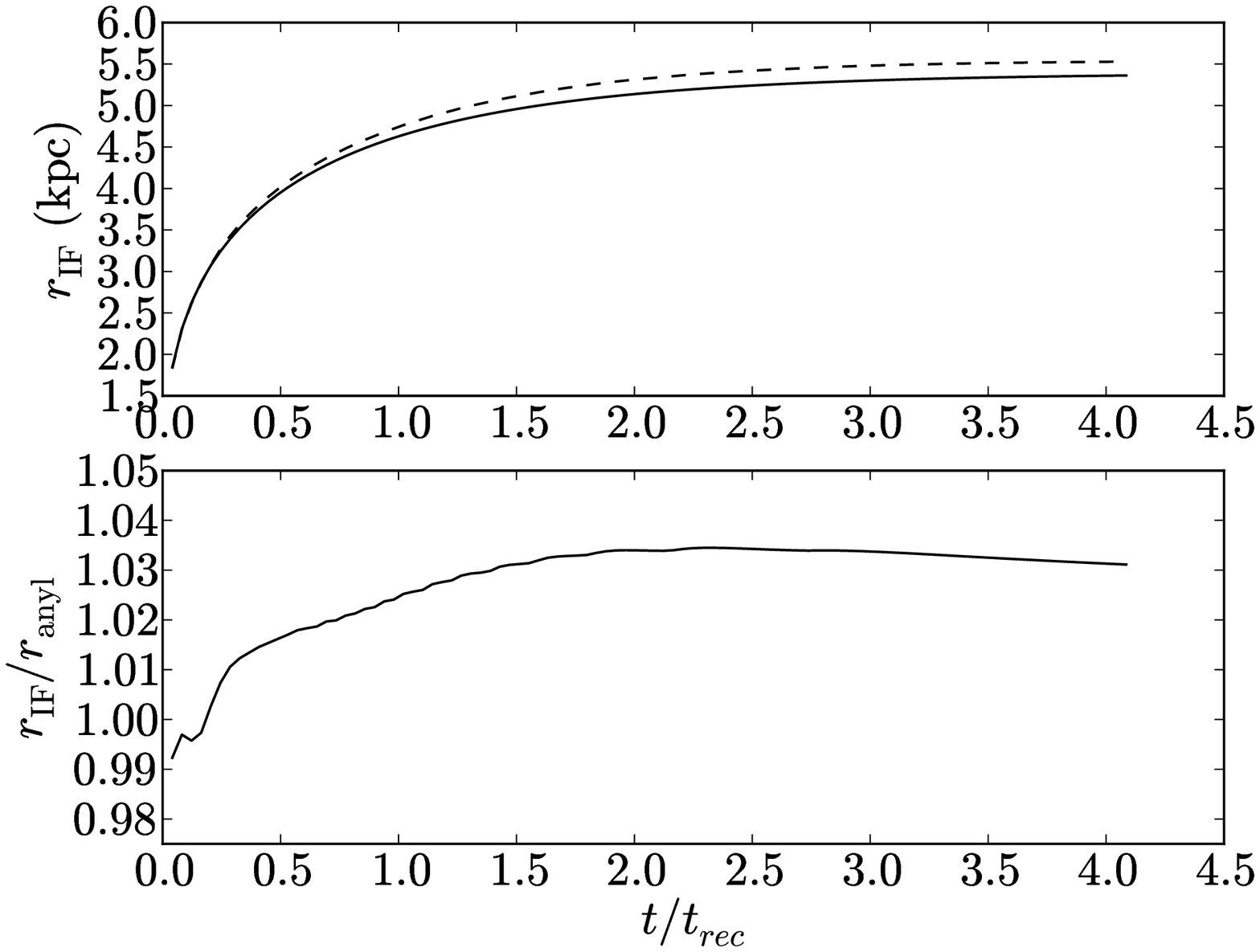}
  \caption{\label{fig:test1_ifront} Test 1 (\hii region
    expansion with a monochromatic spectrum of 13.6 eV).  Top:
    Radially averaged profile of neutral (solid) and ionised (dashed)
    fraction at 10, 30, 100, and 500 Myr.  Bottom: Evolution of the
    calculated (top, dashed) and analytical (top, solid) Str\"{o}mgren
    radius.  The ratio of these radii are plotted in the bottom panel.}
\end{figure}

\begin{figure}
  \plotone{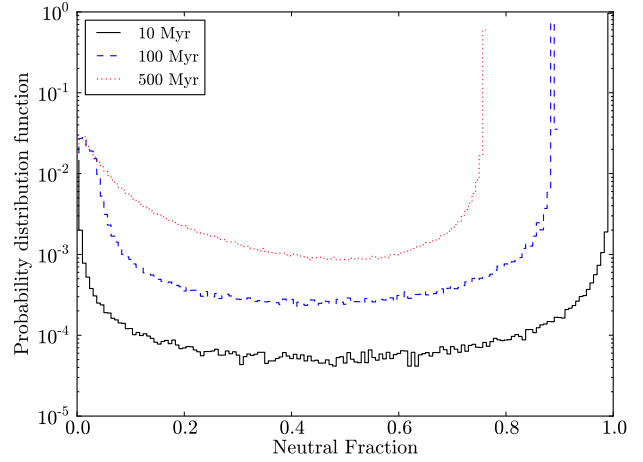}
  \caption{\label{fig:test1_pdf} Test 1 (\hii region expansion with a
    monochromatic spectrum of 13.6 eV). Probability distribution
    function for neutral fraction at 10 Myr (solid), 100 Myr (dashed),
    and 500 Myr (dotted).  Recombination of hydrogen at T = $10^4$ K
    causes the maximum neutral fraction to decrease from 1 to 0.75.}
\end{figure}

\begin{figure}
  \plotone{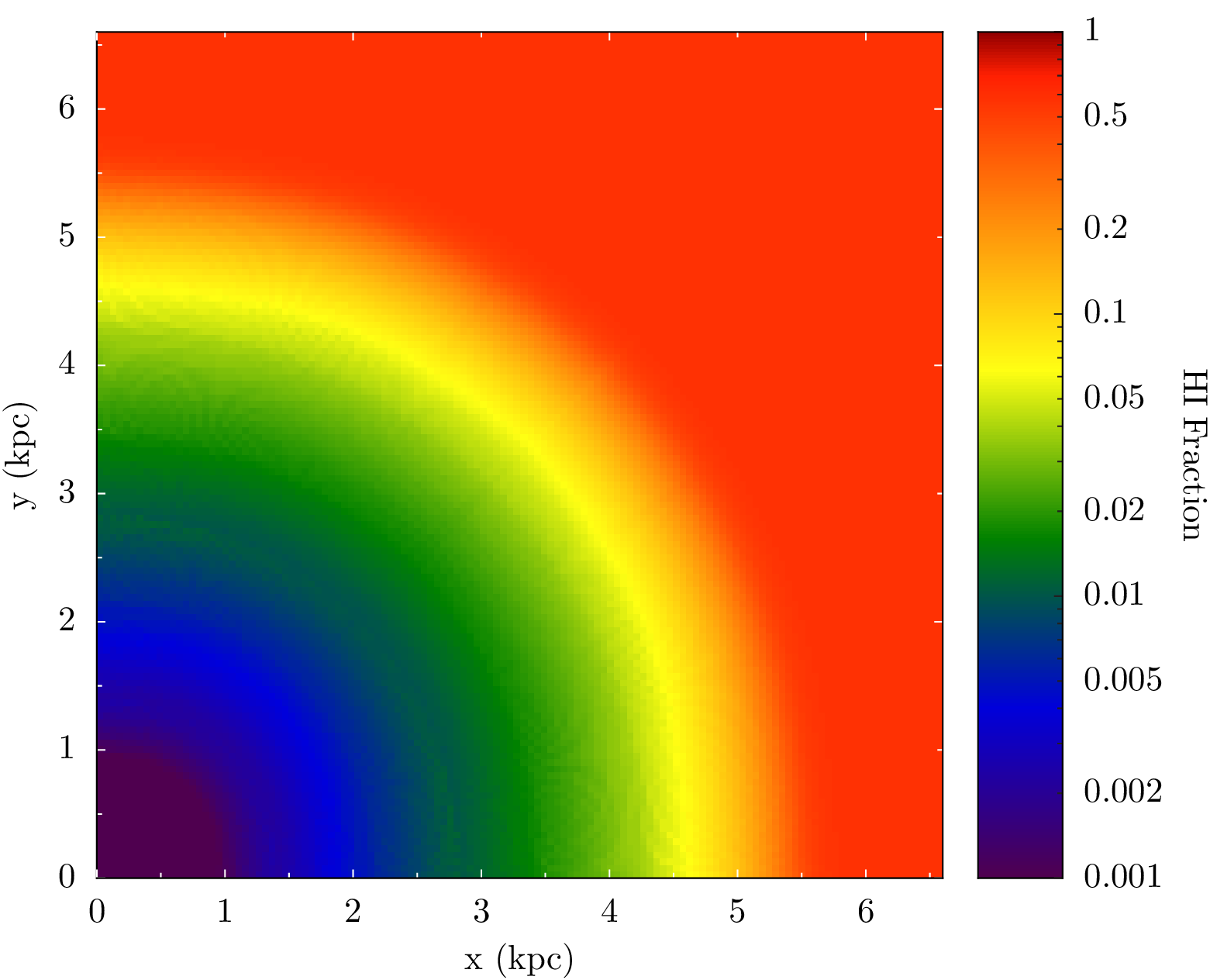}
  \caption{\label{fig:test1_HI} Test 1 (\hii region expansion
    with a monochromatic spectrum of 13.6 eV). Slice of neutral
    fraction at the origin at 500 Myr.}
\end{figure}

The expansion of an ionising region with a central source in a uniform
medium is a classic problem first studied by \citet{Stroemgren39}.
This simple but useful test can uncover any asymmetries or artifacts
that may arise from deficiencies in the method or newly introduced
bugs in the development process.  In this problem, the ionised region
grows until recombinations balance photo-ionisations [equation
(\ref{eqn:Rstr})].  The evolution of the radius $r_s$ and velocity
$v_s$ of the ionisation front has an exact solution of
\begin{equation}
  \label{eqn:rstr}
  r_s(t) = R_s [ 1 - \exp(-t/t_{\rm rec}) ]^{1/3},
\end{equation}
\begin{equation}
  \label{eqn:vstr}
  v_s(t) = \frac{R_s}{3t_{\rm rec}} \frac{\exp(-t/t_{\rm rec})} {[1 -
    \exp(-t/t_{\rm rec})]^{2/3}},
\end{equation}
where $t_{\rm rec} = (\alpha_{\rm B} n_H)^{-1}$ is the recombination
time.

We adopt the problem parameters used in RT06.  The ionising
source emits $5 \times 10^{48}$ ph s$^{-1}$ of monochromatic radiation
at 13.6 eV and is located at the origin in a simulation box of 6.6
kpc.  The ambient medium is initially set at $T=10^4$ K, $n_H =
10^{-3}\;\cubecm$, $x = 1.2 \times 10^{-3}$, resulting in $R_s = 5.4$
kpc and $t_{\rm rec} = 122.4$ Myr.  The problem is run for 500 Myr.
In the original tests, the temperature is fixed at $10^4$ K; however,
our solver is inherently tied to the chemistry and energy solver.  To
mimic an isothermal behavior, we set the adiabatic index $\gamma =
1.0001$, which ensures an isothermal state but not a fixed ionisation
fraction outside of the Str\"{o}mgren sphere.

In Fig. \ref{fig:test1_ifront}, we show (a) the evolution of the
neutral and ionisation fraction as a function of radius at $t = $ 10,
30, 100, and 500 Myr, and (b) the growth of the ionisation front
radius as a function of time and its ratio with the analytical
Str\"{o}mgren radius [equation (\ref{eqn:rstr})].  The ionisation
front has a width of $\sim0.7$ kpc, which is in agreement with the
inherent thickness of $\sim18 \lambda_{\rm mfp} = 0.74$ kpc, given a
13.6 eV mono-chromatic spectrum.  There are small kinks in the neutral
fraction at 1.5 and 3 kpc that corresponds to artifacts created by the
photon package splitting at these radii.  However these do not affect
the overall solution.  One difference between our results and the
codes presented in RT06 is the increasing neutral fraction outside of
the \hii region.  This occurs because the initial ionised fraction and
temperature is set to $1.2 \times 10^{-3}$ and $10^4$ K, which are not
the equilibrium values.  Over the 500 Myr in the calculation, the
neutral fraction increases to 0.2, which is close to its equilibrium
value.  In the right panel of Fig. \ref{fig:test1_ifront}, the
ionisation front radius exceeds $R_s$ by a few percent for most of the
calculation.  This difference happens because the analytical solution
[equation (\ref{eqn:rstr})] assumes the \hii region has a constant
ionised fraction.  The evolution of the ionised fraction as a function
of radius can be analytically calculated \citep[e.g.][]{Osterbrock89,
  Petkova09}, causing the ionisation front radius to be slightly
larger, increasing from 0 to 3\% in the interval 80--350 Myr.  Our
results are in excellent agreement with this more accurate analytical
solution.  We show the distribution of neutral fraction in the domain
for $t$ = 10, 100, and 500 Myr in Fig. \ref{fig:test1_pdf} that agrees
well with the results in RT06.  In Fig. \ref{fig:test1_HI}, we show a
slice of the neutral fraction through the origin.  Other than the ray
splitting artifacts that generate the plateaus at 1.5 and 3 kpc, one
sees spherical symmetry without any noise in our solution.

\subsection{Test 2. \hii region expansion: temperature evolution}

\begin{figure}
  \plotone{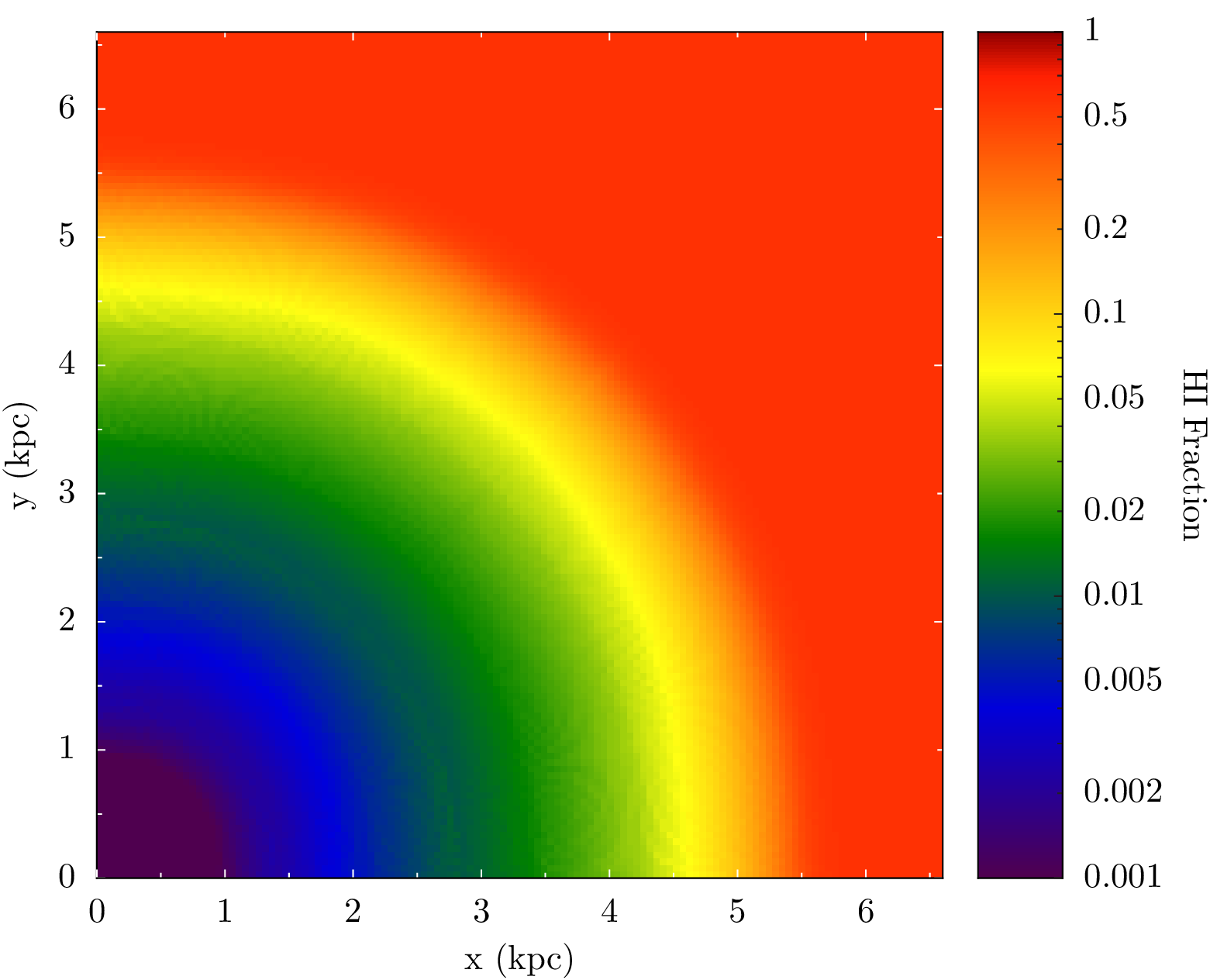}
  \caption{\label{fig:test2_1a} Test 2. (\hii region expansion with a
    $T=10^5$ K blackbody spectrum).  Radially averaged profile of
    neutral (solid) and ionised (dashed) fraction at 10, 30, 100, and
    500 Myr.}
\end{figure}

\begin{figure}
  \plotone{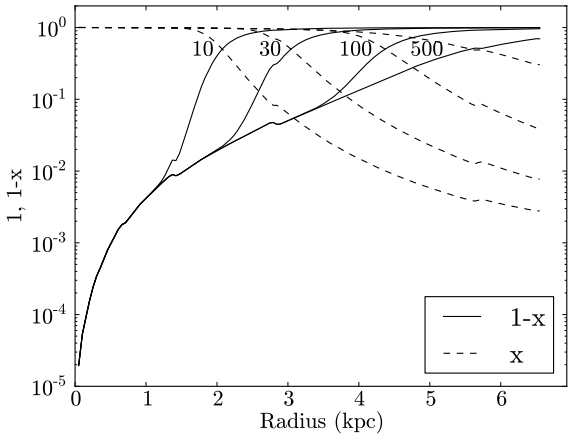}
  \caption{\label{fig:test2_1b} Test 2. (\hii region expansion with a
    $T=10^5$ K blackbody spectrum).  Evolution of the average neutral
    fraction.}
\end{figure}

\begin{figure}
  \plotone{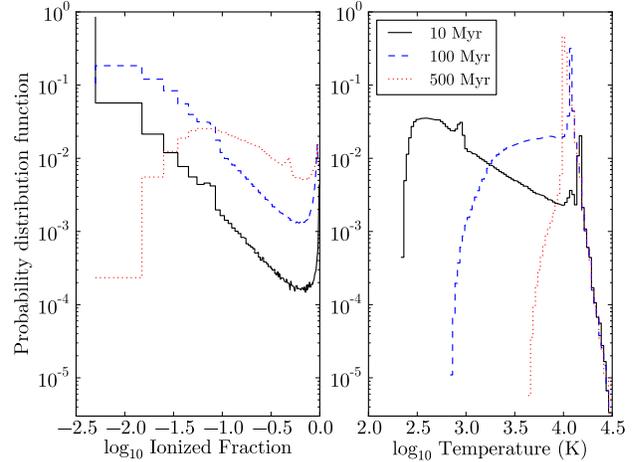}
  \caption{\label{fig:test2_pdf} Test 2. (\hii region expansion with a
    $T=10^5$ K blackbody spectrum).  Probability distribution function
    for ionized fraction (left) and temperature (right) at 10 Myr
    (solid), 100 Myr (dashed), and 500 Myr (dotted).}
\end{figure}

\begin{figure*}
  \epsscale{2}
  \plotone{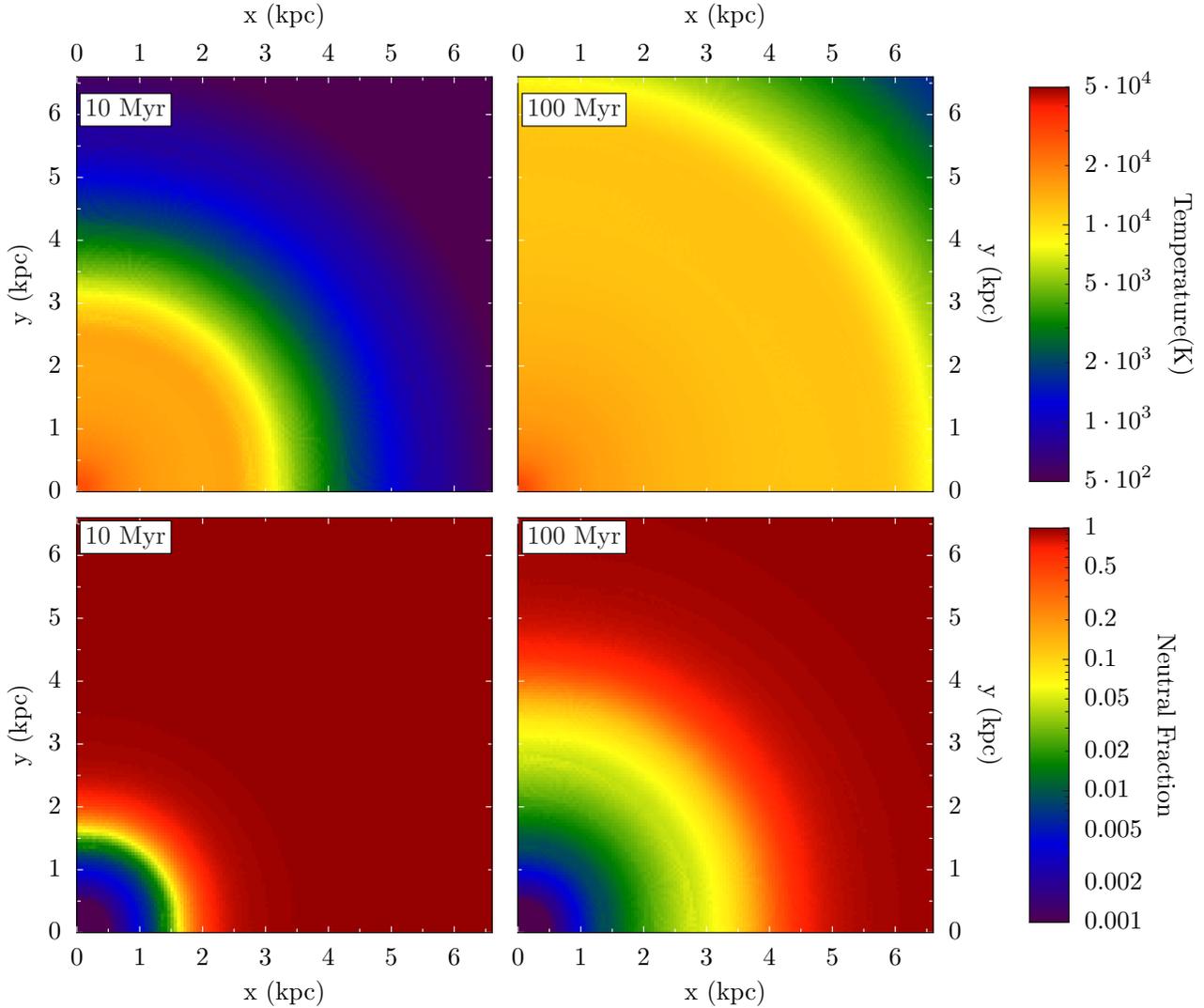}
  \epsscale{1}
  \caption{\label{fig:test2_2} Test 2. (\hii region expansion
    with a $T=10^5$ K blackbody spectrum).  Top: Slices through the
    origin of neutral fraction at 10 and 100 Myr.  Bottom: Slices of
    temperature at 10 and 100 Myr.}
\end{figure*}

\begin{figure}
  \plotone{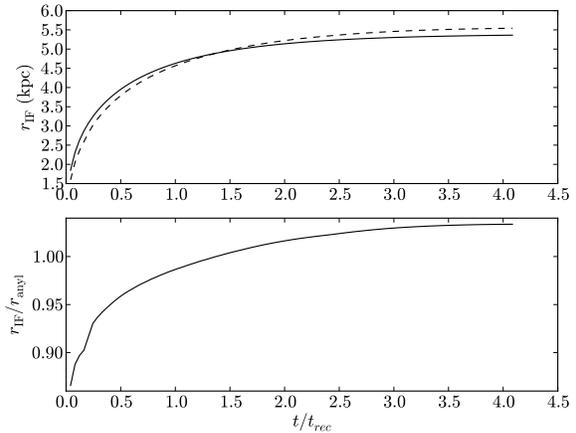}
  \caption{\label{fig:test2_3} Test 2. (\hii region expansion
    with a $T=10^5$ K blackbody spectrum).  Top: Evolution of the
    radius of the simulated ionisation front (dashed) and analytical
    (solid) Str\"{o}mgren radius.  Bottom: The ratio of the calculated
    and analytical Str\"{o}mgren radius.}
\end{figure}

This test is similar to Test 1, but the temperature is allowed to
evolve.  The radiation source now has a blackbody spectrum with a $T =
10^5$ K.  The initial temperature is set at 100 K.  The higher energy
photons have a longer mean free path than the photons at the
ionisation threshold in Test 1.  Thus the ionisation front is thicker
as the photons can penetrate deeper into the neutral medium.  Here we
use 4 energy groups with the following mean energies and relative
luminosities: $E_i = (16.74, 24.65, 34.49, 52.06)$, $L_i/L = (0.277,
0.335, 0.2, 0.188)$.

In Fig. \ref{fig:test2_1a}, we show the radially averaged neutral and
ionised fraction at $t = $ 10, 30, 100, and 500 Myr, and the total
neutral fraction of the domain in Fig. \ref{fig:test2_1b}.  Compared
with Test 1, the ionisation front is thicker, as expected with the
harder spectrum.  Artifacts originating from ray splitting, similar to
Test 1, appear at $r \sim 1$ and 3 kpc as kinks in the neutral
fraction.  The total neutral fraction decreases to 0.67 over 4$t_{\rm
  rec} = 500$ Myr, which is in agreement with the analytical
expectation and other codes in RT06.  Fig. \ref{fig:test2_pdf} shows
the distribution of ionized fraction and temperature in Test 2.  The
overall trends agree with the codes presented in RT06 with the
exception of the ray splitting artifacts that appear as slight rises
in the distribution at $\log x_e \sim -1$ and $\log T \sim 3$. In
Fig. \ref{fig:test2_2}, we show slices of neutral fraction and
temperature through the origin at $t = $ 10 and 100 Myr.  Here one
sees the spherically symmetric \hii regions and a smooth temperature
transition to the neutral ambient medium.  In Fig. \ref{fig:test2_3},
we show the ratio of the ionisation front radius $r_{\rm IF}$ in our
simulation and $R_s$.  Before $1.5t_{\rm rec}$, $r_{\rm IF}$ lags
behind $R_s$, initially by 10\% and then increases to $R_s$; however
afterwards, this ratio asymptotes to a solution that is 4\% greater
than $R_s$.  This behavior is approximately the median result in RT06,
where this ratio varies between 1 and 1.1, and the early evolution of
$r_{\rm IF}$ is under-predicted by almost all of the codes.  If we use
one energy group with the mean energy (29.6 eV) of a $T=10^5$ K
blackbody, we find that $r_{\rm IF}/R_s$ = 1.08, which is
representative of the codes in the upper range of RT06.

\subsection{Test 3. I-front trapping in a dense clump and the
  formation of a shadow}
\label{sec:test3}

\begin{figure}
  \plotone{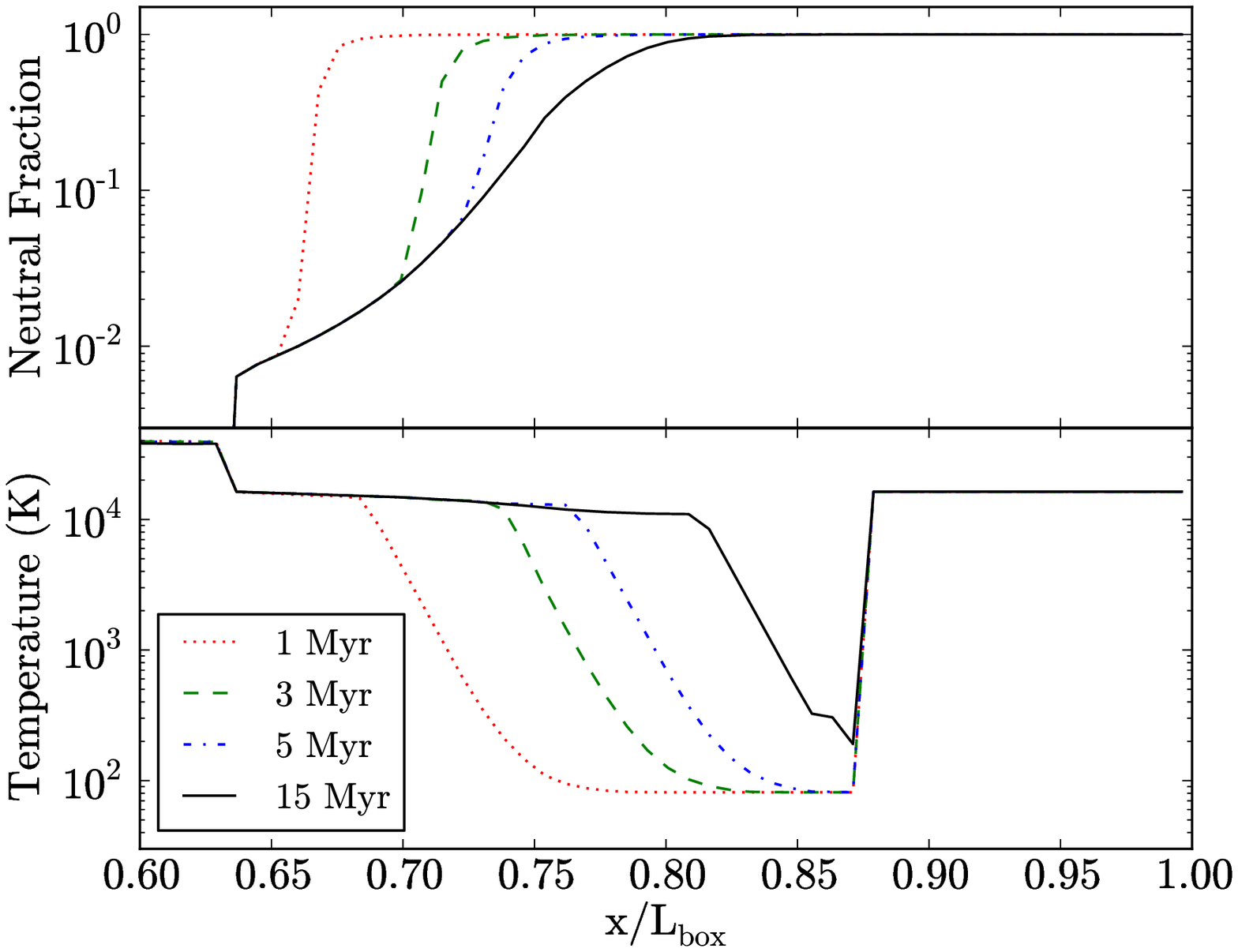}
  \caption{\label{fig:test3_1} Test 3. (I-front trapping in a dense
    clump and shadowing).  Line cut from the point source through the
    middle of the dense clump at $t = 1, 3, 5, 15$ Myr.  of the
    average neutral fraction (top) and temperature (bottom) of the
    clump.}
\end{figure}

\begin{figure}
  \plotone{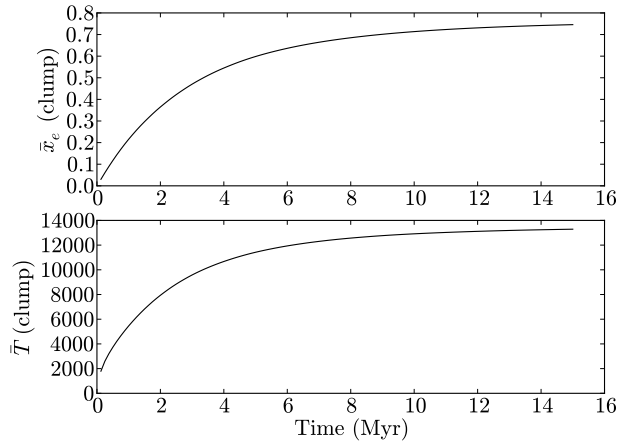}
  \caption{\label{fig:test3_2} Test 3. (I-front trapping in a dense
    clump and shadowing).  Evolution of the average ionised fraction
    (top) and temperature (bottom) of the overdense clump.}
\end{figure}

\begin{figure}
  \plotone{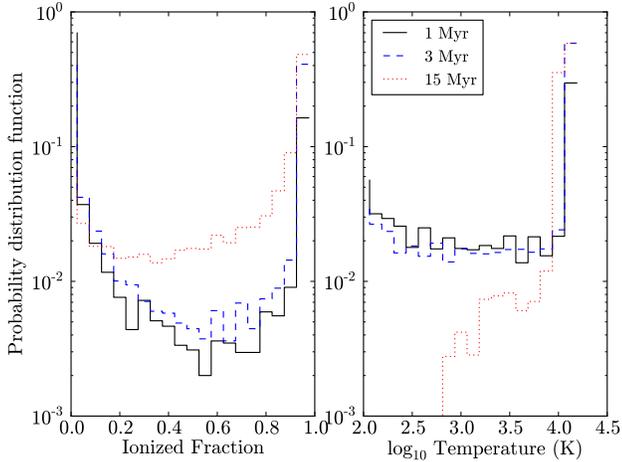}
  \caption{\label{fig:test3_pdf} Test 3. (I-front trapping in a dense
    clump and shadowing). Probability distribution function for
    ionized fraction (left) and temperature (right) at 1 Myr (solid),
    3 Myr (dashed), and 15 Myr (dotted) within the dense clump.}
\end{figure}

\begin{figure*}
  \epsscale{2}
  \plotone{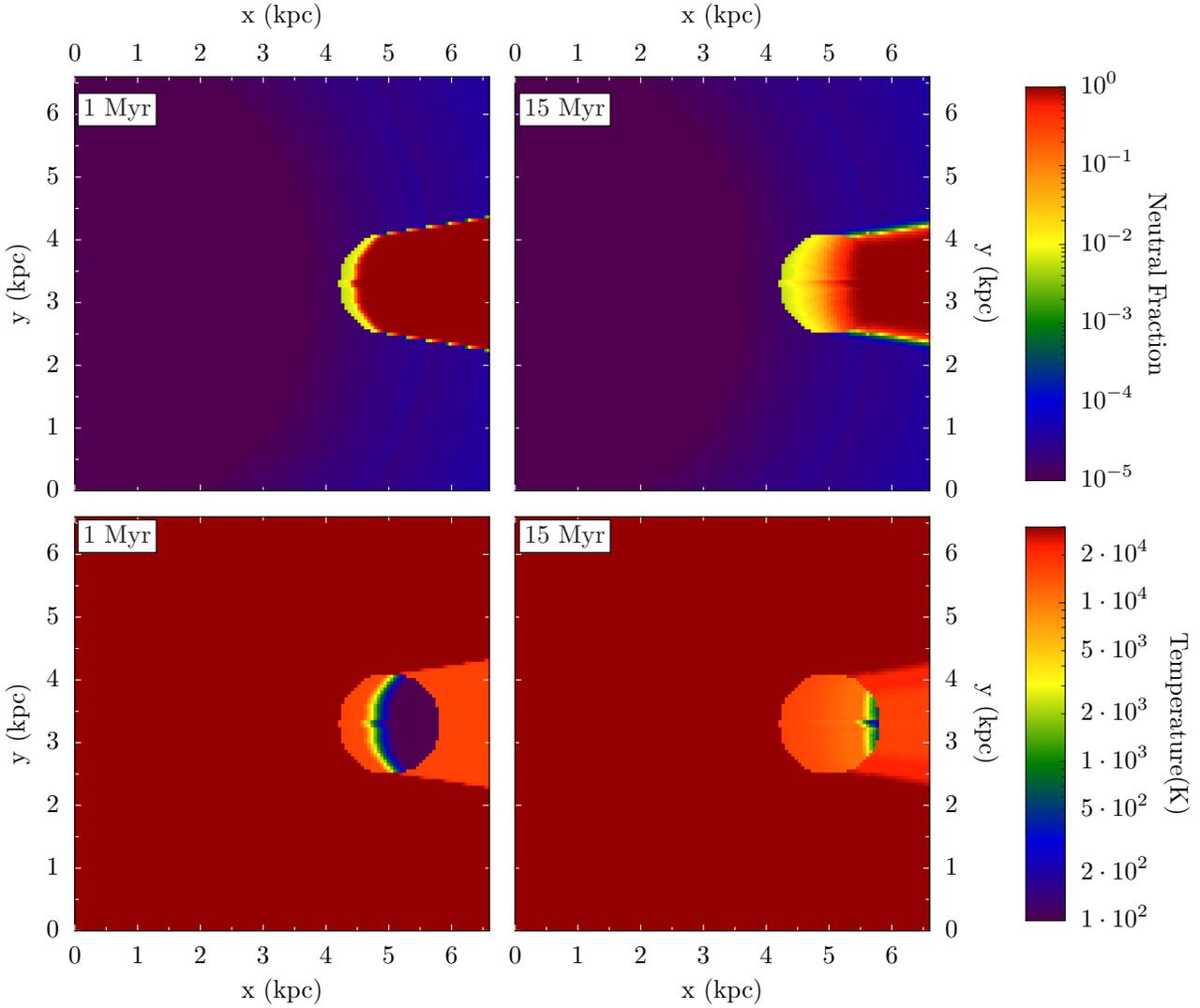}
  \epsscale{1}
  \caption{\label{fig:test3_3} Test 3. (I-front trapping in a dense
    clump and shadowing).  Clockwise from upper left: Slices through
    the origin of neutral fraction (1 Myr), temperature (1 Myr),
    temperature (15 Myr), and neutral fraction (15 Myr).}
\end{figure*}

The diffusivity and angular resolution of a radiative transport method
can be tested with the trapping of an ionisation front by a dense,
neutral clump.  In this situation, the ionisation front will uniformly
propagate until it reaches the clump surface.  Then the radiation in
the line of sight of the clump will be absorbed more than the ambient
medium.  If the clump is optically thick, a shadow will form behind
the clump.  The sharpness of the ionisation front at the shadow
surface can be used to determine the diffusivity of the method.
Furthermore the shadow surface should be aligned with the outermost
neutral regions of the clump, which can visually assess the angular
resolution of the method.

The problem for this test is contained in a 6.6 kpc box with an
ambient medium of $n_H = 2 \times 10^{-4}\;\cubecm$ and $T = 8000$ K.
The clump is in pressure equilibrium with $n_H = 0.04\; \cubecm$ and
$T = 40$ K.  It has a radius of $r_c = 0.8$ kpc and is centred at
$(x,y,z) = (5, 3.3, 3.3)$ kpc.  The ionized fraction of the entire
domain is zero.  In RT06, the test considered a plane parallel
radiation field with a flux $F_0 = 10^6$ ph s$^{-1}$ cm$^{-2}$
originating from the $y=0$ plane.  Our code can only consider point
sources, so we use a single radiation source located in the centre of
the $y=0$ boundary.  The luminosity of $\dot{N}_\gamma = 3 \times
10^{51}$ ph s$^{-1}$ corresponds to the same flux $F_0$ at 5 kpc.  The
location where the ionisation front trapping can be calculated
analytically \citep{Shapiro04}, and with these parameters, it should
halt at approximately the centre of the clump.  We use the same four
energy groups as in Test 2.

In Fig. \ref{fig:test3_1}, we show neutral fraction and temperature of
a one-dimensional cut through the centre of the dense clump at $z =
0.5$ at $t = 1, 3, 5, 15$ Myr.  The ionisation front is halted at a
little more than halfway through the clump, which is consistent with
the analytical expectation.  The hardness of the $T = 10^5$ K
blackbody spectrum allows the gas outside of the ionisation front to
be photo-heated.  Where the gas is ionised, the temperature is between
10,000 and 20,000 K, but decreases sharply with the ionised fraction.
Fig. \ref{fig:test3_2} depicts the average ionised fraction and
temperature inside the dense clump, which both gradually increase as
the ionisation front propagates through the overdensity.  Our results
are consistent with RT06.  The distribution of ionized fraction and
temperature within the clump are shown in Fig. \ref{fig:test3_pdf} and
agree well with the other codes in RT06.  Finally we show slices of
neutral fraction and temperature in the $z = 0.5$ plane in
Fig. \ref{fig:test3_3}.  The neutral fraction slices prominently show
the sharp shadows created by the clump and demonstrates the
non-diffusivity behavior of ray tracing.  The discretisation of the
sphere creates one neutral cell on the left side of the sphere.  This
inherent artifact to the initial setup carries through the
calculation.  We did not smooth the clump surface like in some of the
RT06 codes, in order to remove this artifact.  It is seen in the
neutral fraction and temperature states at all times and is not a
caused by our ray tracing algorithm.

\subsection{Test 4. Multiple sources in a cosmological density field}

\begin{figure}
  \plotone{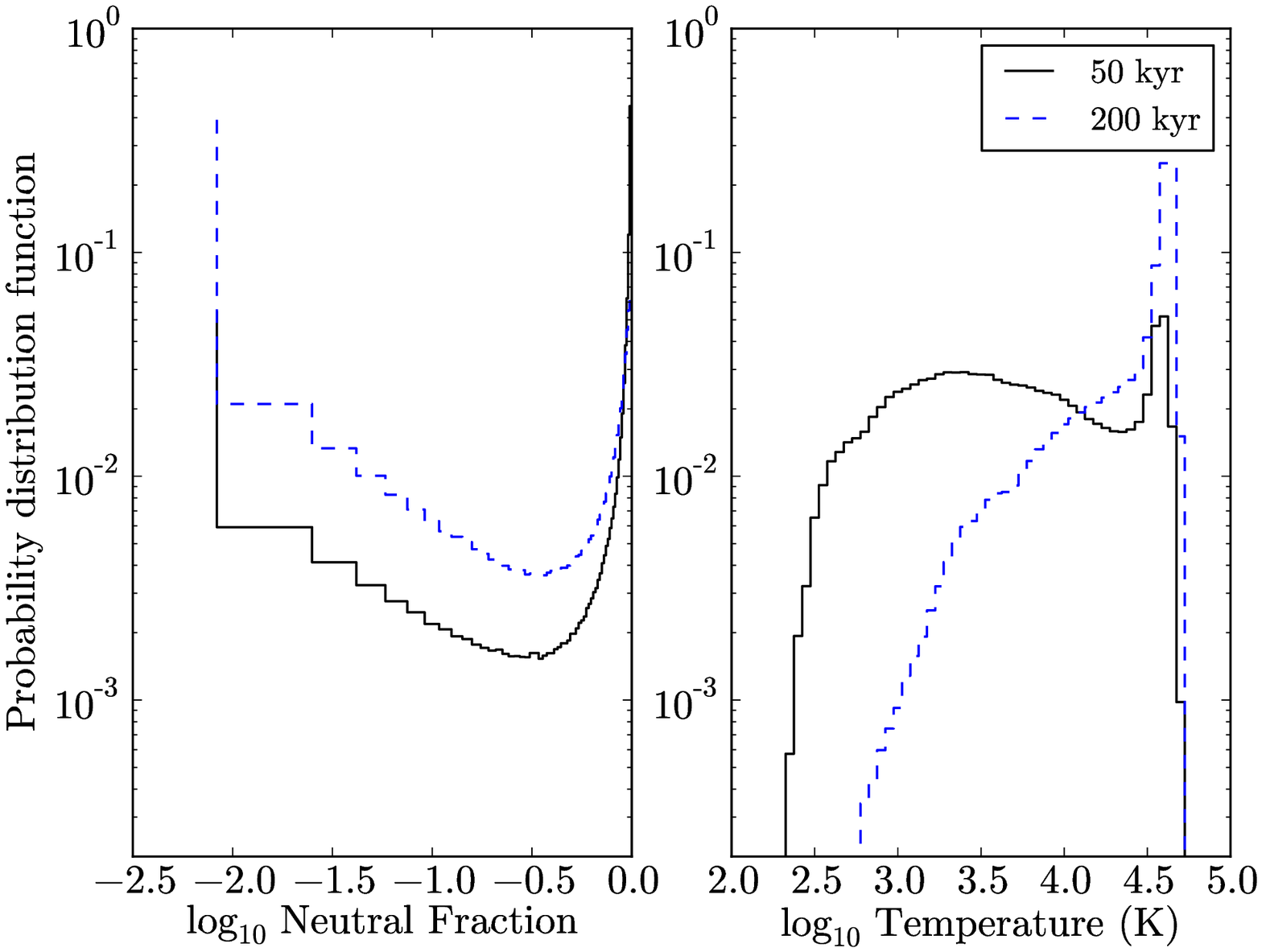}
  \caption{\label{fig:test4_pdf} Test 4. (Multiple cosmological
    sources). Probability distribution function for neutral fraction
    (left) and temperature (right) at 50 kyr (solid) and 200 kyr
    (dashed).}
\end{figure}

\begin{figure}
  \plotone{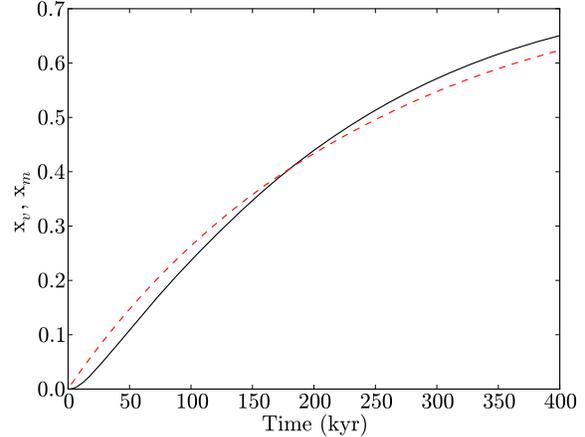}
  \caption{\label{fig:test4_1} Test 4. (Multiple cosmological
    sources).  Evolution of the mass- (dashed) and volume-averaged
    (solid) ionised fraction.}
\end{figure}

\begin{figure*}
  \epsscale{2}
  \plotone{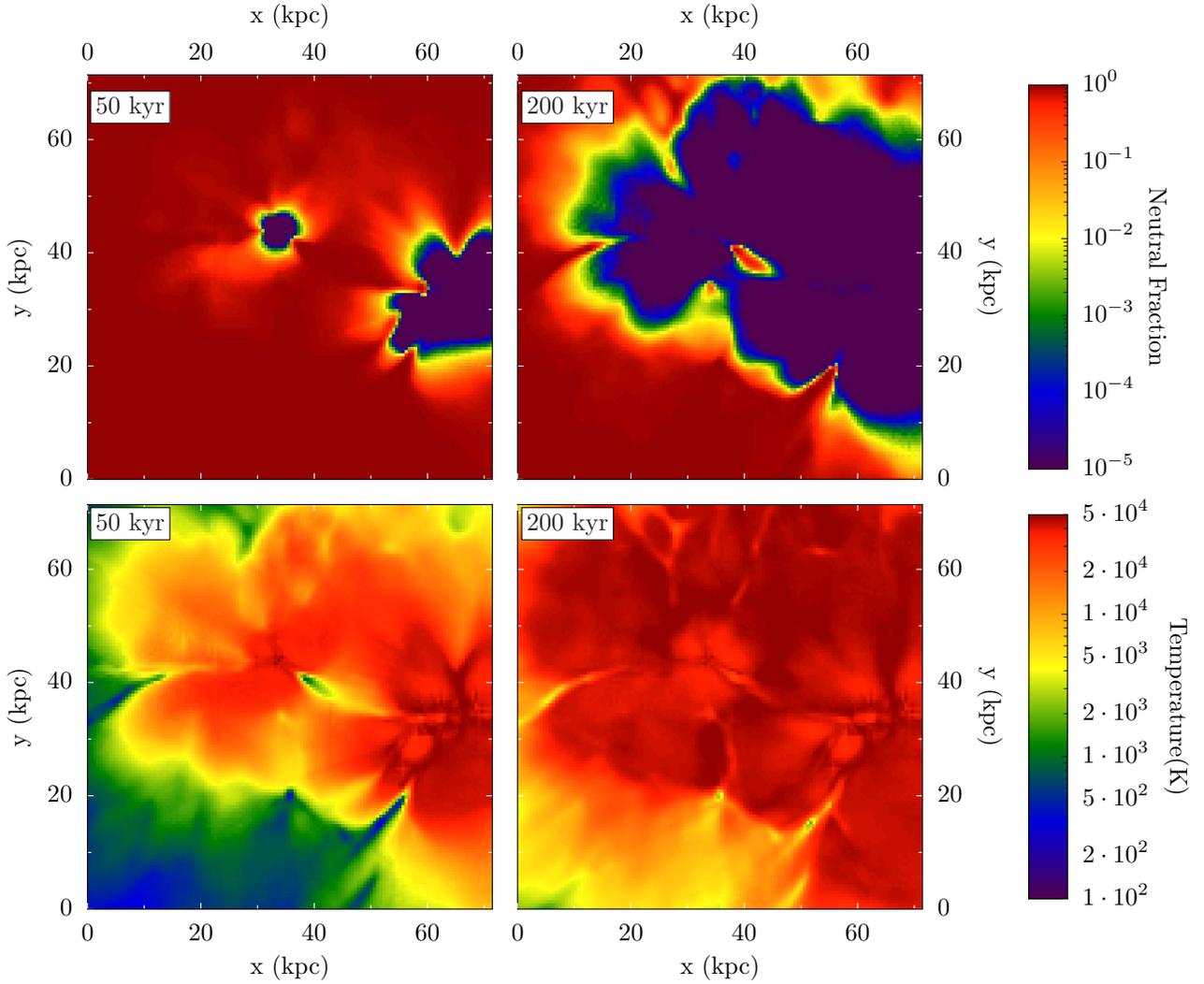}
  \epsscale{1}
  \caption{\label{fig:test4_2} Test 4. (Multiple cosmological
    sources).  Top: Slices through the origin of neutral fraction at
    50 and 200 kyr at the coordinate $z = z_{\rm box}/2$.  Bottom:
    Slices of temperature at 50 and 200 kyr.  No smoothing has been
    applied to the images.}
\end{figure*}

The last test in RT06 involves a static cosmological density field at
$z=9$.  The simulation comoving box size is 0.5 $h^{-1}$ Mpc and has a
resolution of 128$^3$.  There are 16 point sources that are centred
in the 16 most massive haloes.  They emit $f_\gamma = 250$ ionising
photons per baryon in a blackbody spectrum with an effective
temperature $T = 10^5$ K, and they live for $t_s = 3$ Myr.  Thus the
luminosity of each source is
\begin{equation}
  \label{eqn:cosmo_lum}
  \dot{N}_\gamma = f_\gamma \frac{M \Omega_b} {\Omega_m m_H t_s},
\end{equation}
where $M$ is the halo mass, $\Omega_m = 0.27$, and $\Omega_b =
0.043$.  The radiation boundaries are isolated so that the radiation
leaves the box instead being shifted periodically.  The simulation is
evolved for 0.4 Myr.

Fig. \ref{fig:test4_pdf} depicts the distribution of the neutral
fraction and temperature of the entire domain and shows good agreement
with the codes presented in RT06.  We show the growth of the \hii
regions by computing the mass-averaged $x_m$ and volume-averaged $x_v$
ionised fraction in Fig. \ref{fig:test4_1}.  Initially $x_m$ is larger
than $x_v$, and at $t \sim 170$ kyr, the $x_v$ becomes larger.  This
is indicative of inside-out reionisation \citep[e.g.][]{Gnedin00,
  Miralda00, Sokasian03}, where the dense regions around haloes are
ionised first, then the voids are ionised last.  At the end of the
simulation, 65\% of the simulation is ionised.  However by visual
inspection in the slices of electron fraction
(Fig. \ref{fig:test4_2}), there appears to be very good agreement with
C$^2$-ray and FTTE.  By first glance, our result appears to be
different than the RT06 because of the color-mapping.  Our results are
also in good agreement with the multi-frequency version of TRAPHIC
\citep[][see also for better representations of the electron fraction
slices]{Pawlik10}.  In the slices of electron fraction and
temperature, Fig. \ref{fig:test4_2}, the photo-heated regions are
larger than the ionised regions by a factor of 2--3 because of the
hardness of the $T = 10^5$ K blackbody spectrum.

\section{Radiation Hydrodynamics Tests}
\label{sec:radhydro}

We next show results from radiation hydrodynamics test problems
presented in \citet[][hereafter RT09]{Iliev09}.  They involve (1) the
expansion of an \hii region in a uniform medium, similar to Test 2,
(2) an \hii region in an isothermal sphere, and (3) the
photo-evaporation of a dense, cold clump, similar to Test 3.  We turn
off self-gravity and AMR in accordance with RT09.

\subsection{Test 5. Classical \hii region expansion}
\label{sec:test5}

\begin{figure}
  \plotone{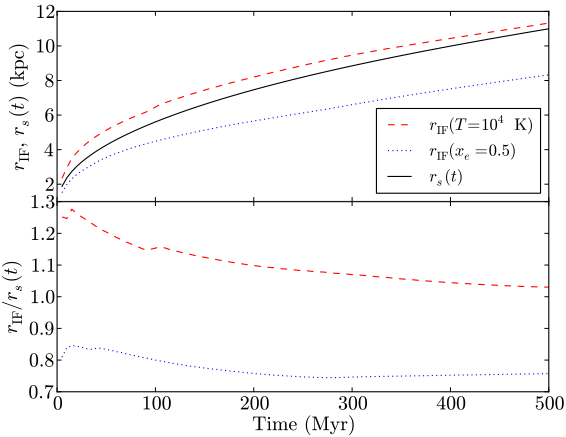}
  \caption{\label{fig:test5_1} Test 5. (\hii region in a uniform
    medium).  Top: Growth of the computed ionisation front radius at
    an ionised fraction $x_e = 0.5$ (dashed) and at a temperature $T =
    10^4$ K (dotted) compared to the analytical estimate [solid;
    equation (\ref{eqn:test5_r})].  Bottom: The ratio of the computed
    ionisation front radii to the analytical estimate.} 
\end{figure}

\begin{figure}
  \plotone{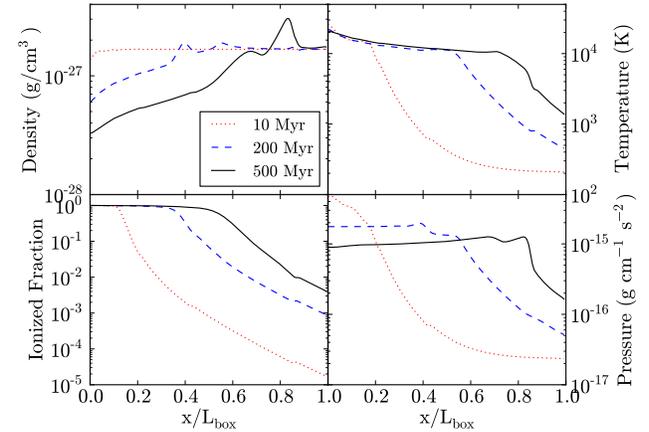}
  \caption{\label{fig:test5_2} Test 5. (\hii region in a uniform
    medium).  Clockwise from the upper left: Radial profiles of
    density, temperature, ionised fraction, and pressure at times $t
    =$ 10, 200, and 500 Myr.}
\end{figure}

\begin{figure*}
  \epsscale{2}
  \plotone{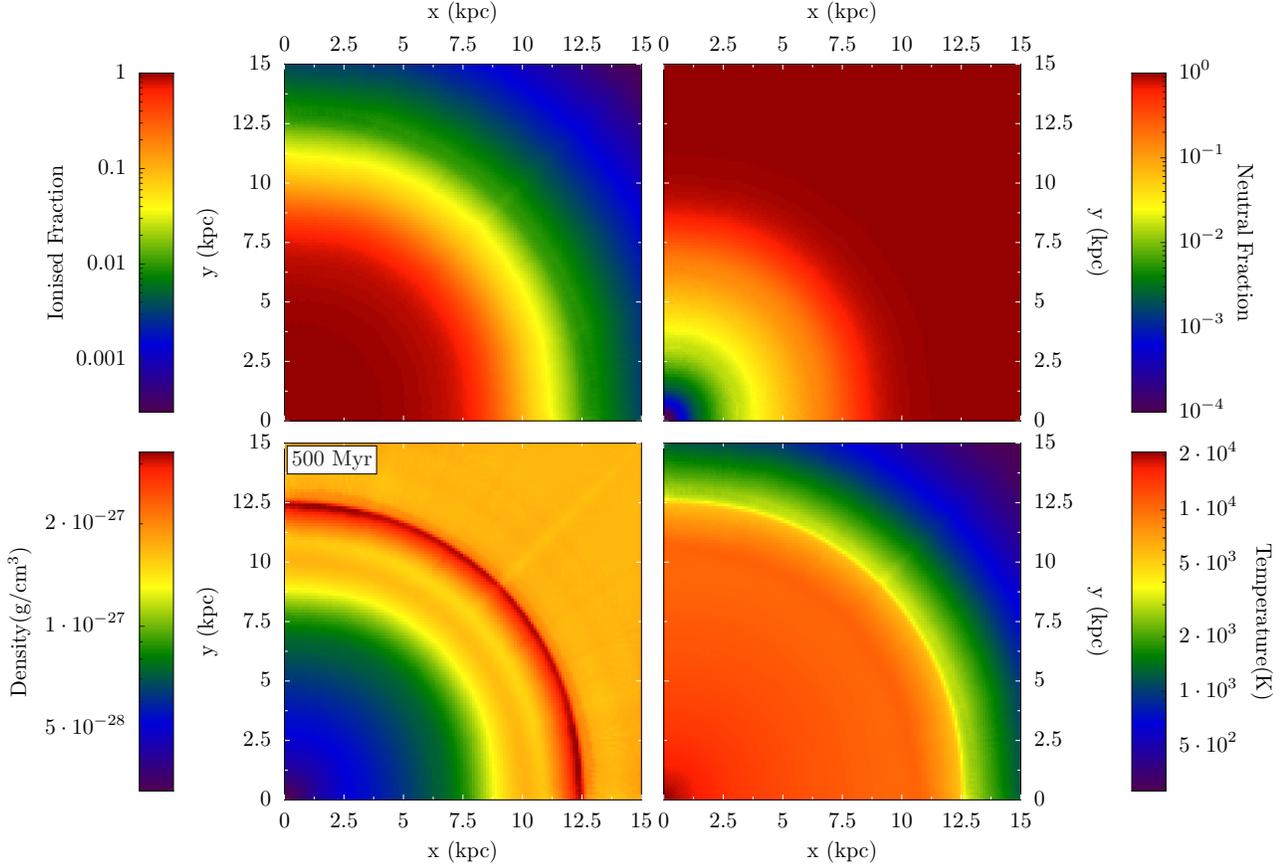}
  \epsscale{1}
  \caption{\label{fig:test5_3} Test 5. (\hii region in a uniform
    medium).  Clockwise from the upper left: Slices through the origin
    of ionised fraction, neutral fraction, temperature, and density at
    time $t =$ 500 Myr.}
\end{figure*}

Here we consider the expansion of an \hii region into a uniform
neutral medium including the hydrodynamical response to the heated
gas.  The ionised region has a greater pressure than the ambient
medium, causing it to expand.  This is a well-studied problem
\citep{Spitzer78} with an analytical solution, where the ionisation
front moves as
\begin{equation}
  \label{eqn:test5_r}
  r_s(t) = r_{s,0} \left(1 + \frac{7c_s}{4R_s}\right)^{4/7},
\end{equation}
where $c_s$ is the sound speed of the ionised gas and $r_{s,0}$ is the
$r_s$ in Equation \ref{eqn:rstr}.  The bubble eventually reaches
pressure equilibrium with the ambient medium at a radius
\begin{equation}
  \label{eqn:test5_final}
  r_f = R_s \left(\frac{2T}{T_0}\right)^{2/3},
\end{equation}
where $T$ and $T_0$ are the ionised and ambient temperatures,
respectively.  These solutions only describe the evolution at late
times, and not the fast transition from R-type to D-type at early
times.

The simulation setup is similar to Test 2 with the exception of the
domain size $L = 15$ kpc.  Here pressure equilibrium occurs at $r_f =
185$ kpc, which is not captured by this test.  However more
interestingly, the transition from R-type to D-type is captured and
occurs around $R_s = 5.4$ kpc.  The test is run for 500 Myr.

The growth of the ionisation front radius is shown in Fig.
\ref{fig:test5_1}, using both $T = 10^4$ K and $x_e = 0.5$ as
ionisation front definitions, compared to the analytical solution
[equation (\ref{eqn:test5_final})].  We define this alternative
measure because the ionisation front becomes broad as the D-type front
creates a shock.  Densities in this shock, as seen in Fig.
\ref{fig:test5_2}, are high enough for the gas to recombine but not
radiatively cool.  Before $2t_{\rm rec} \approx 250$ Myr, the
temperature cutoff overestimates $r_s$ by over 10\%; however at later
times, it provides a good match to the $t^{4/7}$ growth at late times.
With the $x_e = 0.5$ criterion for the ionisation front, the radius is
always underestimated by $\sim20\%$.  This behavior was also seen in
RT09.

Fig. \ref{fig:test5_2} shows the progression of the ionisation front
at times $t$ = 10, 200, and 500 Myr in radial profiles of density,
temperature, pressure, and ionised fraction.  The initial \hii
region is over-pressurised and creates an forward shock wave.  The
high-energy photons can penetrate through the shock and partially
ionises and heats the exterior gas, clearly seen in the profiles.  As
noted in RT09, this heated exterior gas creates an photo-evaporative
flow that flows inward.  This interacts with the primary shock and
creates the double-peaked features in the density profiles at 200 and
500 Myr.  Fig. \ref{fig:test5_3} shows slices through the origin of
the same quantities, including neutral fraction.  These depict the
very good spherical symmetry of our method.  The only apparent
artifact is a very slight diagonal line, which is caused by the
HEALPix pixelisation differences between the polar and equatorial
regions.  This artifact diminishes as the ray-to-cell sampling is
increased.

\subsection{Test 6. \hii region expansion in an isothermal
  sphere}

\begin{figure}
  \plotone{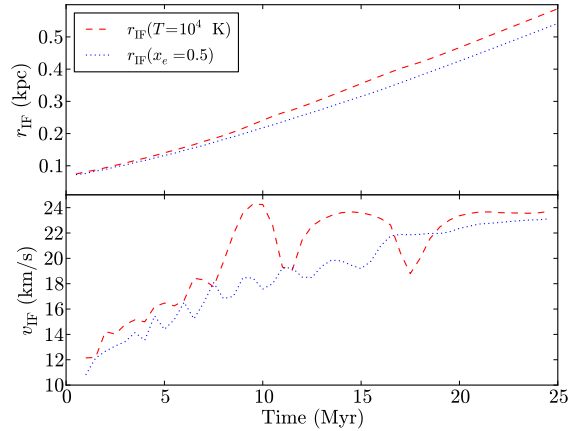}
  \caption{\label{fig:test6_1} Test 6. (\hii region in a 1/$r^2$
    density profile).  Top: Growth of the computed ionisation front
    radius as $T=10^4$ K and $x_e = 0.5$ as definitions for the front.
    Bottom: Velocity of the ionisation front, computed from outputs at
    0.5 Myr intervals.  The velocity is calculated from $r_{\rm IF}$,
    whose coarse time resolution causes the noise seen in $v_{\rm
      IF}$.  It is smooth within the calculation itself.}
\end{figure}

\begin{figure}
  \plotone{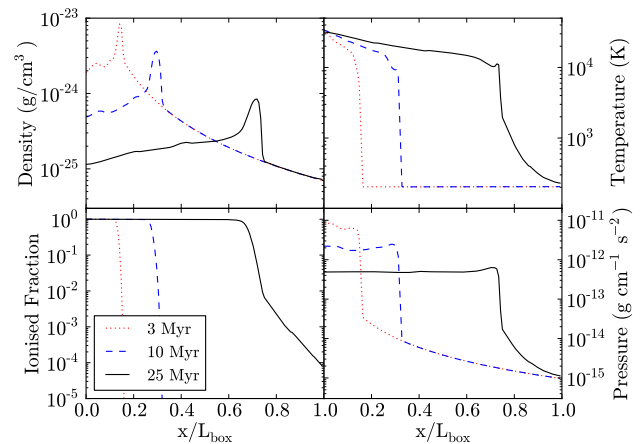}
  \caption{\label{fig:test6_2} Test 6. (\hii region in a 1/$r^2$
    density profile).  Clockwise from the upper left: Radial profiles of
    density, temperature, ionised fraction, and pressure at times $t
    =$ 3, 10, and 25 Myr.}
\end{figure}

\begin{figure*}
  \epsscale{2}
  \plotone{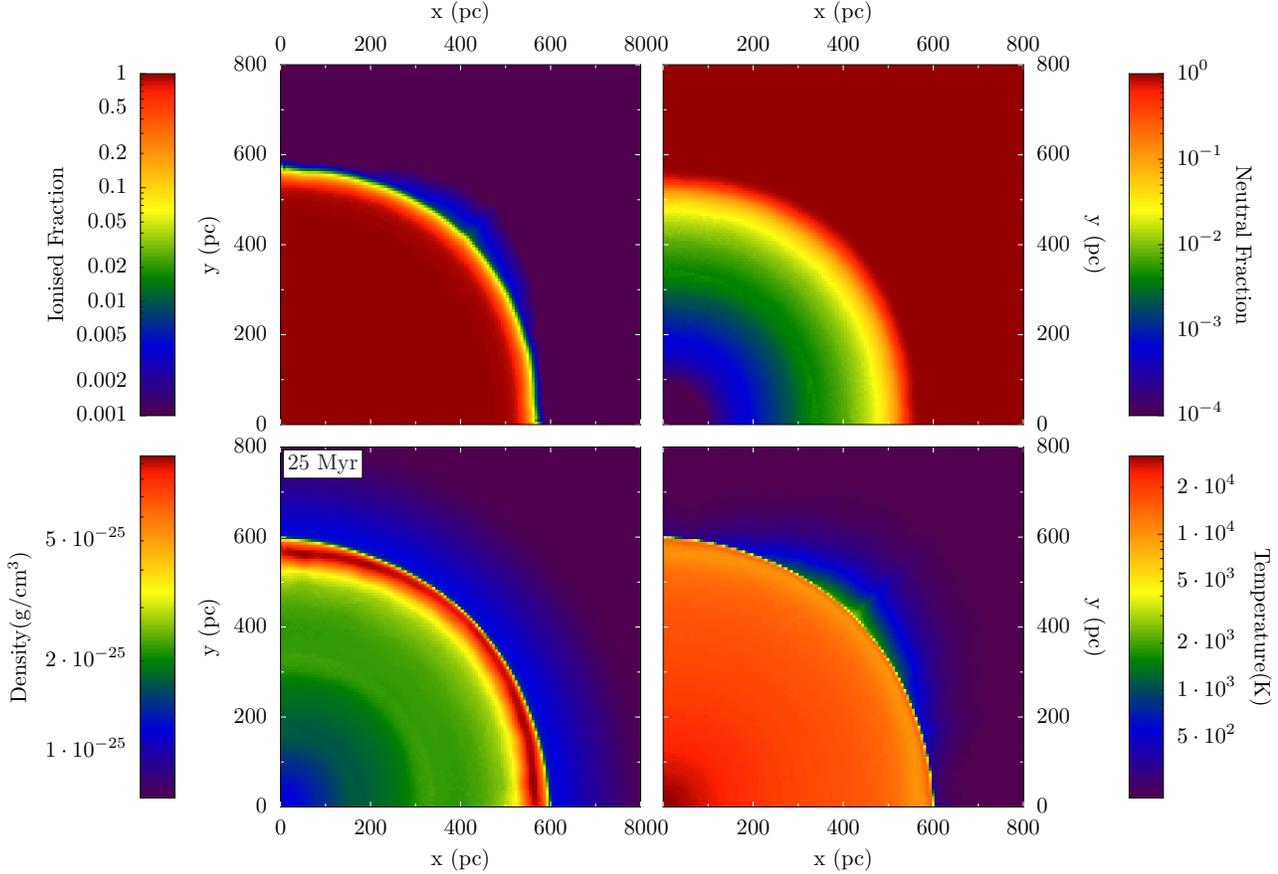}
  \epsscale{1}
  \caption{\label{fig:test6_3} Test 6. (\hii region in a 1/$r^2$
    density profile).  Clockwise from the upper left: Slices through the origin
    of ionised fraction, neutral fraction, temperature, and density at
    time $t =$ 25 Myr.}
\end{figure*}

A more physically motivated scenario is an isothermal sphere with a
constant density $n_c$ core, which is applicable to collapsing
molecular clouds and cosmological haloes.  The radial density profile
is described by
\begin{equation}
  \label{eqn:test6_rho}
  n(r) = \left\{ \begin{array}{l@{\quad}l}
      n_c & (r \le r_0)\\
      n_c (r/r_0)^{-2} & (r > r_0)
    \end{array} \right. ,
\end{equation}
where $r_0$ is the radius of the core.  If the Str\"{o}mgren radius is
smaller than the core radius, then the resulting \hii region
never escapes into the steep density slope.  When the ionisation front
propagates out of the core, it accelerates as it travels down the
density gradient.  There exists no analytical solution for this
problem with full gas dynamics but was extensively studied by
\citet{Franco90}.  After the gas is ionised and photo-heated, the
density gradient provides the pressure imbalance to drive the gas
outwards.

This test is constructed to study the transition from R-type to D-type
in the core and back to R-type in the density gradient.  Thus the
simulation focuses on small-scale, not long term, behavior of the
ionisation front.  The simulation box has a side length $L = 0.8$ kpc
with core density $n_0 = 3.2\;\cubecm$, core radius $r_0 = 91.5$ pc
(15 cells), and temperature $T = 100$ K throughout the box.  The
ionisation fraction is initially zero, and the point source is located
at the origin with a luminosity of $10^{50}$ ph s$^{-1}$ \cubecm.  The
simulation is run for 75 Myr.

Because this problem does not have an analytical solution, we compare
our calculated ionisation front radius and velocity, shown in Fig.
\ref{fig:test6_1}, to the RT09 results.  Their evolution are in
agreement within 5\% of RT09.  As in Test 5, we use an extra
definition of $T=10^4$ K for the ionisation front.  We compute the
ionisation front velocity from the radii at 50 outputs, which causes
the noise seen Fig. \ref{fig:test6_1}.

For the first Myr, the radiation source creates a weak R-type front
where the medium is heated and ionised but does not expand because
$v_{\rm IF} > c_s$.  When $v_{\rm IF}$ becomes subsonic, the medium
can react to the passing ionisation front and creates a shock, leaving
behind a heated rarefied medium.  This behavior is clearly seen in the
radial profiles of density, temperature, ionised fraction, and
pressure in Fig. \ref{fig:test6_2}.  The inner density decreases over
two order of magnitude after 25 Myr.  To illustrate any deviations in
spherical symmetry, we show in Fig. \ref{fig:test6_3} slices of
density, temperature, neutral fraction, and ionised fraction at the
final time.  The only artifact apparent to us is the slight broadening
of the shock near the $x=0$ and $y=0$ planes.  This causes the
ionisation front radius to be slightly smaller in those directions.
In the diagonal direction, the neutral column density through the
shock is slightly smaller, allowing the high-energy photons to
photo-ionise and photo-heat the gas to $x_e = 5 \times 10^{-3}$ and $T
= 2000$ K out to $\sim50$ pc from the shock.  The reflecting
boundaries are responsible for this artifact because this is not seen
when the problem is centred in the domain, removing any boundary
effects.

\subsection{Test 7. Photo-evaporation of a dense clump}

\begin{figure}
  \plotone{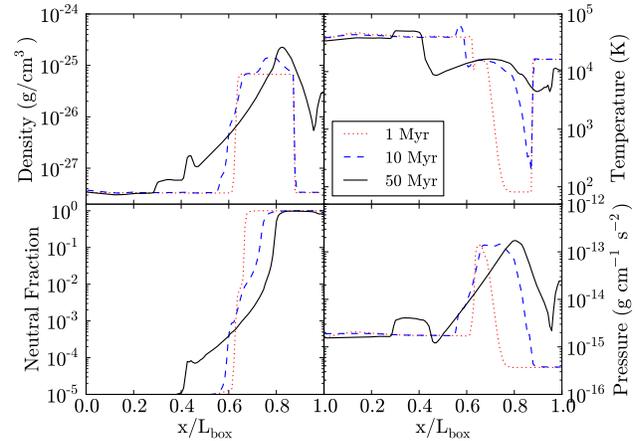}
  \caption{\label{fig:test7_1} Test 7. (Photo-evaporation of a dense
    clump).  Line cuts from the point source through the middle of the
    dense clump at $t = 1, 10, 50$ Myr of (clockwise from the upper
    left) density, temperature, pressure, and neutral fraction.}
\end{figure}

\begin{figure*}
  \epsscale{2}
  \plotone{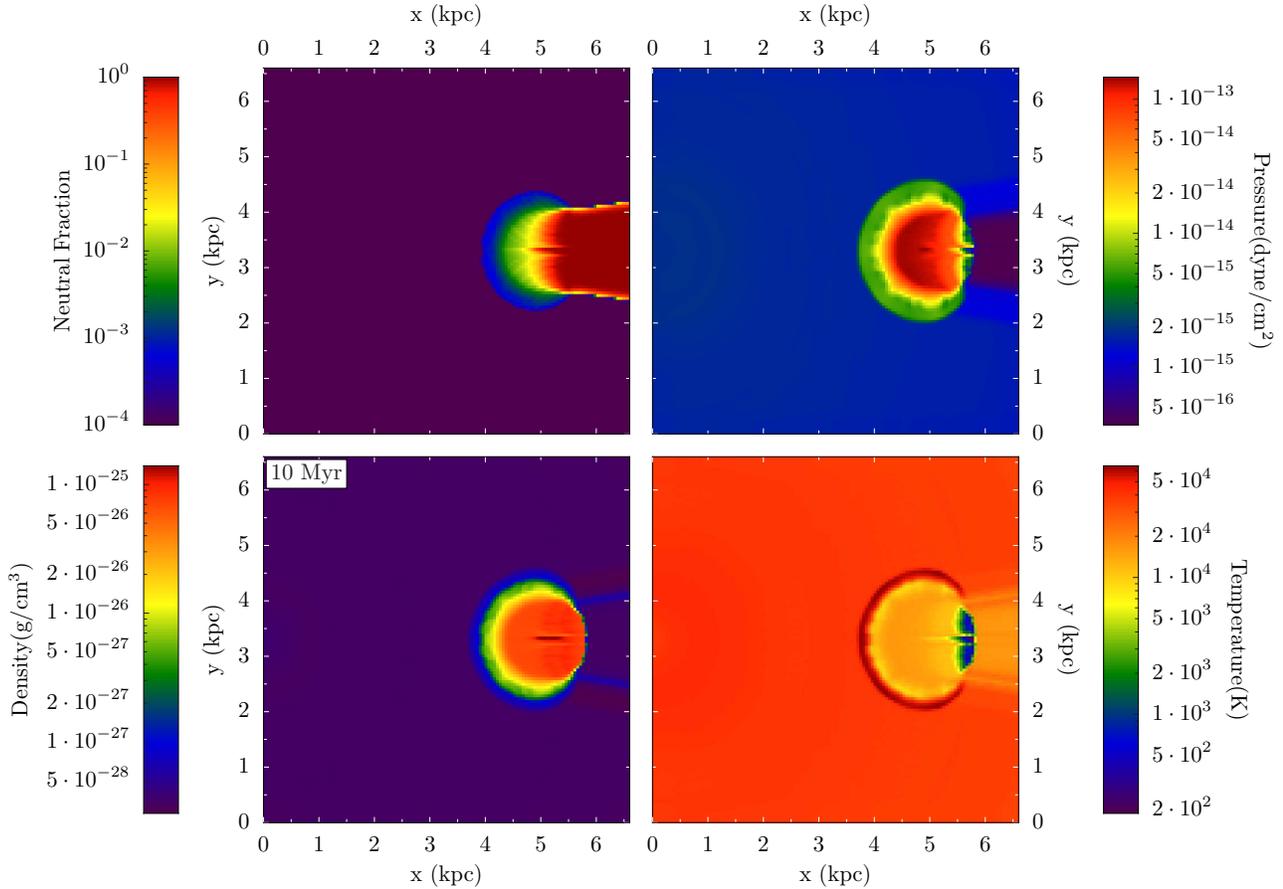}
  \epsscale{1}
  \caption{\label{fig:test7_2} Test 7. (Photo-evaporation of a dense
    clump).  Clockwise from the upper left: Slices through the clump
    centre of neutral fraction, pressure, temperature, and density at
    time $t =$ 10 Myr.}
\end{figure*}

\begin{figure*}
  \epsscale{2}
  \plotone{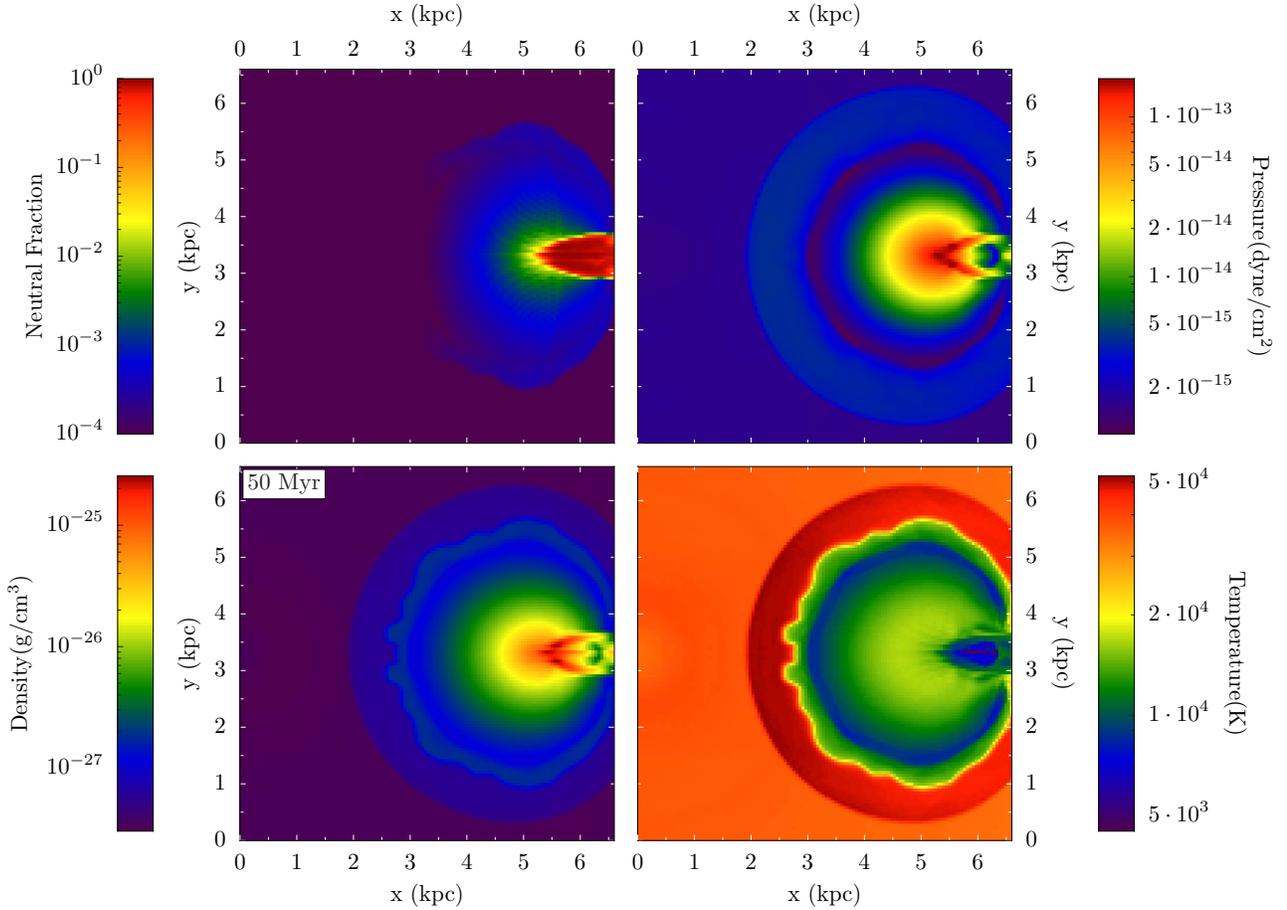}
  \epsscale{1}
  \caption{\label{fig:test7_3} Test 7. (Photo-evaporation of a dense
    clump).  Same as Fig. \ref{fig:test7_1} but at $t = 50$ Myr.}
\end{figure*}

\begin{figure}
  \plotone{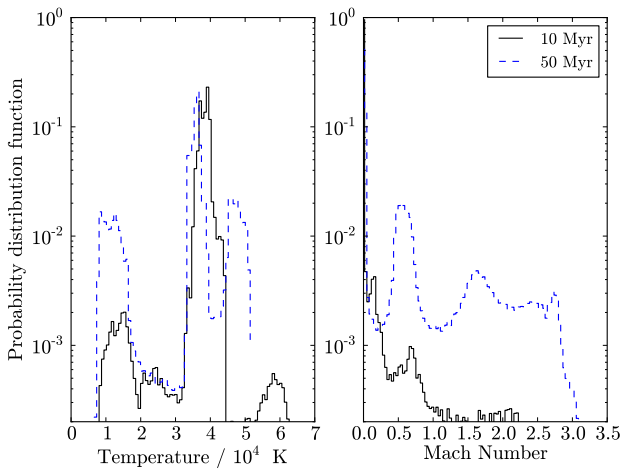}
  \caption{\label{fig:test7_pdf} Test 7. (Photo-evaporation of a dense
    clump).  Probability distribution function for temperature (left)
    and flow Mach number (right) at 10 Myr (solid) and 50 Myr
    (dashed).}
\end{figure}

\begin{figure}
  \plotone{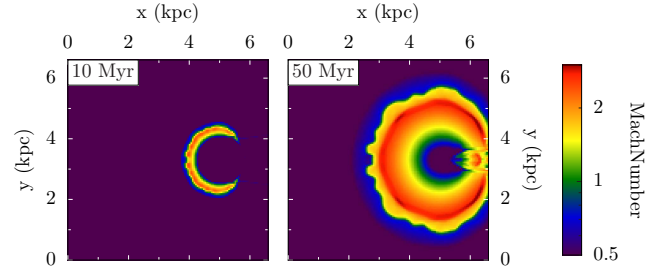}
  \caption{\label{fig:test7_4} Test 7. (Photo-evaporation of a dense
    clump).  Slices through the clump centre of the flow Mach number
    at t = 10 Myr (left) and 50 Myr (right).}
\end{figure}

The photo-evaporation of a dense clump in a uniform medium proceeds
very differently when radiation hydrodynamics is considered instead of
a static density field.  The ionisation front first proceeds as a very
fast R-type front, then it slows to a D-type front when it encounters
the dense clump.  As the clump is gradually photo-ionised and heated,
it expands into the ambient medium.  The test presented here is
exactly like Test 3 but with gas dynamics.  In this setup, the
ionisation front overtakes the entire clump, which is then completely
photo-evaporated.

Fig. \ref{fig:test7_1} shows cuts of density, temperature, neutral
fraction, and pressure in a line connecting the source and the clump
centre at $t = 1$, 10, and 50 Myr.  At 1 Myr, the ionisation front has
propagated through the left-most 500 pc of the clump.  This heated gas
is now over-pressurised, as seen in the pressure plot in Fig.
\ref{fig:test7_1}, and then expands into the ambient medium.  This
expansion creates a photo-evaporative flow, seen in many star forming
regions \citep[e.g. M16;][]{Hester96} as stars irradiate nearby cold,
dense overdensities.  These flows become evident in the density at 10
Myr, seen both in the line cuts and slices (Fig. \ref{fig:test7_2}).
They have temperatures up to 50,000 K.  At this time, the front has
progressed about halfway through the clump, if one inspects the
neutral fraction.  However the high energy photons have heated all but
the rear surface of the clump.  At the end of the test ($t = 50$ Myr),
only the core and its associated shadow is neutral, seen in Fig.
\ref{fig:test7_3}.  The core has been compressed by the surrounding
warm medium, thus causing the higher densities seen at $t = 50$ Myr.
The non-spherical artifacts on the inner boundary of the warm
outermost shell are caused by the initial discretisation of the
sphere, as discussed in \S\ref{sec:test3}.  Next
Fig. \ref{fig:test7_pdf} shows the distributions of temperature and
flow Mach number in the entire domain at 10 Myr and 50 Myr, showing
similar behavior as the codes in RT09.  Lastly in
Fig. \ref{fig:test7_4}, we show slices of the flow Mach number at 10
Myr and 50 Myr, showing the supersonic photo-evaporative flows that
originate from the clump.

\section{Radiation Hydrodynamics Applications}
\label{sec:apps}

We have completed presenting results from the RT06 and RT09 test
suites.  We expand on these test suites to include more complex
situations, such as a Rayleigh-Taylor problem illuminated by a
radiation source, champagne flows, an irradiated blast wave,
collimated radiation, and an \hii with a variable source that further
demonstrate its capabilities and accuracy.  Last we use the new MHD
implementation in \enzo~v2.0 in the problem of a growing \hii region
in a magnetic field.

\subsection{Application 1. Champagne flow from a dense clump}

\begin{figure*}
  \epsscale{2}
  \plotone{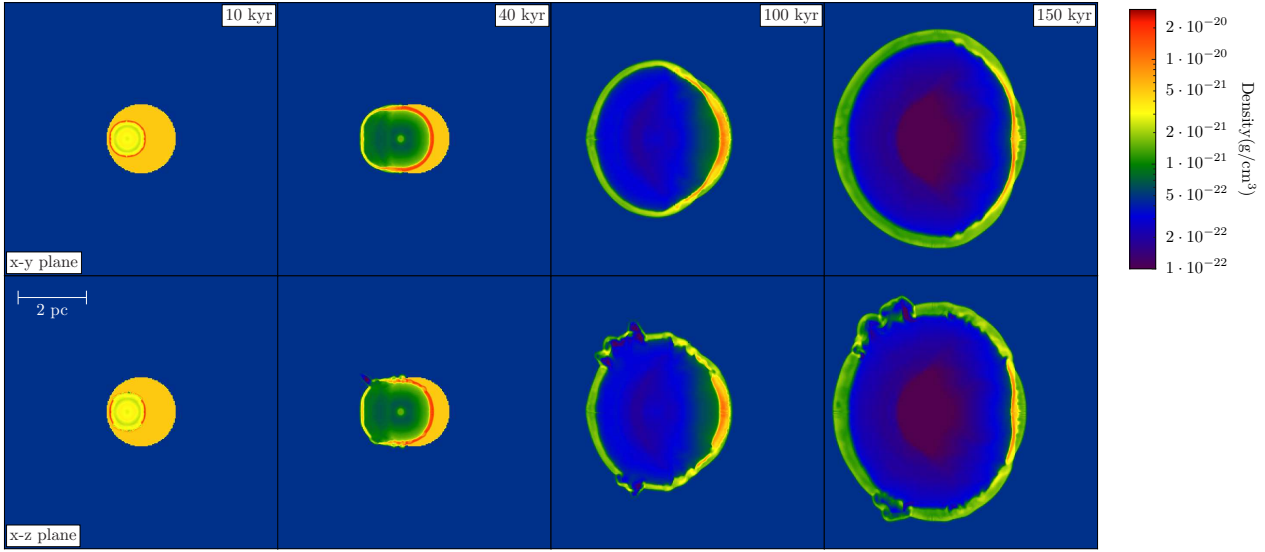}
  \epsscale{1}
  \caption{\label{fig:test8_1} Application 1. (Champagne flow from a dense
    clump).  Slices of density through the initial clump centre in the
    x-y plane (top) and x-z plane (bottom) at $t = 10, 40, 100, 150$
    kyr.  Notice the instabilities that grow from perturbations
    created while the \hii region is contained in the dense
    clump.}
\end{figure*}

Radiation-driven outflows from overdensities, known as champagne
flows, is a long studied problem \citep[e.g.][\S3.3]{Yorke86}.  To
study this, we use the same setup as \citet{Bisbas09} -- a spherical
tophat with an overdensity of 10 and radius of 1 pc in a simulation
box of 8 pc.  The ambient medium has $\rho = 290$ \cubecm~and $T =
100$ K.  The radiation source is offset from the overdensity centre by
0.4 pc.  It has a luminosity of \tento{49} ph s$^{-1}$ and a $T=10^5$
K blackbody spectrum.  The resulting Str\"{o}mgren radius is 0.33 pc,
just inside of the overdense clump.  These parameters are the same
used in \citet{Bisbas09}.  The entire domain initially has an ionised
fraction of \tento{-6}.  We do not consider self-gravity.  The
simulation has a resolution of 128$^3$ on base grid, and we refine the
grid up to 4 times if a cell has an overdensity of $1.5 \times 2^l$,
where $l$ is the AMR level.  The simulation is run for 150 kyr.

We show slices in the x-y and x-z planes of density in Fig.
\ref{fig:test8_1} at $t = 10, 40, 100, 150$ kyr.  In the direction of
the clump centre, the ionisation front shape transitions from
spherical to parabolic after it escapes from the clump in the opposite
direction.  At $t = 10$ kyr, the surface of the \hii region is
just contained within the overdensity.  In the x-z plane, there are
density perturbations only above a latitude of 45 degrees.  We believe
that these are caused by the mismatch between HEALPix pixels and the
Cartesian grid, even with our geometric correction.  After the
ionisation front escapes from the clump in the negative x-direction,
these perturbations grow from Rayleigh-Taylor instabilities as the gas
is accelerated when it exits the clump.  As the shock propagates
through the ambient medium, it is no longer accelerated and has a
nearly constant velocity, as seen in Test 6.  Thus these perturbations
are not as vulnerable to Rayleigh-Taylor instabilities at this point.
The ambient medium and shock are always optically thick, even in the
directions of the bubbles.  \citeauthor{Bisbas09} found that the shock
fragmented and formed globules; however we find the density shell is
stable against such fragmentation.  To investigate this scenario
further, our next tests involve radiation driven Rayleigh-Taylor
instabilities.

\subsection{Application 2. Irradiated Rayleigh-Taylor instability}

Here we combine the classic case of a Rayleigh-Taylor instability and
an expanding \hii region.  The Rayleigh-Taylor instability
occurs when a dense fluid is being supported by a lighter fluid,
initially in hydrostatic equilibrium, in the presence of a constant
acceleration field.  This classic test alone evaluates how subsonic
perturbations evolve.  We consider the case of a single-mode
perturbation.  The system evolves without any radiation until the
perturbation grows considerably and then turn on the radiation source.
These tests demonstrate that \moray~can follow a highly dynamic
system and resolve fine density structures.

We run two cases, an optically-thick and optically-thin case.  In
the former, we take the parameter choices from past literature
\citep[e.g.][]{Liska03, Stone08} by setting the top and bottom halves
of the domain to a density $\rho_1 = 2$ and $\rho_0 = 1$,
respectively.  The velocity perturbation is set in the $z$-direction
by
\begin{eqnarray}
  v_z(x,y,z) & = & 0.01 [1 + \cos(2\pi x / L_x)] \times  \nonumber\\
  & & [1 + \cos(2\pi y / L_y)] \times \nonumber\\
  & & [1 + \cos(2\pi z / L_z)]/8.
\end{eqnarray}

We set the acceleration field $g_z = 0.1$ and the adiabatic index
$\gamma = 1.4$.  We use a domain size of $(L_x, L_y, L_z) = (0.5, 0.5,
1.5)$ with a resolution of (64, 64, 192).  For hydrostatic
equilibrium, we set $P = P_0 - g \rho(z) z$ with $P_0 = 2.5$.  In
order to consider a radiation source with a ionising photon luminosity
of $10^{42}$ ph s$^{-1}$, we scale the domain to a physical size of
(0.5, 0.5, 1.5) pc; time is in units of Myr; density is in units of
$m_h$, resulting in an initial temperature of $(T_0, T_1) = (363,
726)$ K.  The radiation source is placed at the centre of lower
$z$-boundary face and starts to shine at $t = 10$ Myr.

The optically-thin case is set up similarly but with three changes:
(1) a density contrast of 10, (2) a luminosity of $10^{43}$ ph
s$^{-1}$, and (3) the source is born at 6.5 Myr.  The time units are
decreased to 200 kyr so that $(T_0, T_1) = (1.8 \times 10^3, 1.8
\times 10^4)$ K.  Note that in code units, pressure is unchanged.  We
adjust the physical unit scaling because we desire an optically thin
bottom medium with $T > 10^4$ K and $x_e \sim 1$.  Furthermore, the
ionisation front remains R-type before interacting with the
instability.  A possible physical analogue could be a radiation source
heating and rarefying the medium below.

\begin{figure}
  \plotone{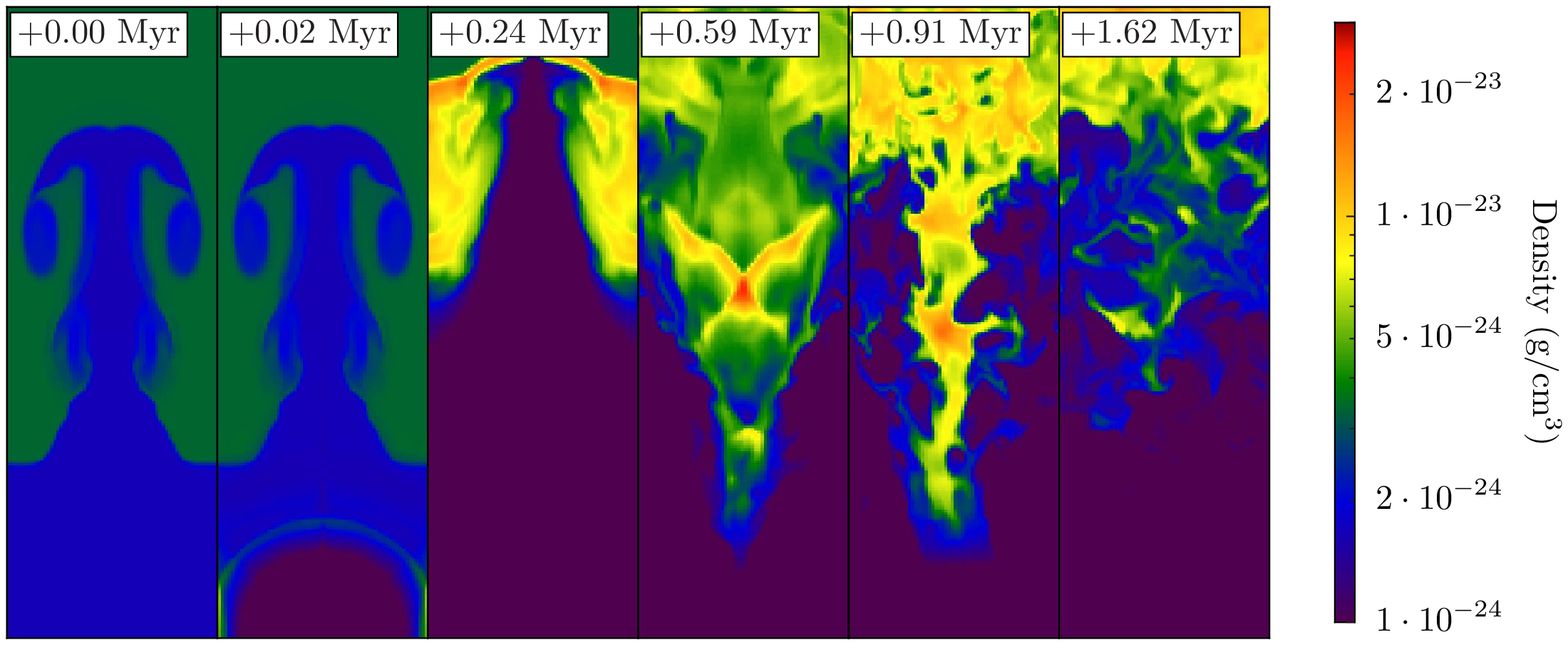}
  \plotone{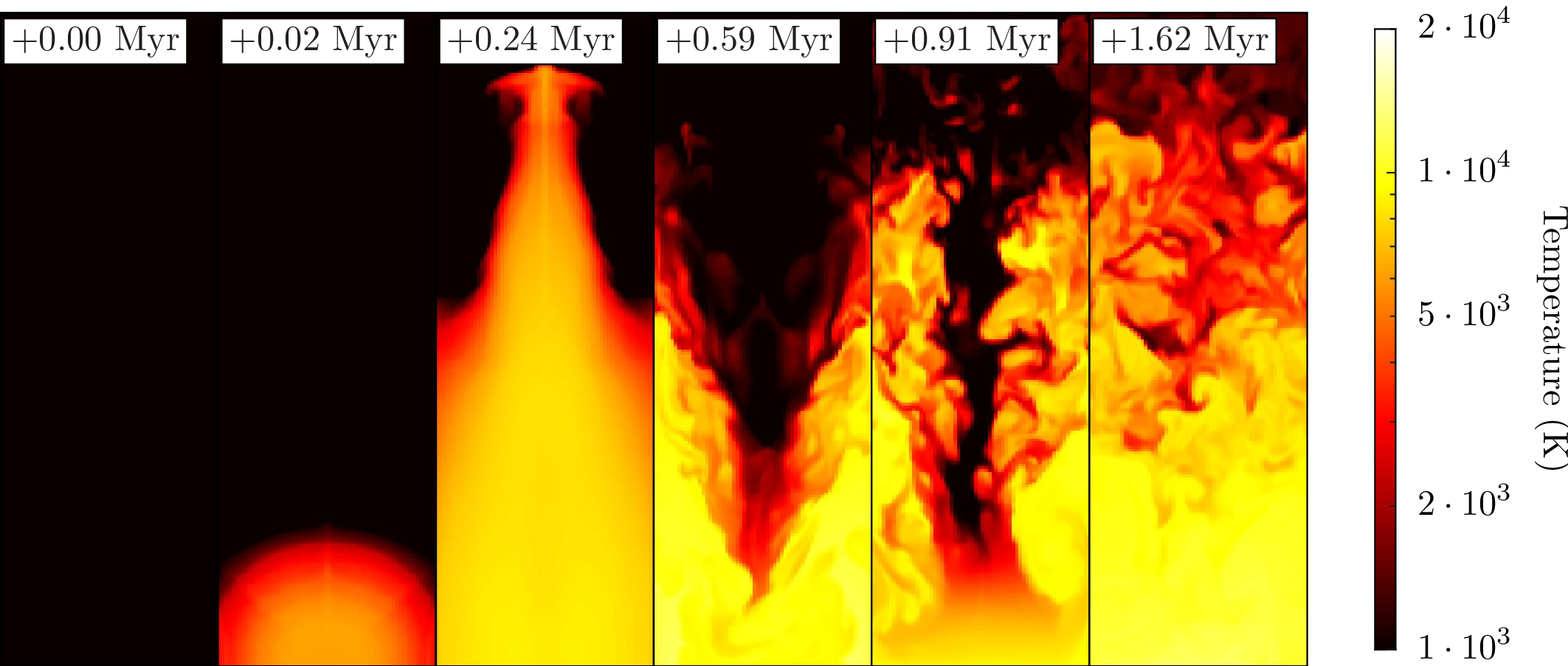}
  \plotone{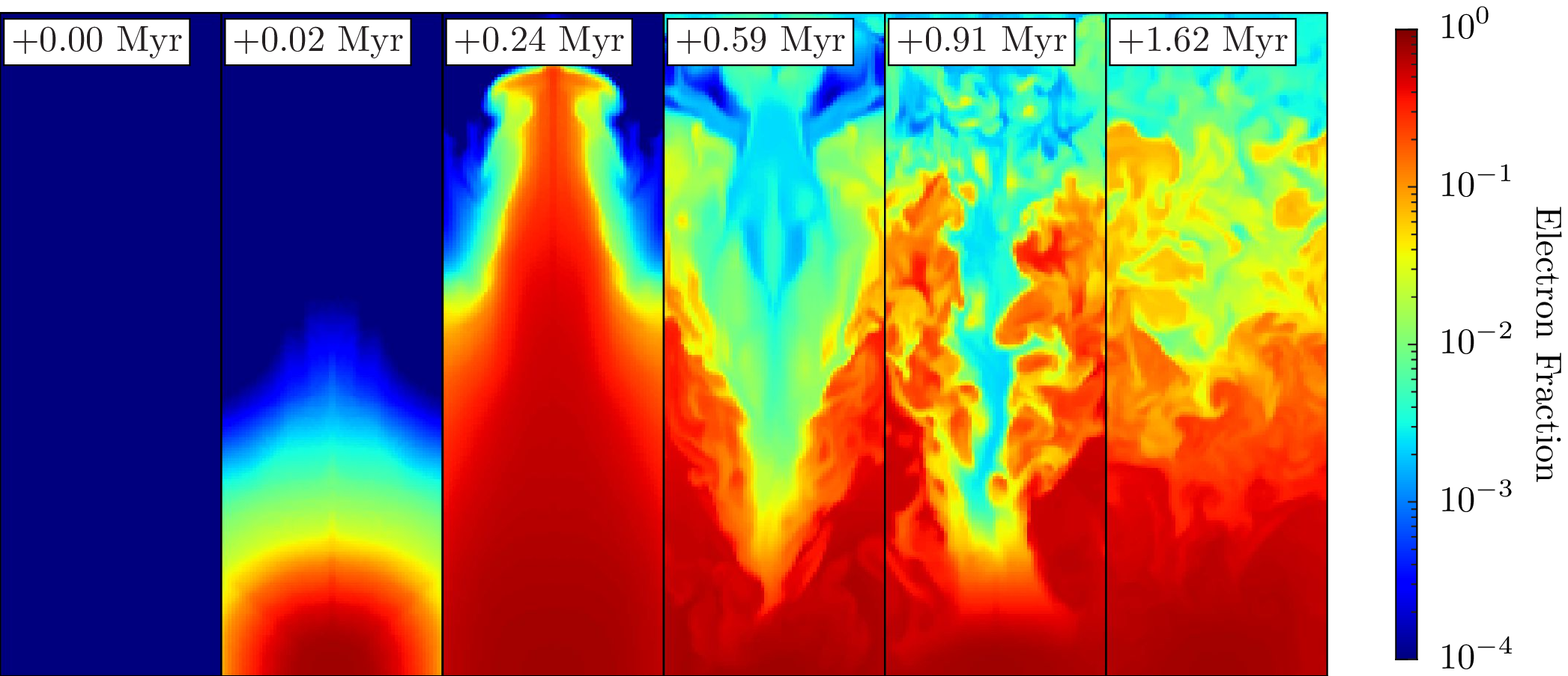}
  \caption{\label{fig:test9_thick} Application 2. (Irradiated Rayleigh-Taylor
    instability; optically thick case).  Slices at $y=0$ of density
    (top), temperature (middle), and electron fraction (bottom).  The
    source turns on at $t=0$.}
\end{figure}

\begin{figure}
  \plotone{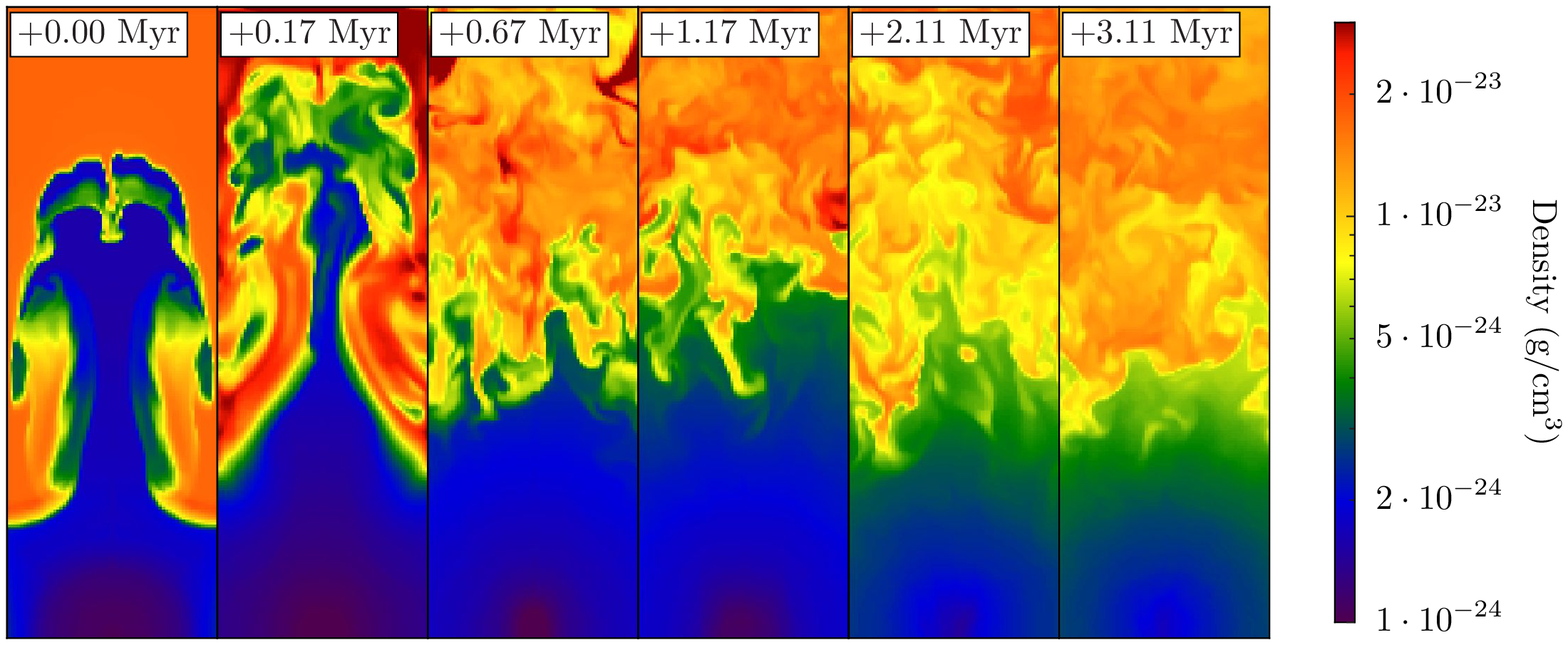}
  \plotone{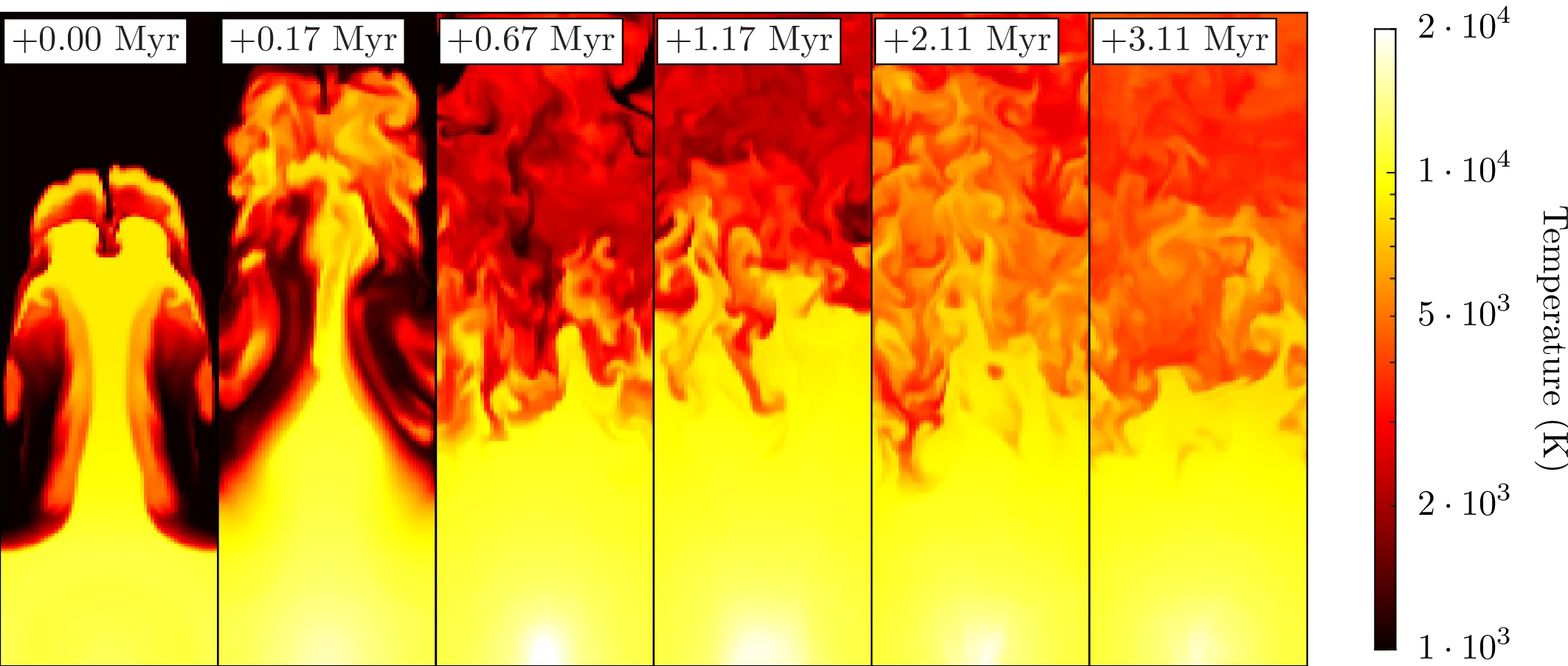}
  \plotone{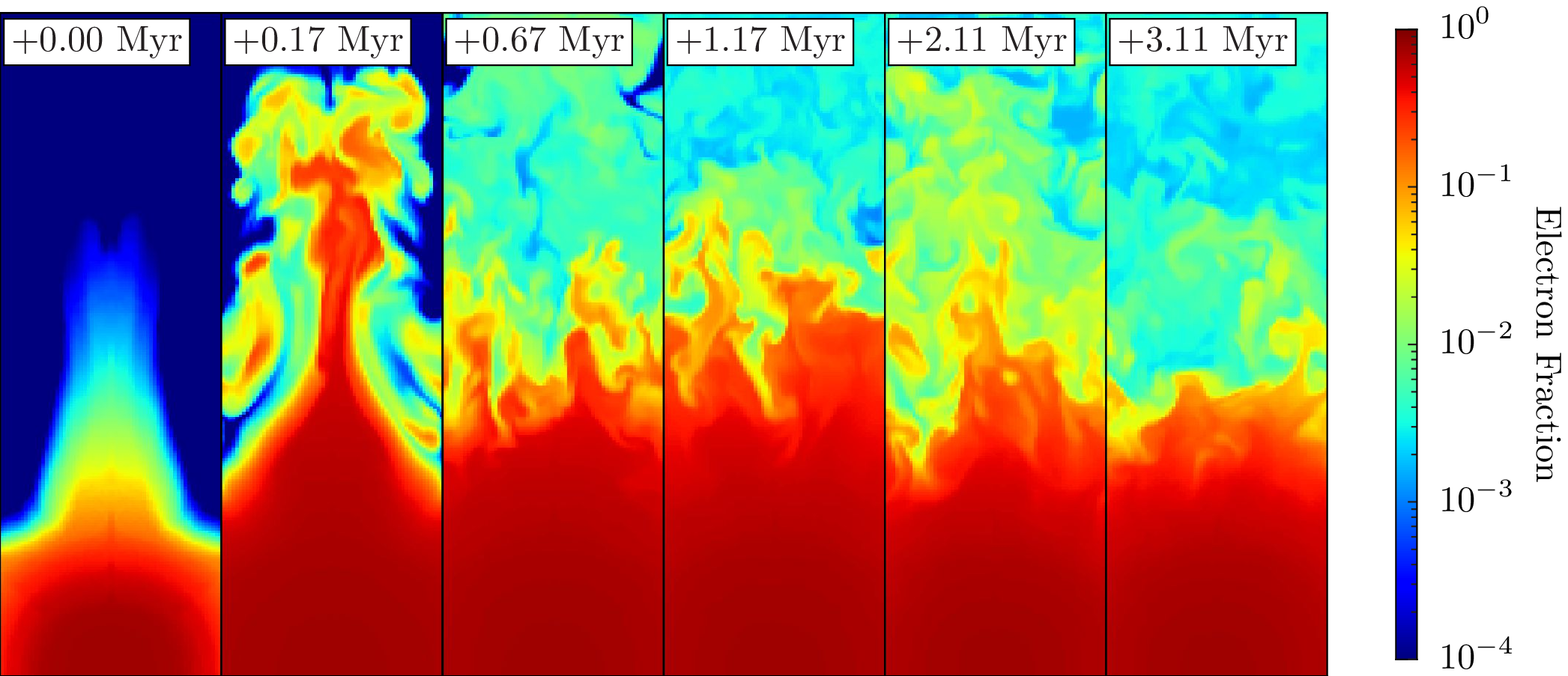}
  \caption{\label{fig:test9_thin} Same as Fig. \ref{fig:test9_thick}
    but for the optically thin case.}
\end{figure}

The $x$ and $y$-boundaries are periodic, and the $z$-boundaries are
reflecting.  These will cause artificial features, in particular,
because of the top reflecting boundary; nevertheless, these tests
provide a stress test on a radiation hydrodynamics solver.  We show
the evolution of the density, temperature, and ionised fraction of the
optically thick and optically thin cases in Figs.
\ref{fig:test9_thick} and \ref{fig:test9_thin}.  The initial state of
the Rayleigh-Taylor instability is shown in the left panels.

In the optically thick case, a D-type front is created, which is
clearly illustrated by the spherical density enhancement at 0.02 Myr.
The shock then passes through the instability at $\sim$0.25 Myr and
reflects off the upper $z$-boundary.  This and complex shock
reflections create a Richtmyer-Meshkov instability \citep[see][for a
review]{RMI}, driving a chaotic jet-like structure downwards.  The
radiation source photo-evaporates the outer parts of this structure.
The interaction between the dense cool ``jet'' and the hot medium
further drives instabilities along the surface, which can be seen when
comparing $t = 0.59$ Myr and $t = 0.91$ Myr slices.  At the latter
time, the jet cannot reach the bottom of the domain before being
photo-evaporated.  Eventually this structure is completely destroyed,
leaving behind a turbulent medium between the hot and cold regions.

The optically thin problem is less violent than the optically thick
case because the R-type front does not interact with the initial
instability as strongly.  The radiation source provides further
buoyancy in the already $T=10^4$ K gas.  The gas first to be ionised
and photo-evaporated are the outer regions of the instability.  The
enhanced heating also drives the upper regions of the instability,
making the top interface turbulent.  It then reflects off the upper
$z$-boundary and creates a warm $T = 5 \times 10^3$ K, partially
ionised ($x_e \sim 10^{-2}$), turbulent medium, seen in the slices $t
\ge 0.67$ Myr.  The slices of electron fraction also show that the
dense gas is optically thick.

\subsection{Application 3. Photo-evaporation of a blastwave}

\begin{figure*}
  \epsscale{2}
  \plotone{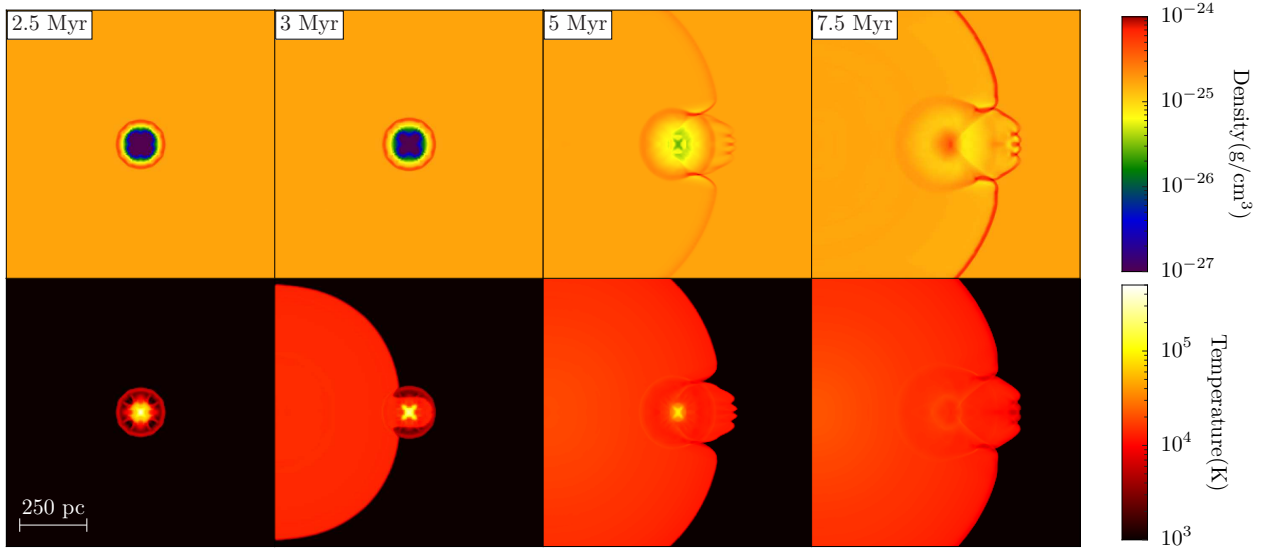}
  \epsscale{1}
  \caption{\label{fig:test10} Application 3 (Photo-evaporation of a
    blastwave).  Slices of density (top) and temperature (bottom) at
    $t = 2.5, 3, 5, 7.5$ Myr in the $x-z$ plane.  As the R-type
    ionisation front propagates through the blastwave centre,
    instabilities grow from the slightly inhomogeneous hot and rarefied
    medium.  Note that the dense shell of the blastwave also creates
    dense inward fingers in the ionisation front shock.}
\end{figure*}

A supernova blastwave being irradiated by a nearby star is a likely
occurrence in massive-star forming regions.  In this test, we set up an
idealised test that mimics this scenario.  The ambient medium has a
density $\rho_0 = 0.1~\cubecm$ and temperature $T_0 = 10$ K.  The
domain size is 1 kpc.  We use 2 levels of AMR with a base grid of
64$^3$ that is refined if the density or total energy slope is greater
than 0.4.  The blastwave is initialised at the beginning of the
Sedov-Taylor phase when the mass of the swept-up material equals the
ejected material.  It has a radius of 21.5 pc, a total energy of
$10^{50}$ erg, and total mass of $100 \Ms$, corresponding to $E = 315$
eV per particle or $E/k_b = 3.66 \times 10^6$ K.  The radiation source
is located at the centre of the left $x$-boundary and has a luminosity
of $10^{50}$ erg s$^{-1}$.  We use a $T=10^5$ K blackbody spectrum
with 2 energy groups (16.0 and 22.8 eV).  The source turns on at 2.5
Myr at which point the blastwave has a radius of 200 pc.  The
simulation is run for 7.5 Myr.

\begin{figure*}
  \epsscale{2}
  \plotone{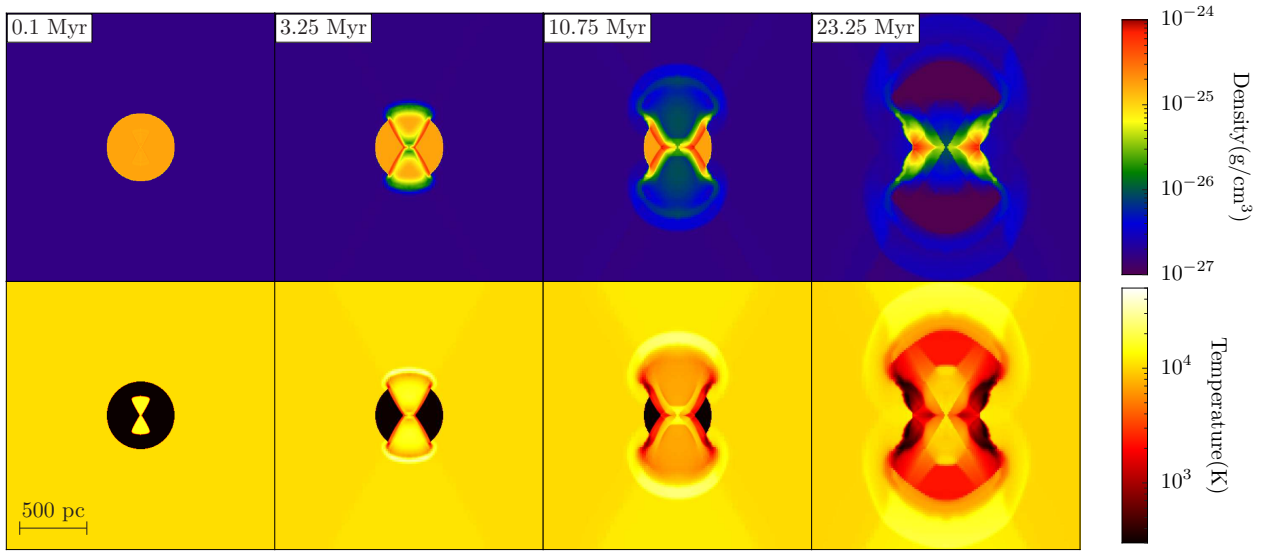}
  \epsscale{1}
  \caption{\label{fig:test11} Application 4 (Collimated radiation from a
    dense clump).  Slices of density (top) and temperature (bottom) at
    $t = 0.1, 3.25, 10.75, 23.25$ Myr.  The conical \hii region
    drives shocks transversely into the overdense sphere and creates
    polar champagne flows.  The ambient medium is heated to $T \sim 3
    \times 10^4$ K as the ionisation front passes the
    constant-pressure cloud surface.  The ionisation front changes
    from D-type to R-type after it enters the ambient medium.}
\end{figure*}

Fig. \ref{fig:test10} shows the ionisation front overtaking and
disrupting the blastwave.  We show the blastwave before the source is
born at 2.5 Myr.  The interior is rarefied ($\rho \sim
10^{-3}~\cubecm$) and is heated to $T \sim 5 \times 10^5$ K by the
reverse shock.  At $t = 3$ Myr, the ionisation front is still R-type,
and it ionises the rear side of the dense shell.  Because the interior
is ionised and diffuse, the ionisation front rapidly propagates
through it until it reaches the opposite shell surface.  Shortly
afterward, the ionisation front transitions from R-type to D-type at a
radius of 0.5 kpc, seen in the formation of a shock in the 5 Myr
density panel.  This transition occurs by the construction of the
problem not by the interaction with the blastwave.  The surfaces of
the blastwave that are perpendicular to the ionisation front have the
highest column density and thus are last to be fully ionised.  The
pressure forces from warm ambient medium and blastwave interior
compress these surfaces, photo-evaporating them in the process,
similar to Test 7.  They survive until the final time $t = 7.5$ Myr.
As the R-type ionisation front interacts with the blastwave interior,
the density perturbations create ionisation front instabilities
\citep{Whalen08_Instab} that are seen on the \hii region surface
at the coordinate $z=0.5$.  Behind the ionisation front, the dense
shell of the blastwave is photo-evaporated, and a smooth overdensity
is left in the initial blastwave centre.

\subsection{Application 4. Collimated radiation from a dense clump}

\begin{figure}
  \plottwo{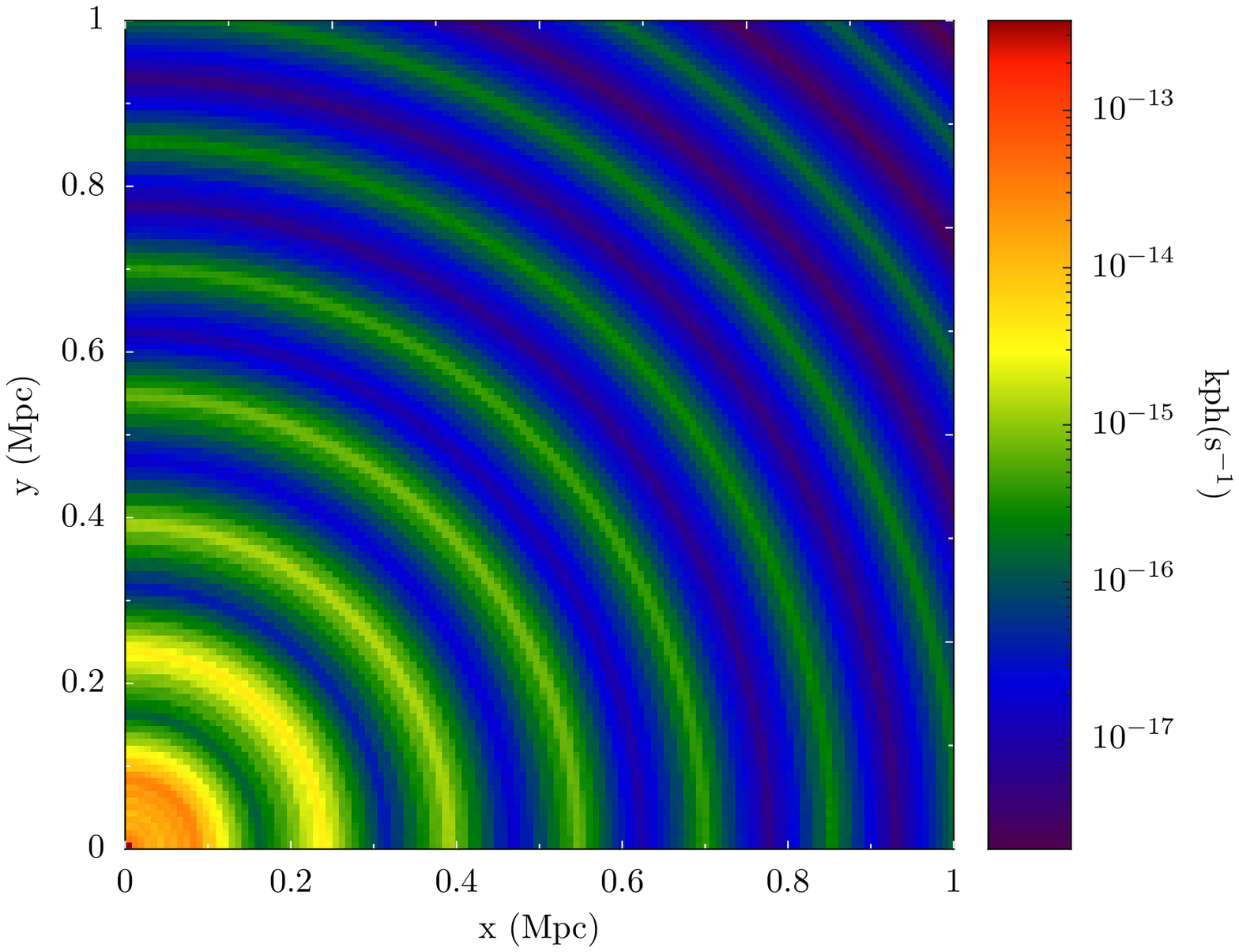}{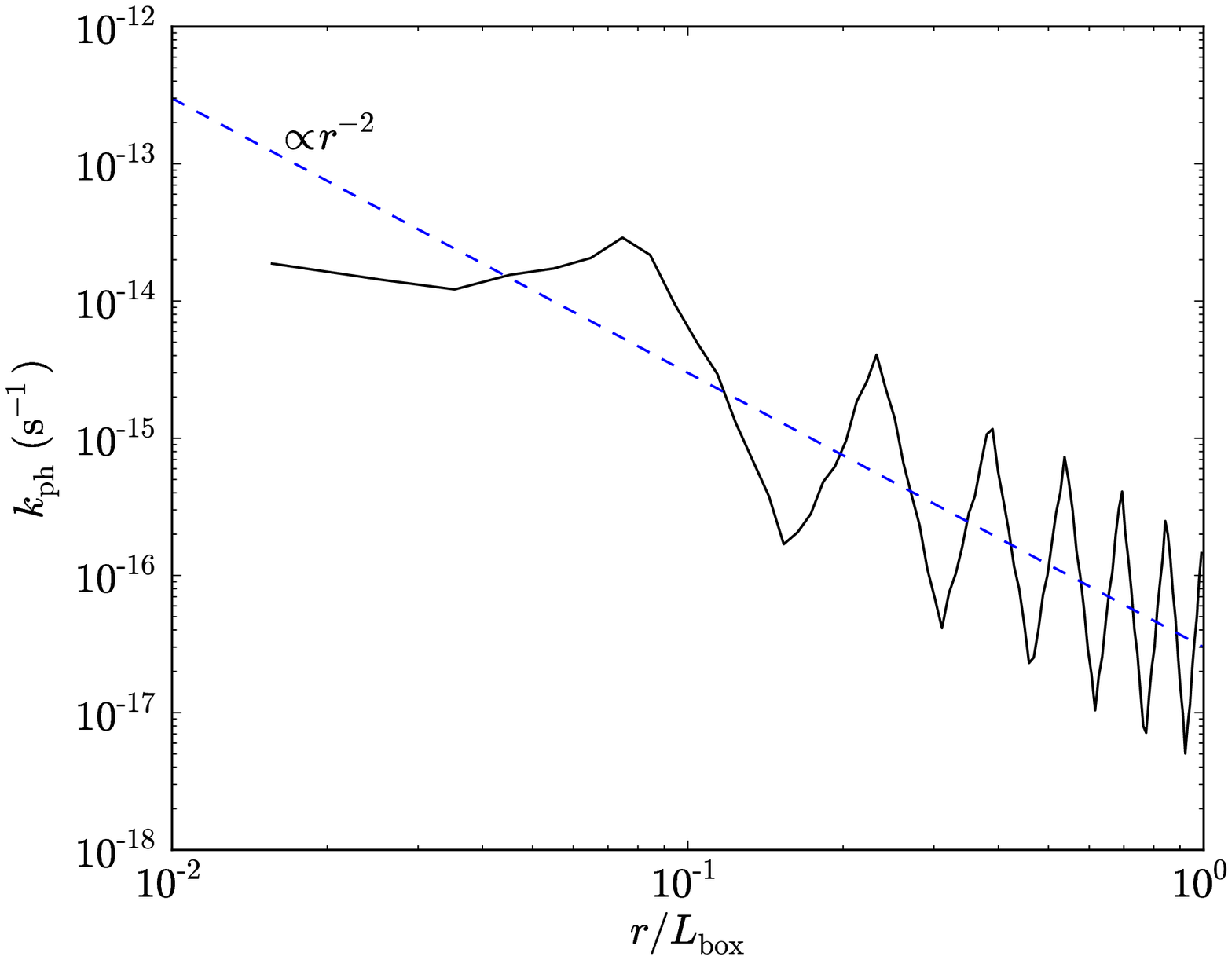}
  \caption{\label{fig:test12_kph} Application 4 (Time variations of the
    source luminosity).  Top: Slice of the photo-ionisation rate
    \kph~through the origin.  The source has a duty cycle of 0.5 Myr,
    and the box has a light crossing time of 3.3 Myr.  The shells of
    high \kph~originate from radiation that was emitted when the
    source was at its peak luminosity, illustrating the
    time-dependence of the radiative transfer equation.  Bottom:
    Radial profile \kph~with the inverse square law overplotted.}
\end{figure}

\begin{figure}
  \plotone{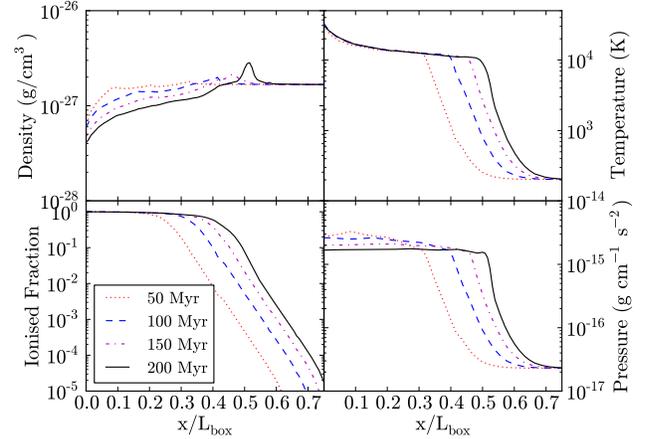}
  \caption{\label{fig:test12_profiles} Application 4 (Time variations of the
    source luminosity).  Radial profiles of (clockwise from upper
    left) density, temperature, pressure, and ionised fraction at $t =
    50, 100, 150, 200, 250$ Myr.  In this problem, the time variations
    in the source have little effect on the overall \hii region
    expansion.  Inside the ionisation front at $t = 50$ Myr, there are
    small density perturbations that are created by the variable
    source that are later smoothed out over a sound crossing time.}
\end{figure}

Some astrophysical systems produce collimated radiation either
intrinsically by relativistic beaming or by an optically-thick torus
absorbing radiation in the equatorial plane.  The latter case would be
applicable in a subgrid model of active galactic nuclei (AGN) or
protostars, for example.  Simulating collimated radiation with ray
tracing is trivially accomplished by only initialising rays that are
within some opening angle $\theta_c$.

We use a domain that is 2 kpc wide and has an ambient medium with
$\rho_0 = 10^{-3}$ \cubecm, $T = 10^4$ K, $x_e = 0.99$.  We place a
dense clump with $\rho/\rho_0 = 100$, $T = 100$ K, $x_e = 10^{-3}$,
and $r = 250$ pc, at the centre of the box.  Radiation is emitted in
two polar cones with $\theta_c = \pi/6$ with 768 (HEALPix level 3)
initial rays, a total luminosity of $10^{49}$ erg s$^{-1}$, and a 17.6
eV mono-chromatic spectrum.  This results in $t_{\rm rec} = 1.22$ Myr
and $R_s = 315$ pc, just outside of the sphere.  The base grid has a
resolution of 64$^3$, and it is refined with the same overdensity
criterion as Application 1.  We run this test for 25 Myr.

We illustrate the expansion of the \hii region created by the
beamed radiation in Fig. \ref{fig:test11}.  Before $t = 3$ Myr, the
\hii region is conical and contained within the dense clump,
depicted in the $t = 0.1$ Myr snapshot of the system.  At this time,
the ionisation front is transitioning from R-type to D-type in the
transverse direction of the cone.  This can be seen in the minute
overdensities on the \hii transverse surface.  When it breaks
out of the overdensity, a champagne flow develops, where the
ionisation front transitions back to a weak R-type front.  The cloud
surface is a constant-pressure contact discontinuity (CD) with a
density jump of 100.  After the front heats the gas at the CD, there
exists a pressure difference of $\sim 100$.  In response, the high
density gas accelerates into the ambient medium and heats it to $3
\times 10^4$ K.  Additionally a rarefaction wave travels towards the
clump centre.  At later times, the transverse D-type front continues
through the clump, eventually forming a disc-like structure at the
final time.  The polar champagne flows proceed to flow outwards and
produces a dense shell with a diffuse ($10^{-28}$ \cubecm) and warm
(5000 K) medium in its wake.

\subsection{Application 5. Time variations of the source luminosity}

\begin{figure*}
  \epsscale{2}
  \plotone{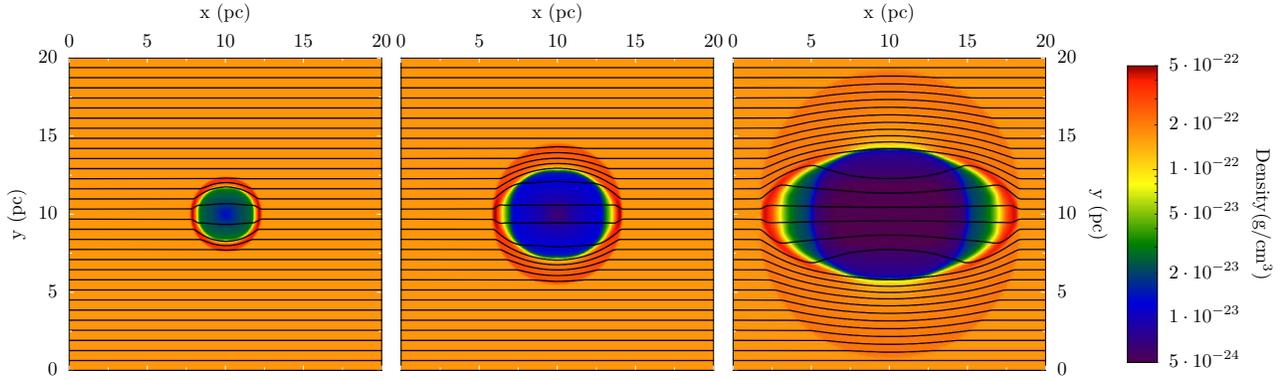}
  \epsscale{1}
  \caption{\label{fig:test14_1} Application 5. (\hii region with MHD).
    Left to right: slices of density at $t = 0.18, 0.53, 1.58$ Myr in
    the x-y plane.  The streamlines show the magnetic field.}
\end{figure*}

\begin{figure*}
  \epsscale{2}
  \plotone{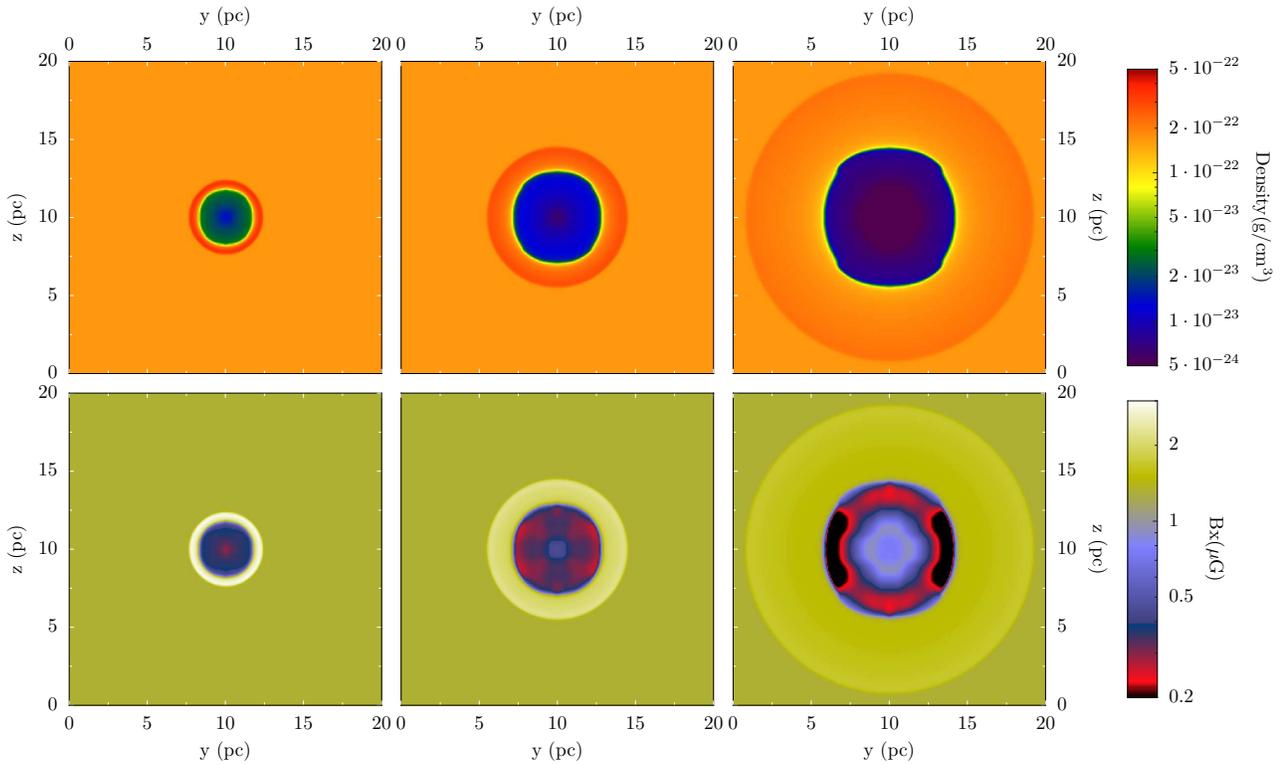}
  \epsscale{1}
  \caption{\label{fig:test14_2} Application 5. (\hii region with
    MHD). Slices of density (top) and the x-component of the magnetic
    field (bottom) in the y-z plane at $t = 0.18, 0.53, 1.58$ Myr
    (left to right).}
\end{figure*}

Our implementation retains the time derivative of the radiative
transfer equation [equation (\ref{eqn:rteqn})] if we choose a constant
ray tracing timestep, which saves the photon packages between
timesteps if $c\;dt_{\rm P} < L_{\rm box}$.  This effect only becomes
apparent when the variation time-scale of the point source is smaller
than the light crossing time of the simulation.  Furthermore, the
timestep should resolve the variation time-scale by at least a few
times.  This property might be important in large box simulations with
variable sources, e.g. AGN radiative feedback.  To test this, we can
use an exponentially varying source with some duty cycle $t_0$.  In a
functional form, this can be described as
\begin{equation}
  \label{eqn:test12}
  L(t) = L_{\rm max} \times \exp[A(t_f/t_0 - 1)]
\end{equation}
where $t_f = 2 \times |t - t_0 \times \mathrm{round}(t/t_0)|$, and
$A=4$ controls the width of the radiation pulse.  To illustrate the
effects of source variability, we remove any dependence on the medium
by considering an optically-thin uniform density $\rho =
10^{-4}~\cubecm$.  We take $L_{\rm box} = 1$ Mpc, which has a light
crossing time of 3.3 Myr.  A source is placed at the origin with
$L_{\rm max} = 10^{55}$ ph s$^{-1}$ and $t_0 = 0.5$ Myr.  We use a
radiative transfer timestep of 50 kyr to resolve the duty cycle by 10
timesteps.  The simulation is run for 6 Myr, so the radiation
propagates throughout the box.

The variability of the source is clearly illustrated in the
photo-ionisation rates, shown in Fig. \ref{fig:test12_kph}.  The
shells of relative maximum \kph~corresponds to radiation that was
emitted when the source was at its peak luminosity.  They are
separated by $ct_0$ Mpc and are geometrically diluted with increasing
radius.  Averaged over shells of the same width, photo-ionisation
rates decrease as $1/r^2$.

Next we test the hydrodynamical response to a varying source by
repeating Test 5.  We set the peak luminosity $L_{\rm max} = 2 \times
10^{49}$ erg s$^{-1}$ that is a factor of 4 more luminous than Test 5,
so the average luminosity is $\sim 5 \times 10^{48}$ erg s$^{-1}$.
The spectrum is mono-chromatic with an energy of 29.6 eV.  We set the
variation time-scale $t_f = cL_{\rm box}/3 = 16.3$ kyr and use a
constant radiative transfer timestep $t_P = t_f/4 = 4.07$ kyr.  The
simulation is run for 200 Myr.  We show the radial profiles of
density, temperature, ionised fraction, and pressure in Fig.
\ref{fig:test12_profiles}.  The variable source has little effect on
the overall growth of the \hii region.  It has the approximately
the same radius as Test 5 at $t = 200$ Myr when run with a
mono-chromatic spectrum (see \S\ref{sec:nu_dep}).  At early times, the
variable source creates density perturbations with an average size of
500 pc inside the ionised region, seen in the $t = 50$ Myr profiles.
They do not create any instabilities and are smoothed out over its
sound crossing time of $\sim 50$ Myr.

\subsection{Application 6. \hii region with MHD}

Another prevalent physical component in astrophysics is a magnetic
field.  We utilise the new magnetohydrodynamics (MHD) framework
\citep{Wang09} in \enzo~v2.0 that uses an unsplit conservative
hydrodynamics solver and the hyperbolic $\nabla \cdot \mathbf{B} = 0$
cleaning method of \citet{Dedner02}.  We show a test problem with an
expanding \hii region in an initially uniform density field and
constant magnetic field.  We use the same problem setup as
\citet{Krumholz07_ART} -- $\rho = 100~\cubecm$, $T = 11$ K, $L_{\rm
  box} = 20$ pc with a resolution of 256$^3$.  This ambient medium is
threaded by a magnetic field $\mathbf{B} = 14.2 \hat{\mathbf{x}} \;
\mu\mathrm{G}$.  The Alfv\'{e}n speed is 2.6 \kms.  The radiation
source is located in the centre of the box with a luminosity $L = 4
\times 10^{46}$ ph s$^{-1}$ with a 17.6 eV mono-chromatic spectrum,
resulting in a Str\"{o}mgren radius $R_s = 0.5$ pc.  The simulation is
run for 1.58 Myr.  The hydrodynamics solver uses an HLL Riemann solver
\citep{HLL} and piecewise linear method (PLM) reconstruction
\citep{PLM} for the left and right states in this problem.

As the \hii region grows in the magnetised medium, shown in
Figs. \ref{fig:test14_1} and \ref{fig:test14_2}, it transforms from
spherical to oblate as it is magnetically confined in directions
perpendicular to the magnetic field.  This occurs at $t > 0.5$ Myr
because the magnetic pressure exceeds the thermal pressure, and the
gas can only flow along field lines.  \citeauthor{Krumholz07_ART}
observed some carbuncle artifacts along the ionisation front; whereas
we see smooth density gradients, which is most likely caused by both
the geometric correction to the ray tracing (\S\ref{sec:meth_fc}) and
the diffusivity of the HLL Riemann solver when compared to Roe's
Riemann solver used in \citet{Krumholz07_ART}, who also use PLM as a
reconstruction method.  The evolution of the magnetic field lines
evolve in a similar manner as their results.

\section{Resolution Tests}

Resolution tests are important in validating the accuracy of the code
in most circumstances, especially in production simulations where the
initial environments surrounding radiation sources are unpredictable.
In this section, we show how our adaptive ray-tracing implementation
behaves when varying spatial, angular, frequency, and temporal
resolutions.

\subsection{Spatial resolution}
\label{sec:dx_dep}

\begin{figure}
  \plotone{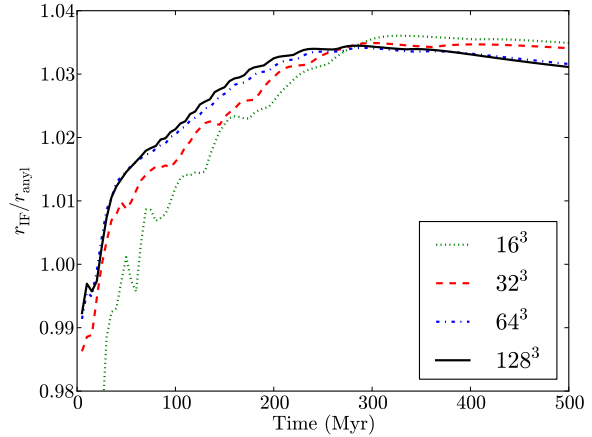}
  \caption{\label{fig:dx_dep1} Growth of the ionisation front radius,
    compared to the analytical radius, in Test 1 with varying spatial
    resolutions.  At resolutions of $16^3$ and $32^3$, the ionisation
    front is underestimated for the first $\sim25$ Myr but converges
    within 0.5\% of the higher resolution runs.}
\end{figure}

Here we use Test 1 (\S\ref{sec:test1}) as a testbed to investigate how
the evolution of the Str\"{o}mgren radius changes with resolution.  We
keep all aspects of the test the same, but use resolutions of 16$^3$,
32$^3$, 64$^3$, and 128$^3$.  In Fig. \ref{fig:dx_dep1}, we show the
ratio $r_{\rm IF}/r_{\rm anyl}$, similar to Fig.
\ref{fig:test1_ifront}, using these different resolutions.  The radii
in the $64^3$ and $128^3$ runs evolve almost identically.  Compared to
these resolutions, the lower $16^3$ and $32^3$ resolution runs only
lag behind by 1\% until 300 Myr, and afterwards it is larger by 0.5\%
than the higher resolution cases.  This shows that our method gives
accurate results, even in marginally resolved cases, which is expected
with a photon conserving method.  Furthermore this demonstrates that
the geometric correction does not significantly affect photon
conservation.

\subsection{Angular resolution}
\label{sec:ang_dep}

\begin{figure}
  \plotone{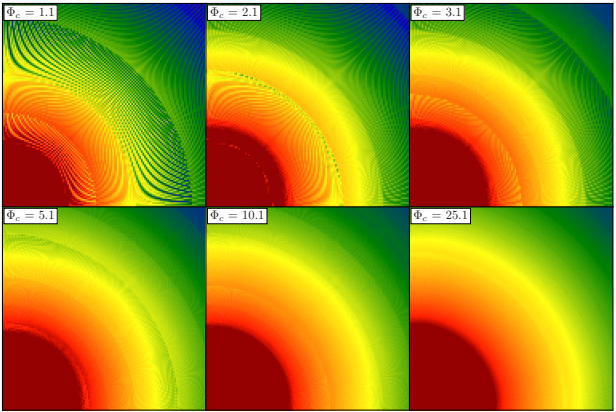}
  \epsscale{1}
  \caption{\label{fig:ang_dep1} Variations in the photo-ionisation
    rates for different ray-to-cell samplings $\Phi_c$.  The colormap
    only spans a factor of 3 to enhance the contrast.  In comparison,
    the photo-ionisation rate actually spans 4 orders of magnitude in
    this test.}
\end{figure}

\begin{figure}
  \plotone{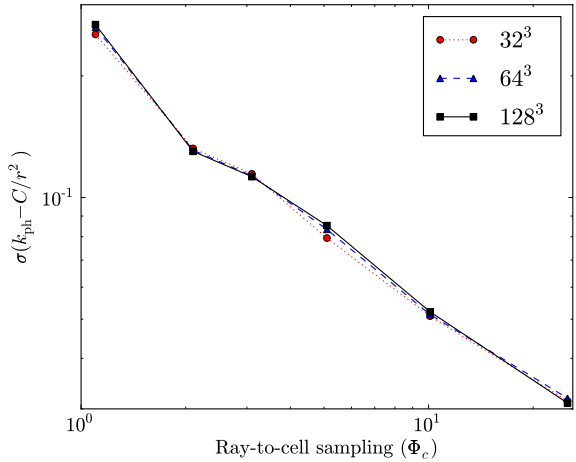}
  \caption{\label{fig:ang_dep2} Standard deviations of the difference
    between the computed photo-ionisation rates and an inverse square
    law as a function of ray-to-cell samplings $\Phi_c$ for different
    spatial resolutions.  There is no dependence on the spatial
    resolution, and the accuracy increases as $\sigma \propto
    \Phi_c^{-0.6}$.}
\end{figure}

The Cartesian grid must been sampled with sufficient rays in order to
calculate a smooth radiation field.  To determine the dependence on
angular resolution, we consider the propagation of radiation through
an optically thin, uniform medium.  The radiation field should follow
a $1/r^2$ profile.  As the grid is less sampled by rays, the deviation
from $1/r^2$ should increase.  This test is similar to Test 1, but the
medium has $\rho = 10^{-3}~\cubecm$, $T = 10^4$ K, and $1 - x_e =
10^{-4}$.  The simulation is only run for one timestep because the
radiation field should be static in this optically-thin test.

We consider minimum ray-to-cell ratios $\Phi_c = (1.1, 2.1, 3.1, 5.1,
10.1, 25.1)$.  Slices of the photo-ionisation rates through the origin
are shown in Fig. \ref{fig:ang_dep1} for these values of $\Phi_c$.
In this figure, we limit the colormap range to a factor of 3 to show
the nature of the artifacts in more contrast.  Unscaled, the rates in
the figures would span 4 orders of magnitude.  When $\Phi_c \le 3.1$,
the cell-to-cell variations are apparent because there are not enough
rays to sufficiently sample the radiation field, even with the
geometric correction factor $f_c$, whose improvements are shown later
in \S\ref{sec:test_fc}.  At $\Phi_c = 5.1$, these artifacts disappear,
leaving behind a shell artifact where the radiation fields do not
smoothly decrease as 1/$r^2$.  At higher values of $\Phi_c$, this
shell artifact vanishes as well.  

One measure of accuracy is the deviation from an 1/$r^2$ field because
this problem is optically-thin.  To depict the increase in accuracy
with ray sampling, we take the difference between the calculated
photo-ionisation rate and a 1/$r^2$ field, and then plot the standard
deviation of this difference field versus angular resolution in Fig.
\ref{fig:ang_dep2}.  We plot this relation for resolutions of $32^3$,
$64^3$, and $128^3$ and find no dependence on spatial resolution,
which is expected because we control the angular resolution in terms
of cell widths, not in absolute solid angles.  We find that the
deviation from an inverse square law decreases as $\sigma \propto
\Phi_c^{-0.6}$.

\subsection{Frequency resolution}
\label{sec:nu_dep}

\begin{figure}
  \plottwo{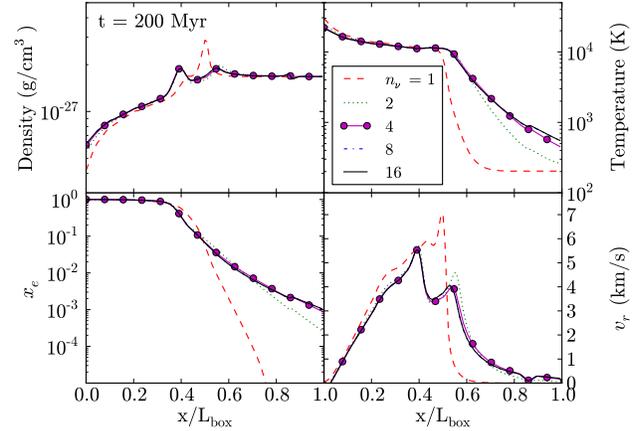}{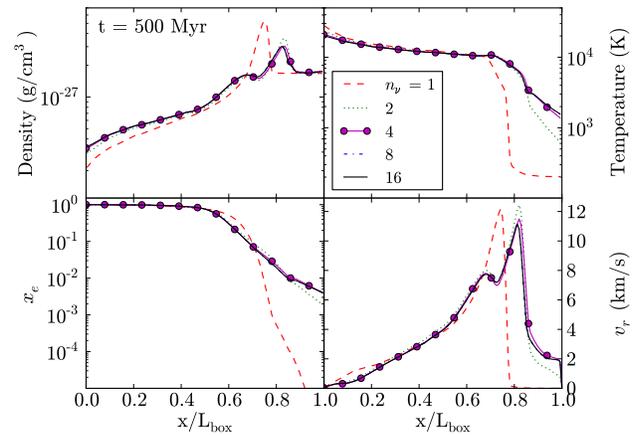}
  \caption{\label{fig:nu_dep} Radial profiles of (clockwise from the
    upper left) density, temperature, radial velocity, and ionised
    fraction for Test 5 with $n_\nu = 1$, 2, 4, 8, and 16 frequency
    bins sampling the $T=10^5$ K blackbody spectrum.  The data are
    shown at $t = 200$ Myr (top) and $t = 500$ Myr (bottom).  The
    double-peaked structure in the shock only appears with a
    multi-frequency spectrum.  The solution converges at $n_\nu \ge
    4$.}
\end{figure}

The ionisation front radius is within 5--10\% of analytical solutions
in Tests 1, 2, and 5 with only one energy group; however a
multi-frequency spectrum can create differences in the reactive flows.
We use Test 5 (\S\ref{sec:test5}; an expanding \hii region with
hydrodynamics) to probe any differences in the solution when varying
the resolution of the spectrum.  In RT09, ZEUS-MP was used to
demonstrate the effect of a multi-frequency spectrum on the dynamics
of the ionisation front in this test.  Instead of a single shock seen
in the mono-chromatic spectrum, the shock obtains a double-peaked
structure in density and radial velocity.  We rerun Test 5 with a
$T=10^5$ K blackbody spectrum sampled by $n_\nu = 1$, 2, 4, 8, and 16
frequency bins.  We use the following energies:
\begin{itemize}
\item $n_\nu = 1$: Mean energy of 29.6 eV
\item $n_\nu = 2$: Mean energies in bins 13.6--30 and $>$30 eV--21.1,
  43.0 eV
\item $n_\nu = 4$: Mean energies in bins 13.6--20, 20--30, 30--40, and
  $>$40 eV--16.7, 24.6, 34.5, 52.1 eV
\item $n_\nu = 8, 16$: Logarithmically spaced between 13.6 and 50 eV
  for the first $n_\nu-1$ bins, and the last bin is the mean energy
  above 50 eV.
\end{itemize}

Fig. \ref{fig:nu_dep} shows the radial profiles of density,
temperature, ionised fraction, and radial velocity at $t = 200$ Myr
and $t = 500$ Myr.  All of the runs with $n_\nu > 1$ show the double
peaked features in density and radial velocities.  The mono-chromatic
spectrum misses this feature completely because all of the radiation
is absorbed at a characteristic column density.  In the
multi-frequency spectra, the higher energy photons are absorbed at
larger column densities and photo-heated this gas.  This heated gas
creates a photo-evaporative flow that collides with the innermost
shock, forming the double peaked density profile.  The $n_\nu \ge 4$
runs are indistinguishable, and the $n_\nu = 2$ spectrum only leads to
a marginally higher density in the outer shock and lower ionised
fractions and temperatures in the ambient medium.  In effect, a
mono-chromatic spectrum can be sufficient if the problem focuses on
large-scale quantities, e.g. ionised filling fractions in reionisation
calculations.  Conversely these effects may be important when studying
the details of small-scale processes, e.g. photo-evaporation.

\subsection{Temporal resolution}
\label{sec:dt_dep}

\begin{figure}
  \plotone{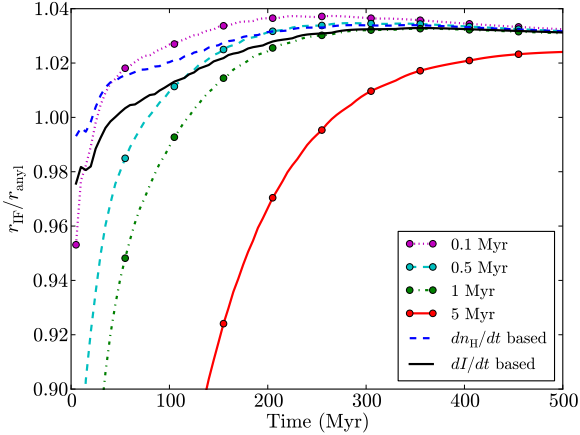}
  \caption{\label{fig:dt_dep1} Growth of the ionisation front radius,
    compared to the analytical radius, in Test 1 with varying
    radiative transfer timesteps.  The $dn_{\rm H}/dt$ and $dI/dt$
    based timesteps provide the best accuracy, combined with
    computational efficiency because they take short timesteps when
    the \hii is expanding rapidly but take long timesteps when
    the photon gradients are small when $r_{\rm IF}$ is large.  At the
    final time, all but the $t = 5$ Myr constant timestep produce
    identical ionisation front radii.}
\end{figure}
\begin{figure}
  \plotone{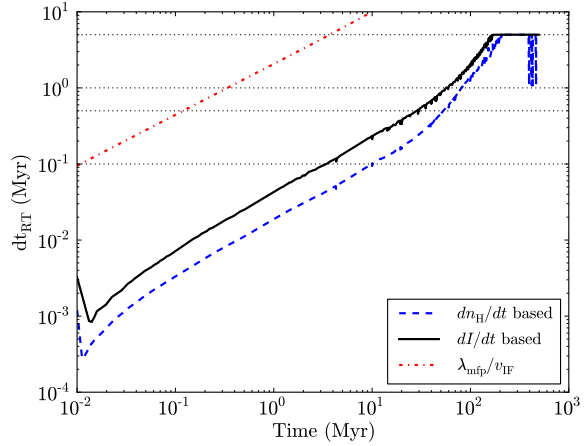}
  \caption{\label{fig:dt_dep2} Variable time-stepping for the methods
    that limit change in neutral fraction (solid) and specific
    intensity (dotted).  The horizontal lines show the constant
    timesteps that were used in the tests.  The crossing time of a
    mean free path by the ionisation front is plotted for reference.}
\end{figure}

The previous three dependencies did not affect the propagation of the
ionisation front greatly.  However in our and others' past experience
\citep[e.g.][]{Shapiro04, Mellema06, Petkova09}, the timestep,
especially too small a one, can drastically underestimate the
ionisation front velocity.  Here we use Test 1 but with $64^3$
resolution to compare different time-stepping methods -- restricted
changes in \hii ($dn_{\rm H}/dt$ based; \S\ref{sec:dt_hi}),
constant timesteps of 0.1, 0.5, 1, and 5 Myr (\S\ref{sec:dt_const}),
and based on incident radiation ($dI/dt$ based; \S\ref{sec:dt_tau}).
The growth of the ionisation front radius is shown in Fig.
\ref{fig:dt_dep1}.  Both the \hii restricted and incident
radiation variable time-stepping methods agree within a few percent
throughout the entire simulation, as does the run with constant $dt_P
= 0.1$ Myr timesteps.  With the larger constant timesteps, the
numerical solution lags behind the analytical one, but they converge
to an accurate \hii radii at late times.  Even $dt_P = 5$ Myr
timestep, which underestimated it by 35\% at 50 Myr, is within a
percent of the analytical solution.

The larger constant timesteps deviate from these more accurate
solutions at early times because the photon energy gradient is large,
and thus so is the ionisation front velocity.  To understand this, the
ionisation front can be considered static in a given timestep.  Here
the ionising radiation can only penetrate into the neutral gas by
roughly a photon mean path $\lambda_{\rm mfp}$.  Only in the next
timestep, the ionisation front can advance.  If the timestep is larger
than $\lambda_{\rm mfp} / v_{\rm IF, anyl}$, then the numerical
solution may fall behind.

The variable time-stepping of the $dn_{\rm H}/dt$ and $dI/dt$ methods
adjust accordingly to the physical situation, as seen in the plot of
timestep versus time (Fig. \ref{fig:dt_dep2}).  They provide high
accuracy when the source first starts to shine.  At later times, the
ionisation front slows as it approaches the Str\"{o}mgren radius, and
large timesteps are no longer necessary.  The $dI/dt$ method has a
similar timestep as the $dn_{\rm H}/dt$ method.  It is larger by a
factor of $\sim 2$ because of our choice of the safety factor $C_{\rm
  RT, cfl} = 0.5$.  This causes its calculated radius to be smaller by
1\% at $t < t_{\rm rec}$, which is still in good agreement with the
analytical value.

\section{Methodology Tests}

Here we show tests that evaluate new features in \moray, such as the
improvements from the geometric correction factor, optically-thin
approximations, treatment of X-ray radiation, and radiation pressure.
Lastly we test for any non-spherical artifacts in the case of two
sources.

\subsection{Improvements from the covering factor correction}
\label{sec:test_fc}

\begin{figure}
  \plotone{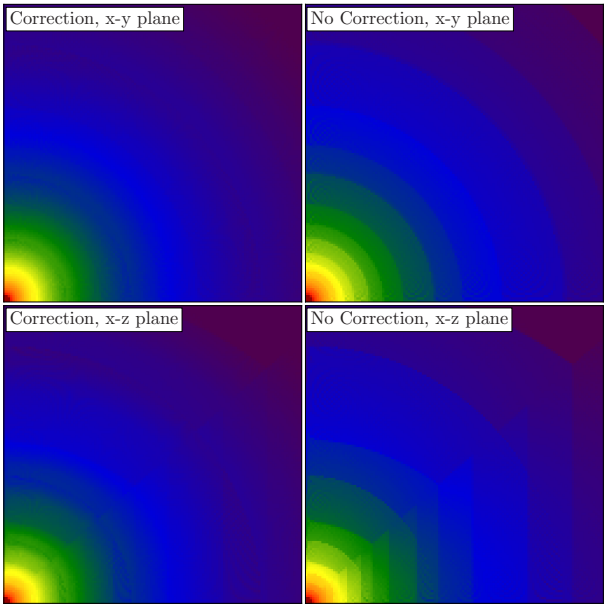}
  \caption{\label{fig:fc_slices} Slices of the photo-ionisation rate
    in the x-y plane (top row) and x-z plane (bottom row) with (left
    column) and without (right column) the geometric correction.  The
    slices are through the origin.  In the x-y plane, it reduces the
    shell artifacts.  In the x-z plane, it reduces the severity of a
    non-spherical artifact delineated at a 45 degree angle, where the
    HEALPix scheme switches from polar to equatorial type pixels.}
\end{figure}

As discussed in \S\ref{sec:meth_fc}, non-spherical artifacts are
created by a mismatch between the HEALPix pixelisation and the
Cartesian grid.  This is especially apparent in optically-thin
regions, where the area of the pixel is greater than the
$(1-e^{-\tau})$ absorption factor.  In this section, we repeat the
angular resolution tests in \S\ref{sec:ang_dep} with $\Phi_c = 5.1$.
Slices of the photo-ionisation rates through the origin are shown in
Fig. \ref{fig:fc_slices}, depicting the improvements in spherical
symmetry and a closer agreement to a smooth $1/r^2$ profile.  Previous
attempts to reduce these artifacts either introduced a random rotation
of the HEALPix pixelisation \citep[e.g.][]{Abel02_RT, Trac07,
  Krumholz07_ART} or by increasing the ray-to-cell sampling.

In the x-y plane without the correction, there exists shell artifacts
where the photo-ionisation rates abruptly drops when the rays are
split.  This occurs because the photon flux in the rays are constant,
so \kph~is purely dependent on the ray segment length through each
cell.  Geometric dilution mainly occurs when the number of rays
passing through a cell decreases.  With the correction, geometric
dilution also occurs when the ray's solid angle only partially covers
the cell.  This by itself alleviates these shell artifacts.  In the
x-z plane without the correction, there is a non-spherical artifact
delineated at a 45 degree angle.  In the lower region, the rays are
associated with equatorial HEALPix pixels, and in the upper region,
they are polar HEALPix pixels.  This artifact is not seen in the x-y
plane because all rays are of a equatorial type.  The geometric
correction smooths this artifact but does not completely remove it.

\subsection{Optically-thin approximation}

\begin{figure}
  \plotone{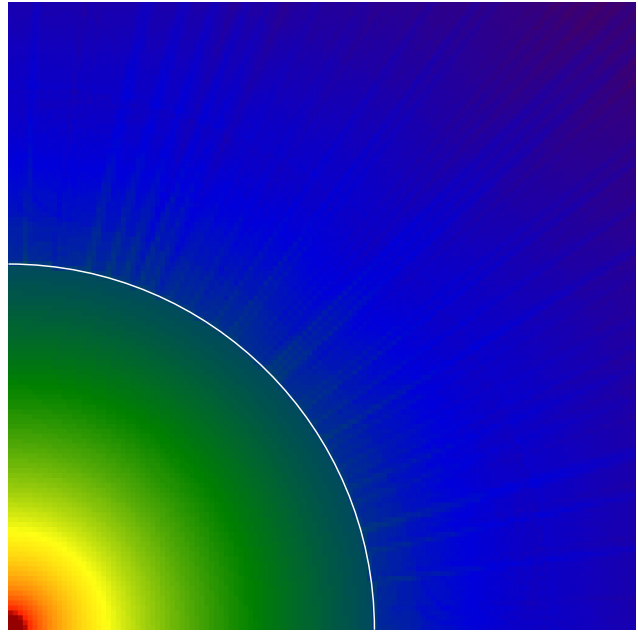}
  \caption{\label{fig:rayopt} Optically-thin approximation to the
    radiation field with one ray per cell in optically thin regions.
    The angular artifacts result from the transition to optically
    thick (white line) at an optical depth $\tau = 0.1$.}
\end{figure}

In practice, we have found it difficult to transition from the
optically-thin approximation to the optically-thick regime without
producing artifacts in the photo-ionisation rate \kph.  We use the
optically thin problem used in the angular resolution test
(\S\ref{sec:ang_dep}) with $\Phi_c = 5.1$ to show these artifacts in
\kph~in Fig. \ref{fig:rayopt}.  The radiation field strictly follows
a $1/r^2$ profile until it reaches $\tau_{\rm thin} \equiv 0.1$, which
is denoted by the white quarter circle in the figure.  Within this
radius, only $\Phi_c = 1$ is required.  Then at this radius, the rays
are then split until a sampling of $\Phi_c$ is satisfied.  Angular
spike artifacts beyond this radius arise because of the interface
between the optically-thin approximation and full ray tracing
treatment.  They originate in cells that intersect the $\tau_{\rm
  thin}$ surface, which are split into the optically thin and thick
definitions.  Unfortunately we have not determined a good technique to
avoid such artifacts.  They occur because of the following reason.
When the first ray with $\tau < 0.1$ exits such a cell, it applies the
optically-thin approximation and marks the cell so no other ray from
the same source contributes to its \kph~and $\Gamma$ field.  However
other rays may exit the cell with $\tau > 0.1$ because the maximum
distance between the far cell faces and the source is not always $\tau
< 0.1$.  Then these rays will split in this cell and add to \kph~and
become attenuated, reducing its photon fluxes.  When the rays continue
to the next cell after this transition, the photon fluxes are not
necessarily equal to each other, creating the angular artifacts seen
in Fig. \ref{fig:rayopt}.  We are continuing to formulate a scheme
that avoids these artifacts because this approximation will be very
advantageous in simulations with large ionised filling factors.

\subsection{X-Ray secondary ionisations and reduced photo-heating}

\begin{figure}
  \plotone{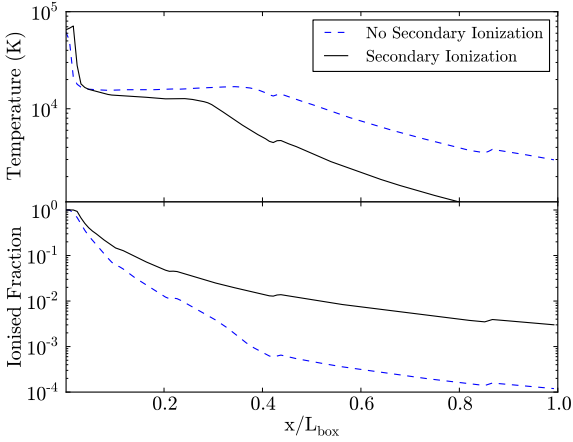}
  \caption{\label{fig:xray_fig} Radial profiles of temperature and
    ionised fraction showing the effects of secondary ionisations from
    a mono-chromatic 1 keV spectrum.  The discontinuity at $r \sim
    0.4$ is caused by artifacts in the ray tracing, which is described
    in Test 1.  The high energy photons can ionise multiple hydrogen
    atoms, increasing the ionised fraction.  In part, less radiation
    goes into thermal energy, lowering the temperature.}
\end{figure}

Here we test our implementation of secondary ionisations from
high-energy photons above 100 eV, described in \S\ref{sec:xrays} and
used in \citet{Alvarez09} in the context of accreting black holes.  We
use the same setup as Test 5 but with an increased luminosity $L =
10^{50}$ erg s$^{-1}$ and a mono-chromatic spectrum of 1 keV.  Fig.
\ref{fig:xray_fig} compares the density, temperature, ionised
fraction, and neutral fraction of the expanding \hii region
considering secondary ionisations and reduced photo-heating and
considering only one ionisation per photon and the remaining energy
being thermalised.

Fig. \ref{fig:xray_fig} shows the main effects of secondary
ionisations from the 1 keV spectrum on the ionisation and thermal
state of the system.  Without secondary ionisations, each absorption
results in one ionisation with the remaining energy transferred into
thermal energy.  But with secondary ionisations, recall that most of
the radiation energy goes into hydrogen and helium ionisations in
neutral gas; whereas in ionised gas, most of the energy is
thermalised.  In this test, only the inner 300 pc is completely
ionised because of the small cross-section of hydrogen at $E_{\rm ph}
= 1$ keV.  Beyond this core, the medium is only partially ionised.
This process expands the hot $T = 10^5$ K core by a factor of 2.  In
the outer neutral regions, the ionisation fraction is larger by a
factor of $\sim 10$, which in turn results in less photo-heating,
lowering the temperature by a factor 2--3.

\subsection{Radiation pressure}

\begin{figure}
  \plottwo{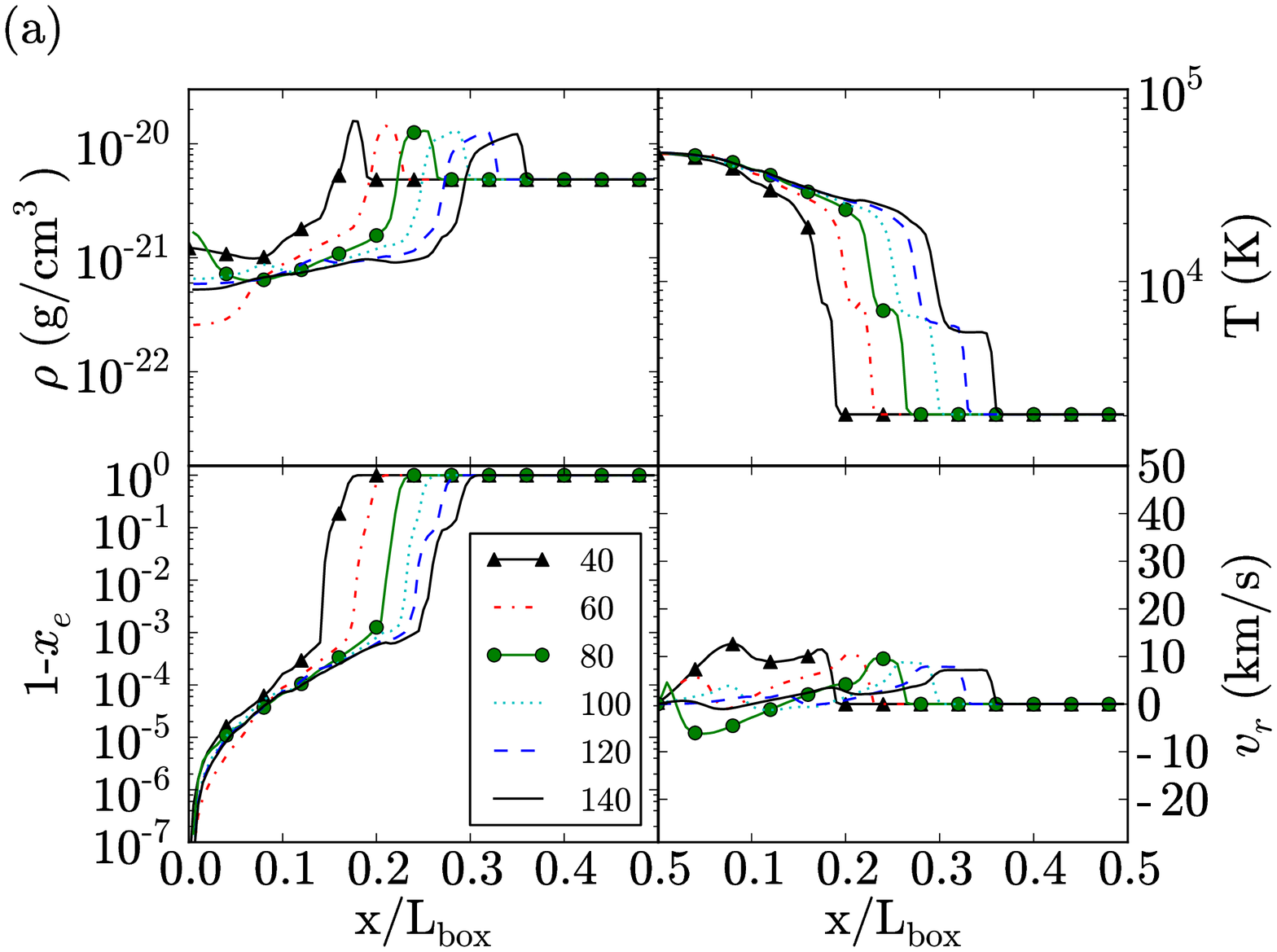}{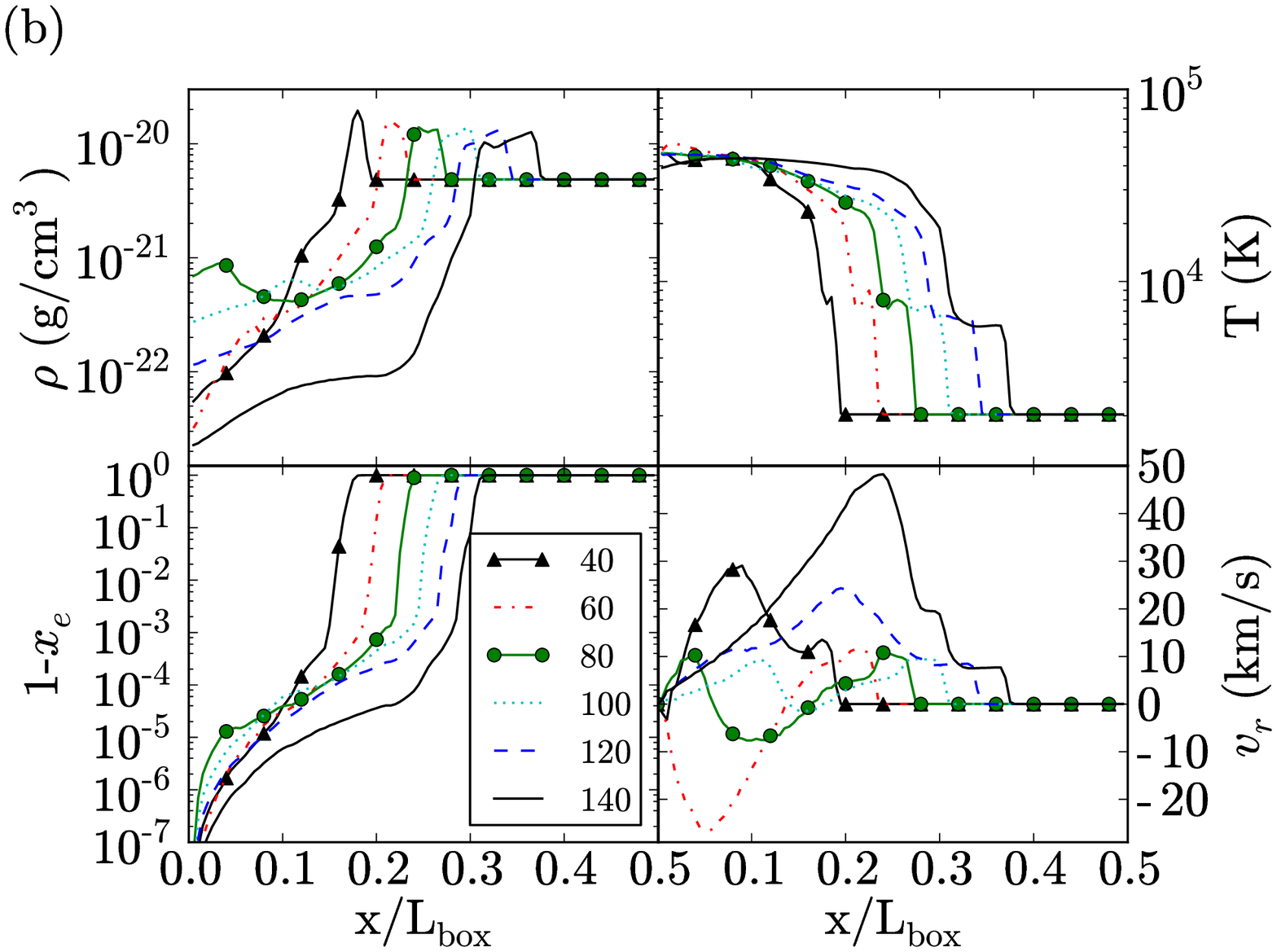}
  \caption{\label{fig:rp_profiles} (a) No radiation pressure.
    Radial profiles of (clockwise from top left) density, temperature,
    radial velocity, and neutral fraction.  Time units in the legend
    are in kyr.  (b) Radial profiles with radiation pressure.  The
    momentum transferred to the gas drives out the gas at higher
    velocities than without radiation pressure.  Afterwards the
    central region is under-pressurised, and the gas infalls toward the
    centre, as seen at t = 60 kyr.  Then the radiation pressure
    continues to force the gas outwards, increasing gas velocities up
    to 50 km/s.}
\end{figure}

Radiation pressure affects gas dynamics in an \hii region when
its force is comparable to the acceleration created by gas pressure of
the heated region.  The imparted acceleration on a hydrogen atom
$\mathbf{a}_{\rm rp} = E_{\rm ph}/c$.  This is especially important
when the ionisation front is in its initial R-type phase, where the
gas has not reacted to the thermal pressure yet.  Thus we construct a
test that focuses on a small scales, compared to the Str\"{o}mgren
radius.  The domain has a size of 8 pc with a uniform density $\rho =
2900\;\cubecm$ and initial temperature $T = 10^3$ K.  The source is
located at the origin with a luminosity $L = 10^{50}$ ph s$^{-1}$ and
a $T=10^5$ K blackbody spectrum.  We use one energy group $E_{\rm ph}
= 29.6$ eV.  The grid is adaptively refined on overdensity with the
same criterion as Test 8.  The simulation is run for 140 kyr.

We compare nearly identical simulations but one with radiation
pressure and one without radiation pressure to quantify its effects.
Radial profiles of density, temperature, neutral fraction, and radial
velocity are shown in Fig. \ref{fig:rp_profiles} for both
simulations at several times.  Without radiation pressure, the
evolution of the \hii is matches the analytical expectations
described in \S\ref{sec:test5}.  At $t=140$ kyr with radiation
pressure, the ionisation front radius is increased by $\sim5\% = 0.16$
pc.  However radiation pressure impacts the system the greatest inside
the ionisation front.  At $t = 40$ kyr, the central density is smaller
by a factor of 20 with radiation pressure, but the temperatures are
almost equal.  A rarefaction wave thus propagates toward the centre,
depicted by the negative radial velocities at $t = 60$ kyr.  This
raises the central density to $10^{-21}$ g \cubecm~at $t = 80$ kyr.
Afterwards, the radiation continues to force gas outwards.  From $t =
100$ kyr to $t = 140$ kyr, the maximum radial velocity of the ionised
gas increases from 10 \kms~to 50 \kms.  This leaves behind an even
more diffuse medium, lowering the gas density by a factor of 10 at $t
= 140$ kyr.  Thus the recombination rates are lower, resulting in
increased ionisation fractions and temperatures in the \hii
region.

\subsection{Consolidated \hii region with two sources}

\begin{figure}
  \plotone{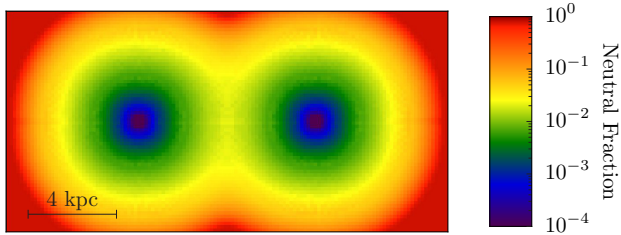}
  \caption{\label{fig:twosrc} Slice of neutral fraction at $t = 500$
    Myr through the sources in the consolidated \hii test.
    There are no artifacts associated with rays being emitted from two
    sources.  Both of the ionised regions are spherically symmetric
    before they overlap.}
\end{figure}

Here we test for any inaccuracies in the case of multiple sources.  We
use the same test problem as \citet[][\S5.1.2]{Petkova09}, which has
two sources with luminosities of $5 \times 10^{48}$ ph s$^{-1}$ and
are separated by 8 kpc.  The ambient medium is static with a uniform
density of $10^{-3}$ \cubecm and $T = 10^4$ K.  This setup is similar
to Test 1.  The domain has a resolution of $128 \times 64 \times 64$
It is 20 kpc in width and is 10 kpc in height and depth.  The problem
is run for 500 Myr.

The \hii regions grow to $r = 4$ kpc where they overlap.  Then
the two sources are enveloped in a common, elongated \hii
region.  To illustrate this, we show the neutral fraction in Fig.
\ref{fig:twosrc}.  Our method keeps spherical symmetry close to the
individual sources, and there are no perceptible artifacts from having
multiple sources.

\section{Parallel Performance}

Last we demonstrate the parallel performance of \moray~in weak and
strong scaling tests.  For large simulations to consider radiative
transfer, it is imperative that the code scales to large number of
processors.

\subsection{Weak Scaling}
\label{sec:weak_sc}

\begin{figure}
  \plotone{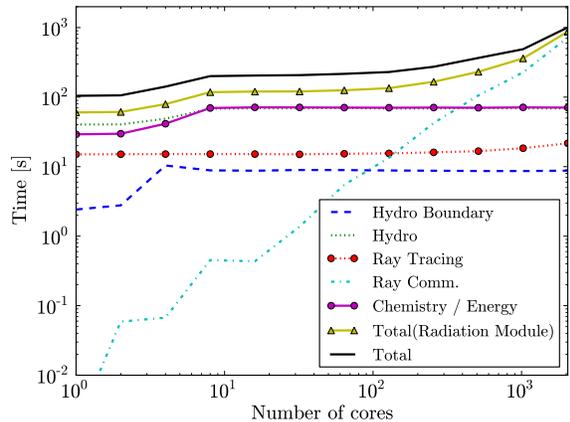}
  \caption{\label{fig:weak} Weak scaling test with one $64^3$ block
    per process.  Each block has one source and is set up similar to
    the radiation hydrodynamics Test 5.  Above 8 processes, all parts
    of the code exhibit good weak scaling except for the
    inter-processor ray communication.  The radiation module timing
    include the ray tracing, communication, chemistry and energy
    solver, and all other overheads associated with the radiation
    transport.  Cache locality of the data causes the decrease in
    performance from 1 to 8 processes.}
\end{figure}

Weak scaling tests demonstrates how the code scales with the number of
processors with a constant amount of work per processor.  Here we
construct a test problem with a $64^3$ block per core.  The grid is
not adaptively refined.  The physical setup of the problem is nearly
the same as Test 5 with a uniform density $\rho = 10^{-3} \;\cubecm$
and initial temperature $T = 100$ K.  Each block has the same size of
15 kpc as Test 5.  At the centre of each grid, there exists a
radiation source with a luminosity $L = 5 \times 10^{48}$ ph s$^{-1}$
and a 17 eV mono-chromatic spectrum.  The problem is run for 250 Myr.
We run this test with $N_p = 2^n$ cores with $n = [0,1 \dots 10,11]$.
The domain has $(N_x, N_y, N_z)$ blocks that is determined with the
MPI routine \texttt{MPI\_Dims\_create}.  For example with $n = 7$, the
problem is decomposed into $(N_x, N_y, N_z) = (4,4,8)$ blocks,
producing a $256 \times 256 \times 512$ grid.  We have run these on
the original (Harpertown CPUs) nodes of the NASA NAS machine,
Pleiades, with 8 cores per node.  Fig. \ref{fig:weak} shows the
performance timings of various parts of the code.  From one to two
cores, the total time only increases by $\sim$1\% due to the overhead
associated with the inter-processor message passing.  From two to
eight cores, the performance decreases in the hydrodynamics solver,
chemistry and energy solver, and obtaining the hydrodynamic boundary
conditions because of cache locality problems of the data being passed
to the CPU.  This occurs because of the CPU architecture, specifically
the L1 cache and core connectivity.  We see less of a penalty in newer
processors, e.g. Intel Nehalam and Westmere CPUs.  Above eight
processes, these routines exhibit near perfect weak scaling to 2048
cores.  Unfortunately there exists a $N_p^{1.5}$ dependence in the ray
communication routines.  It becomes the dominant process above 512
processes.  We are actively pursing a solution to this scaling
problem.  The other parts of the code exhibit excellent weak scaling.
Overall it scales already well to 512 processes, and there is room to
enhance the weak scalability of \moray~in the near future.

\subsection{Strong Scaling}
\label{sec:strong_sc}

\begin{figure}
  \plotone{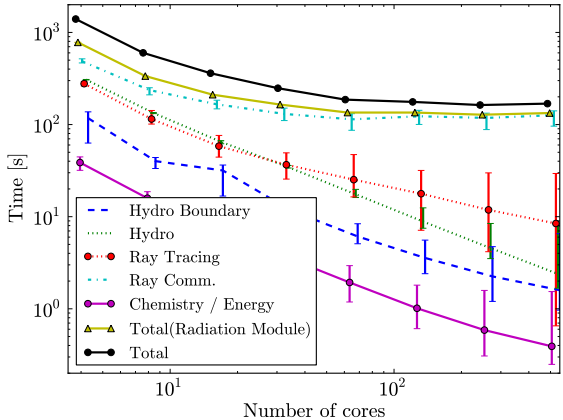}
  \caption{\label{fig:strong} Strong scaling test with a $256^3$ AMR
    calculation of cosmological reionisation with $\sim392^3$ zones.
    The error bars represent the minimum and maximum time spent on a
    single core and measures load balancing.  Each point has been
    slightly offset to make the error bars distinguishable.  The
    hydrodynamics and non-equilibrium chemistry solvers scale well.
    The communication of rays does not scale well in this problem and
    limits the performance, where as the ray tracing scales relatively
    well.}
\end{figure}

Strong scaling tests shows how the problem scales with the number of
processors for the same problem.  The overhead associated with the
structured AMR framework in \enzo~can limit the strong scalability.
One key property of strong scaling is that each processor must have
sufficient work to compute, compared to the communication involved.
In our experience, non-AMR calculations exhibit much better strong
scaling than AMR ones because of reduced inter-processor
communication.  We use an AMR simulation to demonstrate the scalability
of \moray~in a demanding, research application.  Here we use a
small-box reionisation calculation with $L_{\rm box} = 3$ Mpc/h, a
resolution of $256^3$, and six levels of refinement.  We measure the
time spent on the hydrodynamics, non-equilibrium chemistry, ray
tracing, and radiation transport communication in a timestep lasting 1
Myr, at $z=10$.  There are nine radiative transfer timesteps in this
period.  The box has 675 point sources, 15,943 AMR grids, and $6.01
\times 10^7 \approx 392^3$ computational cells in this calculation at
this redshift.  On average, $4.6 \times 10^8$ ray segments are traced
each timestep.  The ionised volume fraction is 0.10.  The calculations
are performed on the Nehalam nodes on Pleiades on 2$^n$ cores, where
$n = [2,...,9]$.

Fig. \ref{fig:strong} shows the strong scaling results of this
calculation.  The hydrodynamics and non-equilibrium chemistry routines
scale very well in this range because they depend on local phenomena.
Note that the dominant process is the radiation transport instead of
the hydrodynamics when compared to the weak scaling tests.  The error
bars in the figure represent the deviations across all cores, and it
shows that load balancing becomes an issue at 128 and 256 cores.  This
is an even larger problem for the ray tracing because the grids that
host multiple radiation sources will have significantly more work than
a distant grid.  For example if a few central grids host several point
sources, the exterior grids must wait until those grids are finished
ray tracing in order to receive the rays.  This idle time constitutes
the majority of the time spent in the ray communication routines
despite our efforts (see \S\ref{sec:parallel}) to minimize idle time.
For this reason, the communication of rays does not scale in this
problem.  One solution is to split the AMR grids into smaller blocks
based on the ray tracing work.  This could impact the performance of
the rest of \enzo~by increasing the number of boundaries and thus
communication.  Nevertheless in simulations that are dominated by
radiation transport, it will be advantageous to construct such a
scheme to increase the feasibility of running larger problems.

\section{Summary}

In this paper, we have presented our implementation, \moray, of
adaptive ray tracing \citep{Abel02_RT} and its coupling to the
hydrodynamics in the cosmology AMR code \enzo, making it a fully
functional radiation hydrodynamics code.  As this method is photon
conserving, accurate solutions are possible with coarse spatial
resolution.  A new geometric correction factor to ray tracing on a
Cartesian grid was described, and it is general to any implementation.
We have exhaustively tested the code to problems with known analytical
solutions and the problems presented in the RT06 and RT09 radiative
transfer comparison papers.  Additionally we have tested our code with
more dynamical problems -- champagne flows, Rayleigh-Taylor
instabilities, photo-evaporation of a blastwave, beamed radiation, a
time-varying source, and an \hii region with MHD -- to
demonstrate the flexibility and fidelity of \moray.  Because
production simulations may not have the resolution afforded in these
test problems, we have tested the dependence on spatial, angular,
frequency, and temporal resolution.  It provides accurate solutions
even at low resolution, except for the large constant timesteps.
However, we have described two methods to determine the radiative
transfer timestep that are based on the variations in specific
intensity or changes in neutral fraction inside the ionisation front.
Both methods give very accurate results and provide the largest
timestep to obtain an accurate solution, ultimately leading to higher
computational efficiency.  On the same topic, we have described a
method to calculate the radiation field in the optically-thin limit
with ray tracing.

Being a ray tracing code, it scales with the number of radiation
sources; nevertheless, it scales well to O($10^3$) processors for
problems with $\sim10^9$ computational cells and $\sim10^4$ sources,
such as reionisation calculations.  We have also shown that the code
shows good strong scaling in AMR calculations, given a large enough
problem.  The combination of AMR and adaptive ray tracing allows for
high-resolution and high-dynamical range problems, e.g. present-day
star formation, molecular cloud resolving cosmological galaxy
formation, and \hii regions of Population III stars.  Furthermore, we
have included Lyman-Werner absorption, secondary ionisations from
X-ray radiation, Compton heating from photon scattering, and radiation
pressure into the code, which extends the reach of \moray~to study
AGN feedback, stellar winds, and local star formation.  Coupling the
radiative transfer with MHD further broadens the applicability of our
code.  The full implementation is included in the latest public
version of \enzo\footnote{http://enzo.googlecode.com}, providing the
community with a full-featured radiation hydrodynamics AMR code.

\section*{Acknowledgments}

We thank Ji-hoon Kim, Richard Klein, and Jeff Oishi for helpful
comments on the manuscript.  J.~H.~W. is supported by NASA through
Hubble Fellowship grant \#120-6370 awarded by the Space Telescope
Science Institute, which is operated by the Association of
Universities for Research in Astronomy, Inc., for NASA, under contract
NAS 5-26555.  Computational resources were provided by NASA/NCCS award
SMD-09-1439.  The majority of the analysis and plots were done with
\texttt{yt} \citep{yt_full_paper}.

\bibliography{ms}

\bsp

\label{lastpage}

\end{document}